\documentclass[preprint,journal]{vgtc}            

\onlineid{0}

\vgtccategory{Research}

\vgtcpapertype{system}

\title{GraphPolaris: A System for Query, Analysis, and\newline Visualization of Graph Databases}

\author{%
  \authororcid{Michael Behrisch}{0000-0002-1102-103X},
  \authororcid{Sjoerd Vink}{0009-0006-1094-3725},
  \authororcid{Leonardo Christino}{0000-0002-8754-8460}, and
  \authororcid{Remco Chang}{0000-0002-6484-6430}
}

\authorfooter{
  \item
  	Michael Behrisch is with Utrecht University and GraphPolaris.
  	E-mail: m.behrisch@uu.nl
  \item
  	Sjoerd Vink is with Utrecht University and GraphPolaris.
  	E-mail: s.a.vink@uu.nl
  \item
  	Leonardo Christino is with GraphPolaris.
  	E-mail: lchristino@graphpolaris.com
  \item
  	Remco Chang is with Tufts University and GraphPolaris.
  	E-mail: remco@cs.tufts.edu
}

\abstract{%
  Graph databases are increasingly adopted as alternatives to tabular, aggregation-focused data models used in business intelligence (BI) systems such as Tableau, Power BI, and Looker.
  They capture complex relationships between entities, processes, and events, enabling analysis of information propagation in networks.
  As a result, graph analysis is central to applications such as fraud detection, social influence analysis, and supply chain resilience.
  Despite these advantages, existing tools do not adequately support interactive analysis of graph databases. 
  Tabular BI systems lack mechanisms for reasoning over nodes and edges, while graph databases require specialized query languages and fragmented workflows that hinder accessibility.
  We present \toolname, a no-code Visual Analytics system that enables users to explore, analyze, and visualize graph databases without programming skills.
  At its core, \toolname features the \textsc{GraphPolaris Query Language} (\GPQL), a formal query grammar that facilitates flexible and composable graph queries, providing a formal foundation for analyzing relationships and graph patterns.
  \GPQL serves as an intermediary between user interactions and the underlying database.
  Its formal foundation enables no-code query construction, database-agnostic query generation, and guarantees that every interaction produces a valid executable query.
  Informed by a formative user study, we designed \toolnames interface and visualizations to lower technical barriers and foster iterative, collaborative exploration of complex networks.
  \revised{We evaluate \toolname through two real-world case studies in telecommunications and supply-chain analysis and a 22-month-long formative mixed-method study, including a MILC-based assessment of its fit to analysts’ graph analytics workflows.}

}

\keywords{Graph Exploration, Multivariate Graphs, Graph Databases, No-Code Analytics}

\teaser{
  \centering
  \includegraphics[width=\linewidth]{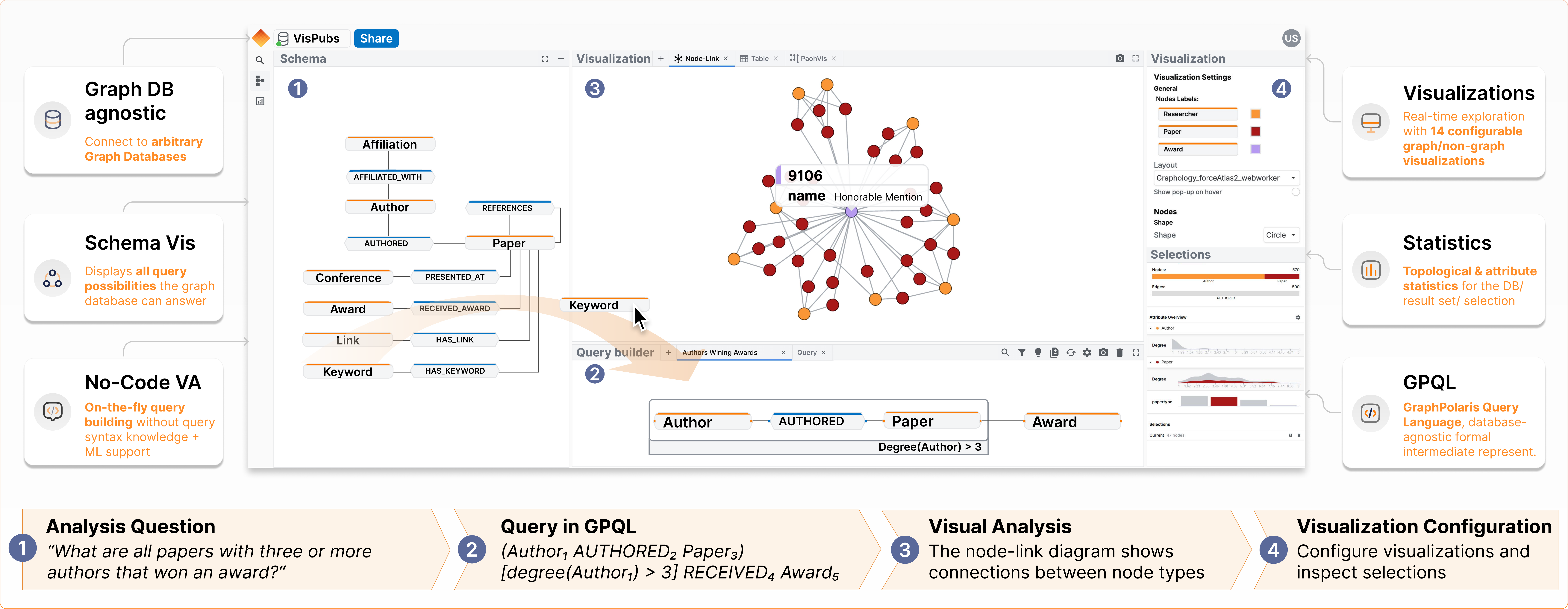}
  \vspace{-1.5em}
  \caption{\toolname democratizes access to graph databases by enabling users to analyze data through a no-code Visual Analytics system. (1) The \textbf{Schema Panel} visualizes the graph schema, providing an overview of the query space supported by the data. (2) The \textbf{Visual Query Builder} features a visual query language for constructing graph queries, and (3) the \textbf{Visualization Panel} displays query results through structural and attribute visualizations. (4) The \textbf{Selection \& Configuration} panel allows drilling down into selections from the Visualization Panel.}
  \label{fig:teaser}
}

\graphicspath{{figs/}{figures/}{pictures/}{images/}{./}} 

\usepackage{tabu}                      
\usepackage{booktabs}                  
\usepackage{lipsum}                    
\usepackage{mwe}  
\usepackage{float}
\usepackage{graphicx}
\usepackage{subcaption}
\usepackage{pdflscape}
\usepackage{caption}
\usepackage[most]{tcolorbox}

\usepackage{tabularx}
\usepackage{longtable}
\usepackage{multirow}

\newcommand{\toolname}{\texttt{GraphPolaris}\xspace}
\newcommand{\toolnames}{\texttt{GraphPolaris'}\xspace}
\newcommand{\GPQL}{\texttt{GPQL}\xspace}
\newcommand{\grammar}[1]{\texttt{#1}}

\newcommand{\revised}[1]{#1}

\definecolor{mred}{rgb}{.80,.12,.30}
\definecolor{grey}{rgb}{0.5,0.5,0.5}
\definecolor{Purple}{rgb}{.75,0,.85}
\definecolor{BlueGreen}{rgb}{.05,.59,.73}
\definecolor{amber}{rgb}{1.0, 0.75, 0.0}
\definecolor{steelblue}{rgb}{0.27, 0.51, 0.71}
\definecolor{violet}{rgb}{0.56, 0.0, 1.0}
\definecolor{myorange}{rgb}{0.94, 0.36, 0.13}

\newcommand{\sectiontitle}[1]{\smallskip\noindent\textbf{#1} \hspace{2pt}}
\definecolor{systemred}{HTML}{9d311e}
\definecolor{systemblue}{HTML}{0374c4}
\definecolor{systemgreen}{HTML}{76d079}

\usepackage{mathptmx}                  

\usepackage[english]{babel}

\addto\extrasenglish{%
}

\usepackage{colortbl}      

\newlength{\namecolwidth}
\newlength{\shortcolwidth}
\usepackage{fontawesome5} 

\newcommand{\iconyes}{\textcolor{orange}{\faCheck}}
\newcommand{\iconno}{\textcolor{grey}{\faTimes}}

\definecolor{greyLow}{rgb}{0.80,0.80,0.80}   
\definecolor{greyMed}{rgb}{0.55,0.55,0.55}   
\definecolor{greyHigh}{rgb}{0.25,0.25,0.25}  

\definecolor{orangeLow}{rgb}{0.97, 0.70, 0.55}   
\definecolor{orangeMed}{rgb}{0.96, 0.53, 0.34}   

\definecolor{orangeBase}{rgb}{0.94, 0.36, 0.13}

\definecolor{orangeHigh}{rgb}{0.80, 0.26, 0.08}  
\definecolor{orangeDark}{rgb}{0.65, 0.18, 0.05}  

\newcommand{\iconlow}{\textcolor{orangeLow}{\faCircle}}   
\newcommand{\iconhigh}{\textcolor{orangeHigh}{\faCircle}} 

\newcommand{\iconscientist}{\textcolor{orangeLow}{\faFlask}}
\newcommand{\iconanalyst}{\textcolor{orangeMed}{\faProjectDiagram}}
\newcommand{\iconexpert}{\textcolor{orangeHigh}{\faUserTie}}

\usepackage[normalem]{ulem} 

\usepackage{tikz}
\newcommand\submittedtext{%
  \footnotesize
  This work has been submitted to the IEEE for possible publication.
  Copyright may be transferred without notice, after which this version
  may no longer be accessible.%
}
\newcommand\submittednotice{%
  \begin{tikzpicture}[remember picture,overlay]
    \node[anchor=south,yshift=10pt] at (current page.south)
    {\fbox{\parbox{\dimexpr0.65\textwidth-\fboxsep-\fboxrule\relax}{%
      \submittedtext}}};
  \end{tikzpicture}%
}

\begin{document}


\maketitle
\submittednotice

\section{Introduction}
\label{sec:introduction}

Traditional business intelligence platforms, such as Tableau, Power BI, and Looker, are built on tabular data models optimized for aggregation-based analysis.
These systems condense large datasets into summarized extracts or totals, enabling analysts to identify high-level trends and patterns~\cite{hogan2021knowledge, zhong2023comprehensive}.
While effective in many business contexts, this aggregation paradigm obscures the fine-grained, connection-driven structures that underlie many real-world phenomena.
For instance, fraud detection in financial services depends on tracing direct and indirect links between transactions and accounts~\cite{perozzi2014focused, zhou2009graph}, yet aggregated views conceal the relational patterns that indicate fraudulent behavior.
Graph data models provide a semantically richer representation by explicitly modeling entities and their relationships, enabling analysis of both structure and attributes.
This representation supports the identification of interaction networks, latent structures, and anomalies that remain hidden in aggregated views.
Applications include social network analysis for identifying influential individuals and communities~\cite{aggarwal2011introduction}, supply chain optimization for sustainability and compliance~\cite{mageto2021big}, knowledge graph construction for search and retrieval~\cite{edge2024local}, and network and IT infrastructure management~\cite{abu2021domain}.

The shift from tabular representations to attributed graph data models introduces challenges in retrieval, exploration, and interpretation.
We assume the property graph model, in which nodes and edges carry arbitrary key-value attributes.
Despite technical expertise in related domains, many analysts encounter fragmented workflows, limited visual guidance, and high entry barriers when working with graph databases.
\revised{Existing tools require coding, rely on disconnected software stacks, and offer inconsistent visualization support, complicating iterative exploration.}
These limitations hinder adoption, restrict the discovery of meaningful relationships, and increase the risk of overlooking critical patterns.
Prior work in VA has advanced graph visualization and interaction techniques~\cite{nobre2019state, mcgee2019state, kerren2014multivariate}, yet a broadly adopted system for exploratory graph analysis similar to Polaris~\cite{stolte2002polaris} remains absent~\cite{li2023knowledge}.
Commercial systems~\cite{neo4j_bloom, tigergraph} and academic platforms~\cite{shannon2003cytoscape, gephi, auber2017tulip} address parts of this gap, but most rely on node-link diagrams, lack live database querying, and do not support standardized collaboration.
Other systems are often domain-specific, which limits their generalizability.

\revised{
We introduce \toolname, a no-code visual analytics system for interactive querying, analysis, and visualization of graph databases.
At its core is the \toolname \texttt{Query Language} (\GPQL), a formal query grammar that abstracts graph query languages, maps directly to interface elements, and is exposed through a Visual Query Builder for direct manipulation on a graphical canvas.
Analysts construct queries by composing nodes, edges, filters, and functions; each interaction yields a valid, executable query that can be transpiled into multiple languages (e.g., Cypher, GQL, GSQL) and enables iterative exploration without writing code.
Similar to Polaris~\cite{stolte2002polaris}, \toolname is thus best understood as a systems contribution to visual analytics, centered on the conceptual contribution of \GPQL and its grammar-driven interaction model that unifies query abstraction, interactive construction, and visualization into a single workflow.
By explicitly focusing on query abstraction and workflow integration, our work complements prior research in graph visualization \cite{nobre2019state} and (graph) visual analytics systems \cite{dunne2012graphtrail, DBLP:journals/tvcg/WenFHGLW26}, rather than proposing a new visual encoding. It addresses a critical gap between expressive graph query languages and accessible visual analytics, enabling analysts to explore complex relational data without query-programming expertise. In this paper, we contribute:

\smallskip
\begin{description}[noitemsep, topsep=0pt, parsep=0pt, partopsep=0pt]
  \item[Conceptual Contribution:] \GPQL, a formal query grammar that abstracts over existing graph query languages and enables vendor-agnostic, direct-manipulation query construction.
  \item[System Contribution:] \toolname, a no-code visual analytics environment that integrates schema exploration, query construction, execution, and enables graph visualization (research).
  \item[Empirical Contribution:] Two real-world case studies and practitioner feedback on applicability, usability, and limitations.
\end{description}
}

\smallskip
\revised{
We evaluate \toolname using two real-world use cases from telecommunications and supply-chain analysis. Telecommunications analysts want to examine whether handover patterns between cell towers correlate with degraded service quality, i.e., increased latency. Supply-chain analysts need to assess direct and indirect tariff impacts, identify structural vulnerabilities such as critical dependencies and single points of failure, and explore mitigation strategies. These cases, together with a 22-month formative deployment and MILC-based analysis, demonstrate \toolname's ability to support graph analytics across different domains. We further report qualitative practitioner feedback on accessibility and usability, as well as limitations related to visual-grammar alignment, schema complexity, and scalability. These findings identify directions for future work.
}


\section{Related Work}
\label{sec:relatedwork}

\toolname builds on three areas: systems for interactive data exploration, grammar-based abstractions for Visual Analytics systems, and graph visualization techniques.

\begin{table}[ht]
    \small
    \centering
    \setlength{\tabcolsep}{2pt}
    \renewcommand{\arraystretch}{1.2}
    \begin{tabularx}{\linewidth}{
        >{\raggedright\arraybackslash}p{0.17\linewidth}  
        >{\centering\arraybackslash}p{0.07\linewidth}    
        >{\centering\arraybackslash}p{0.11\linewidth}   
        >{\centering\arraybackslash}p{0.13\linewidth}    
        >{\raggedright\arraybackslash}p{0.24\linewidth}  
        >{\centering\arraybackslash}p{0.08\linewidth}    
        >{\centering\arraybackslash}p{0.08\linewidth}    
    }
        \toprule
        \textbf{Name} &
        \rotatebox{60}{\textbf{No-Code Interact.}} &
        \rotatebox{60}{\textbf{Formal Query}} &
        \rotatebox{60}{\textbf{Direct Retrieval}} &
        \rotatebox{60}{\textbf{Visualizations}} &
        \rotatebox{60}{\textbf{Collaboration}} &
        \rotatebox{60}{\textbf{ML Support}} \\
        \midrule
        Neo4j Bloom \newline \iconanalyst, \iconexpert
          & (\iconyes) & Cypher   & Neo4j
          & Node-Link (NL) diagram, histogram
          & \iconlow  & \iconno \\
        Linkurious \newline \iconanalyst, \iconexpert
          & (\iconyes) & Cypher, Gremlin & Agnostic
          & NL, (time) histogr.
          & \iconhigh & \iconlow \\
        Graphistry \newline \iconscientist, \iconanalyst
          & \iconno  & Cypher, (G)SQL & Neo4j, TigerGraph
          & NL, scatterpl., histogr.
          & \iconlow & \iconhigh \\
        GraphTrail \newline \iconscientist
          & \iconno  & \iconno  & \iconno
          & NL, matrix, hybrid bar chart, tag cloud
          & \iconno  & \iconno \\
        Cytoscape \newline \iconscientist
          & \iconno  & \iconno  & \iconno
          & NL, bar/pie/line charts on nodes
          & \iconno  & \iconlow \\
        Vistorian \newline \iconscientist
          & \iconno & \iconno  & \iconno
          & NL, matrix/lists, (time-)arc, map
          & \iconno  & \iconno \\
        \rowcolor{gray!8}
        GraphPolaris \newline \iconscientist, \iconanalyst, \iconexpert
          & \iconyes & GPQL     & Agnostic
          & 14 types, incl. NL, matrix, arc, map, calendar, histogr.
          & \iconhigh & \iconhigh \\
        \bottomrule
    \end{tabularx}
    \vspace{-0.33em}
    \caption{\textbf{Positioning GraphPolaris relative to commercial and research systems}: \toolname combines non-templated no-code querying, a formal query definition language, database-agnostic execution, research-grade visualization techniques, and support for both  \iconscientist~scientists, \iconanalyst~analysts, and \iconexpert~domain experts alike.}
    \vspace{-2.5em}
    \label{tab:graph_systems_comparison}
\end{table}

\subsection{Systems for Interactive Data Exploration}

Data exploration systems support the analysis and interpretation of datasets through interactive interfaces.
We define these systems as tools that span the full exploration process, from querying graph databases to visualizing results.
One of the most influential examples in this field, and a major source of inspiration for \toolname, is Polaris~\cite{stolte2002polaris}.
Polaris introduced a visual specification paradigm that shaped a large class of business intelligence tools.
Broader surveys~\cite{DBLP:conf/ieeevast/ZhangSBMSPWLK12, DBLP:journals/tvcg/BehrischSSSMWMP19} have cataloged the strengths and limitations of these systems, which remain largely focused on tabular data.

\vspace{0.5em}\noindent\textbf{Academic Systems for Graph Exploration.}
Several research systems aim to fill this gap (see: \autoref{tab:graph_systems_comparison}). Tools such as Vistorian~\cite{molinero2017understanding} offer interactive network visualizations tailored for historical and dynamic networks. VisKonnect~\cite{latif2021visually} maps relationships among historical figures through event co-occurrence, while Ahmad et al.~\cite{ahmad2021towards} visualize patient histories in biomedical datasets. Cytoscape~\cite{shannon2003cytoscape}, a widely used tool in life sciences, integrates molecular interaction networks with associated data.
Tulip~\cite{auber2004tulip} is an open-source framework for large graph visualization and analysis, widely used in research for scalable interactive exploration.
PivotGraph~\cite{wattenberg2006visual} aggregates nodes/edges to simplify exploration of large graphs.
GraphTrail~\cite{dunne2012graphtrail} supports provenance-aware graph analysis workflows with dynamic filtering and summarization.
\revised{Envisage~\cite{DBLP:journals/tvcg/WenFHGLW26} focuses on graph pattern matching and querying by supporting underspecified query intent, repetitive substructures, and flexible attribute constraints through a visual query builder.} LinkQ~\cite{li2024linkq} leverages large language models to translate natural language questions into knowledge graph queries, complementing visual query previews with ground-truth KG data to improve reliability.
Despite their utility, most of these systems are domain-specific, often rely on batch processing, and lack live database connectivity, making them ill-suited for general-purpose or real-time graph exploration.

\vspace{0.5em}\noindent\textbf{Commercial Systems for Graph Exploration.}
Commercial systems have attempted to address this gap (see: \autoref{tab:graph_systems_comparison}). Neo4j Bloom~\cite{neo4j_bloom}, a visual browser for the Neo4j database, enables analysts to construct Cypher queries and visualize results as node-link diagrams. Other tools, such as TigerGraph~\cite{tigergraph} and Gephi~\cite{gephi}, also facilitate graph exploration. Beyond these, a diverse ecosystem of commercial graph tools exists, including \href{https://www.graphistry.com}{Graphistry}, \href{https://linkurious.com}{Linkurious}, \href{https://www.yworks.com}{yFiles}, \href{https://kineviz.com}{Kineviz}, and \href{https://puppygraph.com}{PuppyGraph}, which offer scalable Visual Analytics or domain-specific features.
However, these systems pose a significant technical barrier for non-technical practitioners, requiring analysts to construct complex queries through code. Moreover, the lack of immediate feedback during iterative query construction makes exploration cumbersome.
Most commercial solutions offer limited analytical capabilities, restricting analysts to a small set of fixed visualizations and providing little support for iterative sense-making. They rarely integrate mechanisms for feedback (e.g., refining queries based on intermediate insights) or verification (e.g., checking whether results align with analytical intent), both essential for iterative Visual Analytics workflows.

Despite advancements, graph exploration systems still struggle with scalability, support for dynamic graphs, and flexibility in representing diverse graph structures and relationships~\cite{lissandrini2022knowledge, li2023knowledge}. These limitations remain central obstacles in making graph analytics broadly accessible.
By addressing this gap between research-oriented systems with rich analytical capabilities and commercial systems that prioritize flexibility over usability, \toolname contributes a unifying approach that combines expressive power with practical usability. By enabling no-code graph querying and instant result visualizations, \toolname bridges the divide between analytical depth and usability.

\subsection{Grammars for Visual Analytics Systems}
Underlying many of the widely adopted systems is a set of abstractions or representations that formalize interactions in the analysis process. From a systems perspective, these abstractions help make data exploration more manageable by reducing it to a series of well-defined operations. This makes the query space enumerable, enabling mapping between user interactions and analytical operations.

\vspace{0.5em}\noindent\textbf{Bridging User Interactions and Data Queries.}
Polaris' VizQL~\cite{hanrahan2006vizql} is a prime example, which translates analyst interactions into MDX queries and renders the result as visualizations without the need for manual coding.
Looker's LookML~\cite{looker} is a modeling language that is used to create reusable definitions that Looker uses to generate SQL queries dynamically.
PowerBI's DAX~\cite{powerbi} is a language that enables analysts to define custom calculations and queries on data within tabular models.
These languages act as intermediaries between the system’s interface and the underlying database, enabling user-driven exploration through interactions while translating those into database operations.

\vspace{0.5em}\noindent\textbf{Querying Graph Databases.}
Graph query languages extract relevant subgraphs by matching patterns, often formalized as conjunctive regular path queries for expressive graph querying~\cite{barcelo2013querying,calvanese2000containment,deutsch2022graph}. Over time, many graph query languages have been proposed, but only recently have GQL and SQL/PGQ introduced a shared pattern-matching core based on path queries governing entity traversal~\cite{ISO39075:2024,deutsch2022graph,francis2023researcher}. Although these languages support filtering, aggregation, and multi-graph operations, their syntactic complexity makes them hard to use in no-code settings. \GPQL addresses this gap by mediating between intent expressed in the interface and executable graph queries.

\vspace{0.5em}\noindent\textbf{Grammars for Graph-Based Visual Analytics Systems.}
\revised{Several graph-specific abstractions inform GPQL's design. GraphQ~IR~\cite{nie2022graphq} formalizes a query grammar as an intermediate representation bridging natural language and graph query languages (SPARQL, Cypher, Lambda-DCS) for neural semantic parsing, but is not designed for interactive, no-code query construction or live database connectivity. HiRegEx~\cite{li2024hiregex} is a declarative, regex-based grammar for querying \emph{hierarchical} data exclusively, and does not generalize to property graphs. Earlier path query languages such as G+~\cite{DBLP:conf/sigmod/CruzMW87} and GraphLog~\cite{DBLP:conf/pods/ConsensM90} used regular expressions over graph paths but lacked visualization-coupled interaction. Conjunctive regular path queries (CRPQs)~\cite{barcelo2013querying, angles2017foundations} provide the theoretical foundation for expressive graph querying; however, their expressiveness (negation, optional edges, variable-length paths) makes them challenging to expose directly in a no-code interface. \GPQL is therefore an opinionated, composable sublanguage of Cypher/GQL: every Visual Query Builder interaction maps to a valid, immediately executable query, deliberately trading some expressiveness for guaranteed executability~(\autoref{sec:motivation-gpql}).}
\section{\revised{Practitioner Context}: A Formative Study}
\label{sec:user-requirements}

To \revised{contextualize} \toolname's design in real-world analytical practice, we conducted a formative survey (N=18) with professionals working with graph-structured data or visualizations. 
\revised{The survey was not intended to determine whether a visual, textual, or hybrid query interface was preferable. Rather, it identified recurring practices and barriers relevant to the system's design.}

\subsection{Study Design}
The survey comprised 17 multiple-choice, Likert-scale, and open-ended questions on practitioner experience, exposure to graph analytics tasks, and perceived challenges.
We draw on prior work on knowledge graph usage~\cite{li2023knowledge} to structure the questionnaire.
\revised{Eligible participants were required to have experience with graph visualization or databases and were recruited through the authors’ professional and social networks, resulting in a self-selected participant pool that skewed toward technically experienced practitioners; this outcome reflects that graph database analytics remains a niche domain largely populated by expert users.}
\revised{We therefore interpret the findings as practitioner context rather than as representative evidence about non-technical users.}
Participation was voluntary, anonymous, and could be withdrawn at any time.
No personal data was collected.
The survey required approximately 10–15 minutes \revised{to complete}.
We applied thematic analysis to capture common practices and challenges among technically proficient practitioners.
From these responses, we derive four user-centered design priorities (see \autoref{sec:user-requirements}) that guide the design of \toolname.
\revised{The full questionnaire and participant characteristics are provided in \autoref{appendix:survey}.}


\subsection{Study Results}
\begin{figure}[ht]
    \centering
    \includegraphics[width=1.0\linewidth]{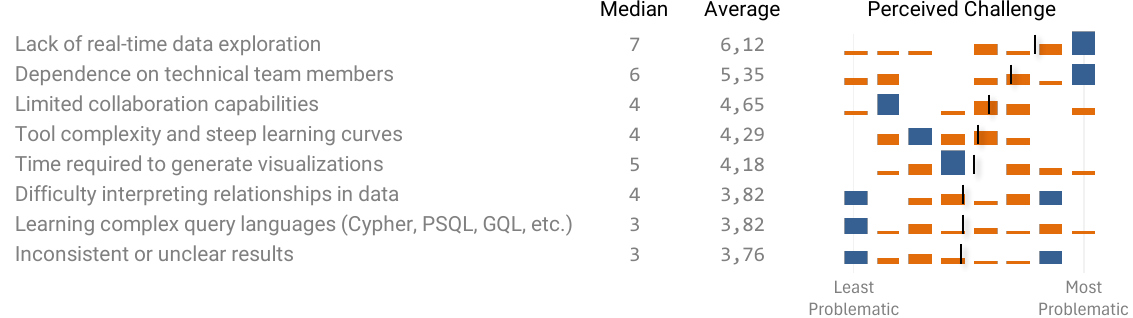}
    \vspace{-1em}
    \caption{Practitioners express the greatest pain points in data analytics tools as the absence of real-time capabilities and heavy reliance on technical team members, while issues like collaboration, tool complexity, and time-to-visualize are also notable but less severe; challenges involving query languages and inconsistent results appear less critical by comparison (N=18).}
    \vspace{-1.5em}
    \label{fig:requirements_challenges}
\end{figure}

\begin{figure}[ht]
    \centering
    \includegraphics[width=1.0\linewidth]{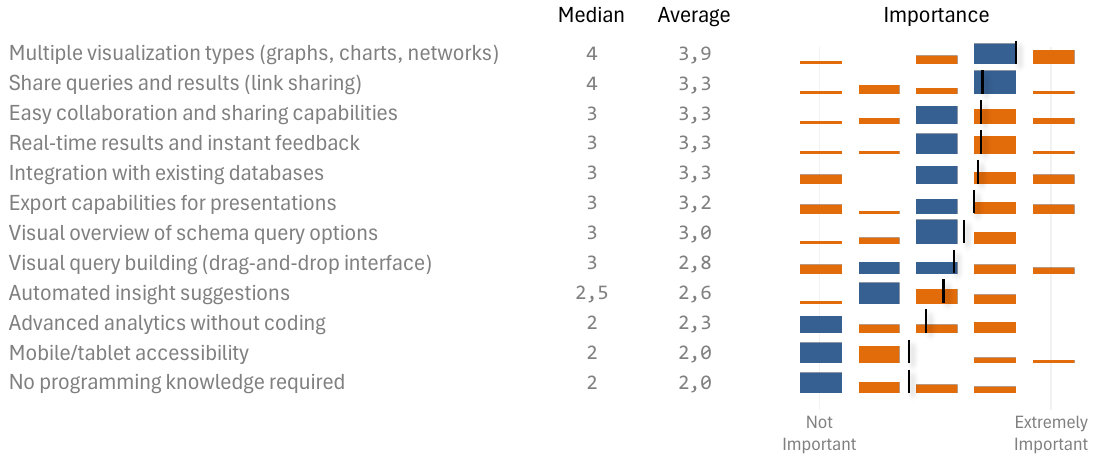}
    \vspace{-1.5em}
    \caption{Analysts assign the highest feature importance to having multiple visualization types (median 4; mean 3.9) and the ability to share queries and results (median 4; mean 3.3), while real-time feedback, collaboration, and integration with databases are also valued but to a slightly lesser extent. Features like automated insights, advanced analytics without coding, mobile access, and not requiring programming knowledge are considered less critical priorities for practitioners (N=18).}
    \label{fig:requirements_importance}
    \vspace{-0.25em}
\end{figure}

For this work, we focus on two main perspectives: (1) key challenges in modern graph analytics (\autoref{fig:requirements_challenges}), and (2) participants’ assessments of requirements for emerging graph analytics tools (\autoref{fig:requirements_importance}).

To extract the core graph analytics pain points, we asked \revised{participants} to rank the challenges from least problematic (1) to most problematic (8). 
The data reveals that real-time data exploration (mean: 7, avg: 6.12) and minimizing dependency on technical team members (mean: 6, avg: 5.35) are the top priorities for practitioners. Future data exploration tools must focus on enhancing data democratization, immediacy, and autonomy \cite{DBLP:conf/pacis/SamarasingheLS22,liautaud2000business} and must place their emphasis on interactive (visual) data exploration without latency, making these points a critical area for both technical and interface innovation. Secondly, reducing reliance on technical experts for performing analyses will democratize data access and foster more collaborative, agile decision-making.
Conversely, obstacles such as tool complexity (mean: 4, avg: 4.29) and the need to learn complex query languages (mean: 3, avg: 3.82) are less critical but highlight opportunities for improved usability. 

When we asked the participants to rate the graph analytics tool features regarding their perceived importance on a scale of 1 (not important) to 5 (extremely important), we found that supporting a variety of visualization types (median: 4, avg: 3.9) and enabling result sharing (median: 4, avg: 3.3) have the highest priorities, signaling that flexible presentation and dissemination features drive effective tool adoption.
Features like real-time feedback (median: 3, average: 3.3), collaboration (median: 3, avg: 3.3), integration with existing databases (median: 3, average: 3.3), and export capabilities (median: 3, avg: 3.2) are also important.
Our participants placed limited importance on automated insight generation (median: 2.5, avg: 2.6), mobile accessibility (median: 2, avg: 2.0), no-code analytics (median: 2, avg: 2.3), and visual query building (median: 3, avg: 2.8).
While this is likely due to our sample consisting largely of experienced data scientists who are comfortable writing code, it is surprising that no-code analytics features (query-building and machine learning) are not prioritized.
For true data democratization, future graph analytics tools must become more accessible and support flexible, understandable workflows for non-technical business practitioners and non-data scientists.

Our thematic analysis also discovered persistent challenges in the graph analytics tool offering and its capabilities, which manifested in three recurring entrance and application barriers.
First, participants reported that disproportionate effort goes into building and maintaining complex software stacks rather than conducting actual analysis. The participants' workflows often relied on disconnected tools, requiring manual data conversion and scripting, which makes iterative exploration impractical and frequently enforces reliance on technical specialists. For instance, ten participants state that they are creating graphs from relational or tabular data in programming libraries (e.g., NetworkX in Python, igraph, SNAP). Only seven participants note that they use graph databases in conjunction with programming libraries. 
Second, 7 participants described difficulty in producing meaningful visualizations of complex relationships, citing inconsistent configuration options, limited algorithmic coverage, and a tendency to revert to static, tabular tools under time pressure.
Finally, again, 7 participants reported that current tools do not satisfy their analytics needs, or they spend more time learning the tools rather than analyzing data.
\revised{Together, these responses indicate fragmented workflows and visualization difficulties even among technically experienced practitioners.
They motivate integrated analytical support, but do not establish usability or effectiveness for non-technical users.}
This, however, leads to a dangerous escalation chain of insecurities in the data exploration, decisions based on incomplete information due to analysis barriers, and ultimately offloading analysis challenges to more technical team members or avoiding relationship data entirely.

\subsection{User-Centered Design Priorities} \label{sec:user-requirements}
Our formative survey reveals a gap between practitioners’ analytical goals and current graph tools, especially for retrieving, exploring, and interpreting graph databases.
Although participants were highly proficient in programming and general data analysis, most had limited experience with graph databases and seldom performed interactive querying. Instead, their workflows typically involved converting tabular data into graph formats using multiple disconnected tools. These fragmented processes, coupled with limited visual guidance and a lack of integrated exploration support, constrain interactive analytics and collaboration. To address these gaps, we derived four user-centered design priorities that guide the design of \toolname:

\begin{itemize}
    \item[R1] \textbf{Rapid visual query construction}: \toolname should let analysts quickly \textit{build and refine} graph queries as visual hypotheses, where each interaction (selecting entities, relationships, filters) yields a syntactically valid, executable query \cite{sacha2014knowledge, tableau-whitepaper}.
    \item[R2] \textbf{Immediate visual feedback}: Query results should be rendered as soon as execution completes, supporting iterative exploration, comparison of alternatives, and uninterrupted analytical flow \cite{DBLP:journals/tvcg/BachFATKFC23}.
    \item[R3] \textbf{Multiple views for structure and attributes}: \toolname must offer configurable visualizations for both topology (e.g., node-link diagrams, adjacency matrices) and attributes (e.g., bar charts, scatterplots), as well as hybrid encodings, allowing practitioners to choose and combine views to match their goals \cite{auber2017tulip}.
    \item[R4] \textbf{Integrated analytical workflows}: The system should support the full analytic process, from querying and exploration to collaboration and sharing of insights, without requiring analysts to leave the environment \cite{DBLP:journals/tvcg/YuS17}.
\end{itemize}

\section{The GraphPolaris Query Language} \label{chap:gp_grammar}

\begin{table*}[ht]
    \footnotesize
    \centering
    \begin{tabularx}{\textwidth}{l>{\raggedright\arraybackslash}m{0.135\textwidth}X}
        \toprule
        \textbf{Term} & \textbf{Production Rule} & \textbf{Semantic Definition} \\
        \toprule
        \texttt{Q} & \texttt{P$^+$ | $\omega$(P) } & A query \texttt{Q} resolves one or more path patterns \texttt{P} with an optional terminal function $\omega$ applied to the result of the query. \\
        \midrule
        \texttt{P} & \texttt{E(RE)$^*$ | F$_p$(P) } & A path pattern \texttt{P} consists of an entity \texttt{E} followed by a relation \texttt{R} and another entity \texttt{E}. Path patterns can be joined to represent recursive relations. An optional predicate \texttt{[F$_p$]} can be applied to filter the query results. Stylistically, for ease of reading, we write \texttt{F$_p$(P)} as \texttt{(P){[F$_p$]}} in the rest of the paper.\\ 
        \midrule
        \texttt{E} & \texttt{E$_i$ ~|~ F$_e$(E$_i$)} & An entity \texttt{E} corresponds to a node type in the graph. The subscript $i$ represents an automatically generated ID for referencing purposes (see \autoref{sec:path-patterns} for more detail). It may include an optional predicate to filter based on attributes or topological conditions. Stylistically, we write \texttt{F$_e$(E$_i$)} as \texttt{E$_{i[F_e]}$} in the rest of the paper. \\
        \midrule
        \texttt{R} & \texttt{R$_i$ ~|~ F$_r$(R$_i$)} & A relation \texttt{R} connecting two entities. It can be constrained by a predicate \texttt{F$_r$} to filter based on attributes or topological conditions. Similar to an entity, each relation is assigned an automatically generated ID, $i$. Stylistically, we write \texttt{F$_r$(R$_i$)} as \texttt{R$_{i[F_r]}$} in the rest of the paper. \\ 
        \midrule
        \texttt{F$_e$,F$_r$,F$_p$} &  & A user-defined function, UDF, can be applied to an entity (\texttt{F$_e$}), relation (\texttt{F$_r$}), or path (\texttt{F$_p$}). It restricts the set of elements based on logical conditions. \\ 
        \bottomrule
    \end{tabularx}
    \caption{\textbf{Breaking down each term} in the proposed \toolname \texttt{Query Language}, as detailed in \autoref{chap:gp_grammar}.}
    \label{table:query_grammar}
    \vspace{-1em}
\end{table*}

This section defines the syntax and semantics of the \toolname \texttt{Query Language} (\GPQL), a formal query grammar that underpins the system.
\GPQL serves as an intermediary abstraction between the interface and graph databases.
\revised{It enables database-agnostic query generation, ensures a consistent mapping between interactions and executable queries, and supports analysis-specific visualization recommendations.}


\subsection{Motivation for GPQL}
\label{sec:motivation-gpql}


\revised{
While full-fledged graph query languages such as Cypher, SPARQL, and GQL offer extensive expressive power, their richness makes it infeasible to expose their unified feature set in a no-code VA interface.
\GPQL is therefore an \emph{opinionated abstraction}: a proper sublanguage in the sense of expressive power~\cite{DBLP:conf/dlog/ArecesR98,DBLP:journals/corr/abs-1805-10415}, analogous to how a calculator implements a restricted but complete fragment of arithmetic, i.e., every valid input produces a valid result, yet not all of mathematics is expressible through it.
Every opinionated query interface, from VizQL~\cite{tableau-whitepaper} to LookML~\cite{looker}, enforces such an abstraction, trading expressive power for guaranteed executability, which is a fundamental property of grammar-driven VA systems.

\GPQL aligns the structure and semantics of domain-expert interactions in the Visual Query Builder (see \autoref{sec:visual_qb}): every interaction corresponds to a syntactically valid \GPQL expression that is systematically compiled into database-native query languages.
\GPQL is, thus, a proper sublanguage fragment of Cypher and GQL: every \GPQL query is intertranslatable with a Cypher query with the same semantic value over property-graph models, whereas some Cypher queries, such as those requiring negation, optional edges, or search-style constructs, fall outside the compositional fragment that \GPQL exposes directly.
We assess this design trade-off empirically in \autoref{sec:grammar-level-limitiations}.
}

\subsection{Formal Syntax} \label{chap:illustrative_grammar}
We designed \GPQL to define patterns for querying graph databases.
It includes entities (\texttt{E}), relations (\texttt{R}), path patterns (\texttt{P}), and predicates and functions (\texttt{F}, \texttt{$\omega$}), which together form the specification of queries (\texttt{Q}). The full query grammar is presented in \autoref{table:query_grammar}.

\vspace{0.5em}\noindent\textbf{Entities, Relations, and Paths.}
A query consists of one or more path patterns, each representing a structured traversal through the graph. A path pattern must start and end with an entity and may contain a sequence of alternating entities and relations. Path patterns are composable, meaning they can be combined to form more complex structures while always resulting in a graph.
For instance, a simple query might be expressed as \grammar{E$_{1}$ R$_{1}$ E$_{2}$}, where \texttt{E$_{1}$} represents the source of the path, \texttt{E$_{2}$} the destination, and \texttt{R$_{1}$} the edge between them. 
\autoref{sec:path-patterns} presents additional examples and discusses path patterns in more detail. 

\vspace{0.5em}\noindent\textbf{Predicates and Functions.}
User-defined functions (UDFs) can be applied to queries in two ways: as predicates for filtering and as the \textit{terminal function} of a query. 
Predicates operate at different levels while maintaining the same data type as their input, including entity predicates (\texttt{F$_e$}), relation predicates (\texttt{F$_r$}), and path predicates (\texttt{F$_p$}). 
For example, an entity predicate \grammar{Person$_{1[name=`Mary\textrm']}$} filters all \texttt{Person} nodes where the \texttt{name} attribute equals `Mary,' returning a subset of those nodes. 
In contrast, a terminal function applies to the query's overall result, allowing the return of arbitrary data types. 
This enables the application of graph algorithms for tasks like centrality analysis or shortest path detection, as well as machine-learning models for classification, clustering, or anomaly detection based on query results. 
\autoref{sec:udf} describes UDFs in more detail. 

\subsection{Path Patterns in Graph Queries}
\label{sec:path-patterns}

Another example is a multi-hop path, \grammar{(E$_{1}$ R$_{1}$ E$_{2}$) (E$_{2}$ R$_{2}$ E$_{3}$)}, where \texttt{E$_{1}$} is connected with \texttt{E$_{3}$} via \texttt{E$_{2}$} along two edges \texttt{R$_{1}$} and \texttt{R$_{2}$}.
Note that each entity and relation in the above examples has an ID automatically assigned by the \toolname engine.
The unique IDs allow for the expression of more complex graph structures. 
For example, \grammar{(E$_1$ R$_1$ E$_2$) (E$_1$ R$_2$ E$_3$)} represents a ``forked'' path,  \grammar{(E$_1$ R$_1$ E$_2$) (E$_3$ R$_2$ E$_2$)} a convergence point, and \grammar{(E$_1$ R$_1$ E$_2$) (E$_2$ R$_2$ E$_3$) (E$_3$ R$_3$ E$_1$)} a cycle.
We assign unique IDs to relations as well, enabling cross-referencing of their attributes, essential for expressing conditions across multiple edges, such as temporal constraints. For instance, \grammar{(E$_1$ R$_1$ E$_2$)(E$_3$ R$_2$ E$_2$)[R$_1$.year == R$_2$.year]} filters for paths in which the converging relations share the same year.

At runtime, each symbol in the above examples will be bound to a node or edge name in the data schema before query execution.
For example, for a publication dataset, \grammar{E$_1$ R$_1$ E$_2$} could be bound to \grammar{Author$_{1}$ WROTE$_{1}$ Paper$_{2}$}.
The variable binding allows for further expressivity in query construction. 
For example, \grammar{Person$_{1}$ IS\_FRIEND$_{1}$ Person$_{2}$} and \grammar{Person$_{1}$ IS\_FRIEND$_{1}$ Person$_{1}$} represents two different types of relationships with node type \texttt{Person}.  
The first query finds all friends of \texttt{Person$_{1}$}, whereas the second query is a self-referential friendship, meaning cases where a person is recorded as their own friend.

\subsection{User-Defined Functions}
\label{sec:udf}
A UDF is a function \texttt{F} that operates on an input graph \texttt{G} (possibly with additional parameters) and produces an output. \revised{We list all implemented UDFs in \autoref{appendix:udf_terminalfunctions}. 
UDFs fall into two categories based on their input–output relationship. Predicates preserve the input type and return a subset of it, whereas type-changing functions lose composability and can only be used as terminal functions.}

\vspace{0.5em}\noindent\textbf{Predicates.}
A predicate is a UDF that takes a graph \texttt{G}, a set of nodes \texttt{E}, or a set of edges \texttt{R} as input, and returns a subset of the input as output (\grammar{F(T) $\subseteq$ T, where T = E | R | G}). Predicates serve as filters, selecting elements based on specified conditions. These conditions can be evaluated on raw data attributes or derived attributes, such as those computed from graph topology (e.g., node degree).

Data attribute predicates filter nodes and edges based on conditions applied to their stored attributes. These conditions follow the structure \textit{attribute-operator-value}. For example, in \grammar{Person$_{1[age > 50]}$}
the attribute is age, the operator is > (greater than), and the value is 50. The value in a predicate can also be derived from another attribute within the same node/edge or from a related node/edge. For instance, in the query \grammar{(Person$_1$ IS\_FRIEND$_1$ Person$_2$)[Person$_1$.age > Person$_2$.age]}, the condition ensures that \grammar{Person$_1$} must be older than \grammar{Person$_2$} for the relationship to be included in the query result.

Predicates can also be applied to derived attributes from the topology. For example, in the query 
\grammar{(Person$_{1}$ IS\_FRIEND$_1$ Person$_{2}$)[Person$_1$.degree > 5]}, the predicate filters the results to include only cases where \grammar{Person$_{1}$} has more than five friends. 
It should be noted that all predicates can have multiple conditions as well, following first-order logic. 
For example, the term \grammar{Person$_{1[age > 50~and~name=`Mary\textrm']}$} finds all \grammar{Person} nodes named `Mary' who are over the age of 50.

\vspace{0.5em}\noindent\textbf{Terminal Functions.}
\label{sec:terminalfunctions}
A terminal function ($\omega$) is applied to the resulting graph from executing a query, but does not return a graph output. 
Instead, terminal functions produce a non-graph result, such as numbers, arrays, or other data structures.
Since applying $\omega$ breaks composability, it must be the last action in the query execution workflow.

The use of terminal functions enhances the expressiveness of the \toolname grammar by enabling the integration of custom computational logic within path patterns. Analysts can incorporate graph algorithms, machine-learning models, or other analytical functions that extend beyond the predefined \GPQL constructs. An example of such a UDF is community detection, which processes the input graph and returns the original graph alongside an array of cluster labels, where each node is assigned a number indicating its cluster membership. This enables the application of structural analysis directly within query patterns while maintaining compatibility with the broader graph processing pipeline.
For example, in \grammar{(Person$_{1}$ IS\_FRIEND Person$_{2}$)[average\_clustering(Person$_{1}$ IS\_FRIEND Person$_{2}$, *parameters)]}, a terminal function computes the average clustering coefficient of a subgraph extracted via a query.
This function processes the graph but returns a single numeric value rather than another graph, ensuring that it is the final step in the query execution.


\subsection{Illustrative Examples}
The following examples demonstrate \GPQL's expressiveness:

\smallskip
\noindent\textit{Find all employees who work for companies headquartered in the same country as their university}. This query filters employees whose company headquarters matches the country of their university, ensuring that employees studied and work in the same region. This can be represented as: \grammar{((Employee$_1$ WORKS\_FOR Company$_2$) (Employee$_1$ STUDIED\_AT University$_3$))[Company$_2$.hq = University$_3$.country]}.

\smallskip
\noindent\textit{Find all co-authors of influential researchers, who wrote papers in the same years}. This query identifies researchers who have collaborated with highly influential authors (influence is determined by a centrality measure). This can be expressed as: \grammar{((Author$_1$ WROTE$_1$ Paper$_2$)(Author$_3$ WROTE$_2$ Paper$_2$))[Author$_1$\texttt{.centrality} $\geq$ 0.1 and WROTE$_1$\texttt{.year} = WROTE$_2$\texttt{.year}]}.

\smallskip
\noindent\textit{Compute the median age of employees who have switched jobs at least twice}. This query filters employees based on the number of job transitions and applies a terminal function to compute their median age, which can be represented as: \grammar{$\omega_{\text{median(Employee${1}$.age)}}$((Employee$_1$ WORKS\_FOR Company$_2$)[Employee$_1$.degree > 2])}.

\smallskip



We discuss additional examples in the use-case \autoref{chap:applications}. 
\section{The GraphPolaris System} \label{chap:system}




\toolname is a Visual Analytics system that supports the full graph exploration lifecycle from query formulation to insight communication (\autoref{fig:teaser}). \revised{As depicted in \autoref{fig:mapping}}, \GPQL underpins the querying interface, while flexible visualization, configuration panels, and insight sharing extend the analyst experience beyond query construction. Guided by our formative requirements study (\autoref{sec:user-requirements}), this section first describes how practitioners express analytical intent and construct queries with the Visual Query Builder (R1, R2), then how \toolname supports exploration, visualization-driven insight, and collaboration (R3, R4). 
\revised{Due to space constraints, we present only a high-level overview here; \autoref{chap:implementation} provides additional implementation details, including deployment configuration, performance-related design choices, and integration with graph database backends.
}

\subsection{Query Construction Through the Visual Query Builder}
\label{sec:visual_qb}

Here, we focus on how practitioners interactively construct graph queries using the Schema Panel and Visual Query Builder.
Analysts begin by exploring the schema, then assemble visual blocks (referred to as \textit{pills}) to specify \textit{entities}, \textit{relationships}, \textit{filters}, and \textit{UDFs}.
Each pill combination corresponds to a valid \GPQL expression.

\vspace{0.5em}\noindent\textbf{Exploring the Schema of the Graph Database.}
\label{sec:schemapanel}
The Schema Panel (Panel 1 in \autoref{fig:teaser}) provides a structured overview of the data model. It displays entity types (\includegraphics[height=0.75em]{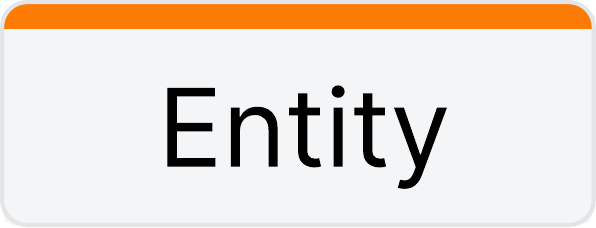}), their relationships (\includegraphics[height=0.75em]{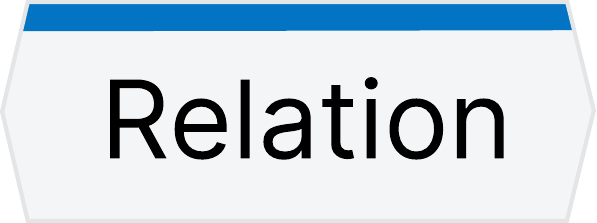}), and associated attributes (in the tooltip) in a node-link layout. This overview does not enumerate query instances but reveals the structure from which valid queries can be derived, enabling practitioners to conceptualize the space of analytical questions supported by the dataset~\cite{suh2022grammar}.
By offering this overview, the Schema Panel addresses the common challenge that analysts are unaware of the data and relationships available to them.
 
We use \textit{pills} to represent 
entities, their relations, and their associated attributes (in gray, attached to the respective node or edge).
Here, attributes denote the properties stored on nodes or edges in the graph database (e.g., a \texttt{Person} node with an \textit{age} attribute, or a \texttt{Transaction} edge with an \textit{amount} attribute).
\revised{This approach helps analysts intuitively form and refine queries based on the data’s structure. For example, \autoref{fig:teaser} shows a schema describing authors, papers, and conferences, supporting questions such as ``\textit{Which authors have presented at which conferences?}'' or ``\textit{Which papers were co-authored by researchers with the same affiliation?}''.
With this overview, the analyst can start forming questions and queries by dragging pills into the Visual Query Builder.}


\setcounter{figure}{4}
\begin{figure*}[t]
    \centering
    \includegraphics[width=0.95\linewidth]{figs/workflow.pdf}
    \caption{\textbf{\toolname's workflow supports end-to-end graph exploration}. Analysts begin by inspecting the graph database schema to discover available entities and relationships. They then express analytical intent through the Visual Query Builder, where interactions are compiled into database-native queries. Query results are rendered in interactive visualizations that enable exploration, refinement, and communication of insights.}
    \vspace{-1.5em}
    \label{fig:mapping}
\end{figure*}

\vspace{0.5em}\noindent\textbf{Assembling Queries Through Visual Composition.} \label{chap:visual_qb}
\toolname enables no-code query construction by mapping \GPQL to the interactive elements of the Visual Query Builder (Panel 2 in \autoref{fig:teaser}).
Analysts formulate analytical questions by dragging pills from the Schema Panel to build path patterns and apply filters or transformations visually.
In this way, analysts can focus on the substance of their inquiry rather than the syntax of graph query languages, while the system guarantees that each interaction results in an executable query that can be visualized.

The Visual Query Builder operationalizes this process through four types of pills, each corresponding directly to constructs in \GPQL (\autoref{chap:gp_grammar}).
\textit{Entity pills} (\includegraphics[height=0.75em]{figs/entity.pdf}) correspond to entities (\texttt{E}), while \textit{relationship pills} (\includegraphics[height=0.75em]{figs/relation.pdf}) represent relations (\texttt{R}).
\textit{Logic pills} (\includegraphics[height=0.75em]{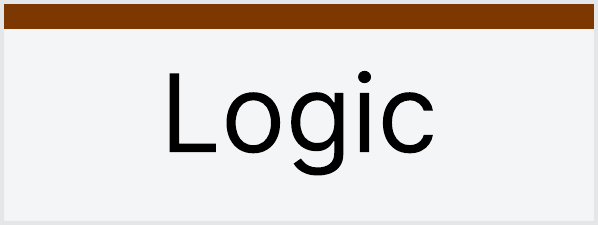}) capture predicates applied to entities, relations, or paths (\texttt{F$_e$, F$_r$, F$_p$}).
Finally, \textit{meta pills} (\includegraphics[height=0.75em]{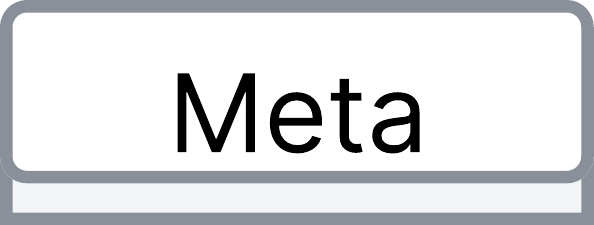}) correspond to terminal functions ($\omega$).
Type-checking ensures that predicates and functions return valid subsets or compatible data types, maintaining consistency between grammar and interface.
Once assembled, the \GPQL expression is transpiled into a vendor-specific query, executed on the connected database, and returned for immediate visualization and iterative refinement.

\autoref{fig:teaser} shows a query created using the Visual Query Builder\revised{, laid out with the \texttt{DAGRE} algorithm, which we use for its good visual readability.}
The query consists of a path pattern formed by three \textit{entity pills}, with a \textit{meta pill} applying a degree predicate greater than 3 to the author node. 
\revised{This query pattern is internally represented as:
\grammar{(Author$_{1}$ \ WROTE$_{1}$ Paper$_{2}$)[Author$_1$.\text{degree} > 3] (Paper$_{2}$ RECEIVED\_AWARD$_{2}$ Award$_{3}$)}.
Notice here that we can express the relationship \texttt{Paper$_{2}$ RECEIVED\_AWARD$_{2}$ Award$_{3}$} in the Visual Query Builder without the need of explicitly adding the relation pill \texttt{RECEIVED\_AWARD} (because there is only one possible relation between \texttt{Paper} and \texttt{Award}), improving query construction speed.}

\subsection{Display Types for Structural and Attribute Analysis}
\label{sec:graphvis}

Once a query is executed, the resulting subgraph or table is visualized in the Visualization Panel (Panel 3 in \autoref{fig:teaser}).
Multiple graph visualizations are feasible, each having its distinct advantages and disadvantages for accomplishing graph analytics tasks \cite{DBLP:journals/jvis/ChenGZDW19, nobre2019state}. 
\revised{
To satisfy practitioner requirement R3, \toolname currently supports a set of 14 visualization types tailored to graph topology, attributes, and statistical summaries, which we describe below.
}

\vspace{-0.5em}
\setcounter{figure}{3}
\begin{figure}[H]
    \centering
    \includegraphics[width=1.0\linewidth]{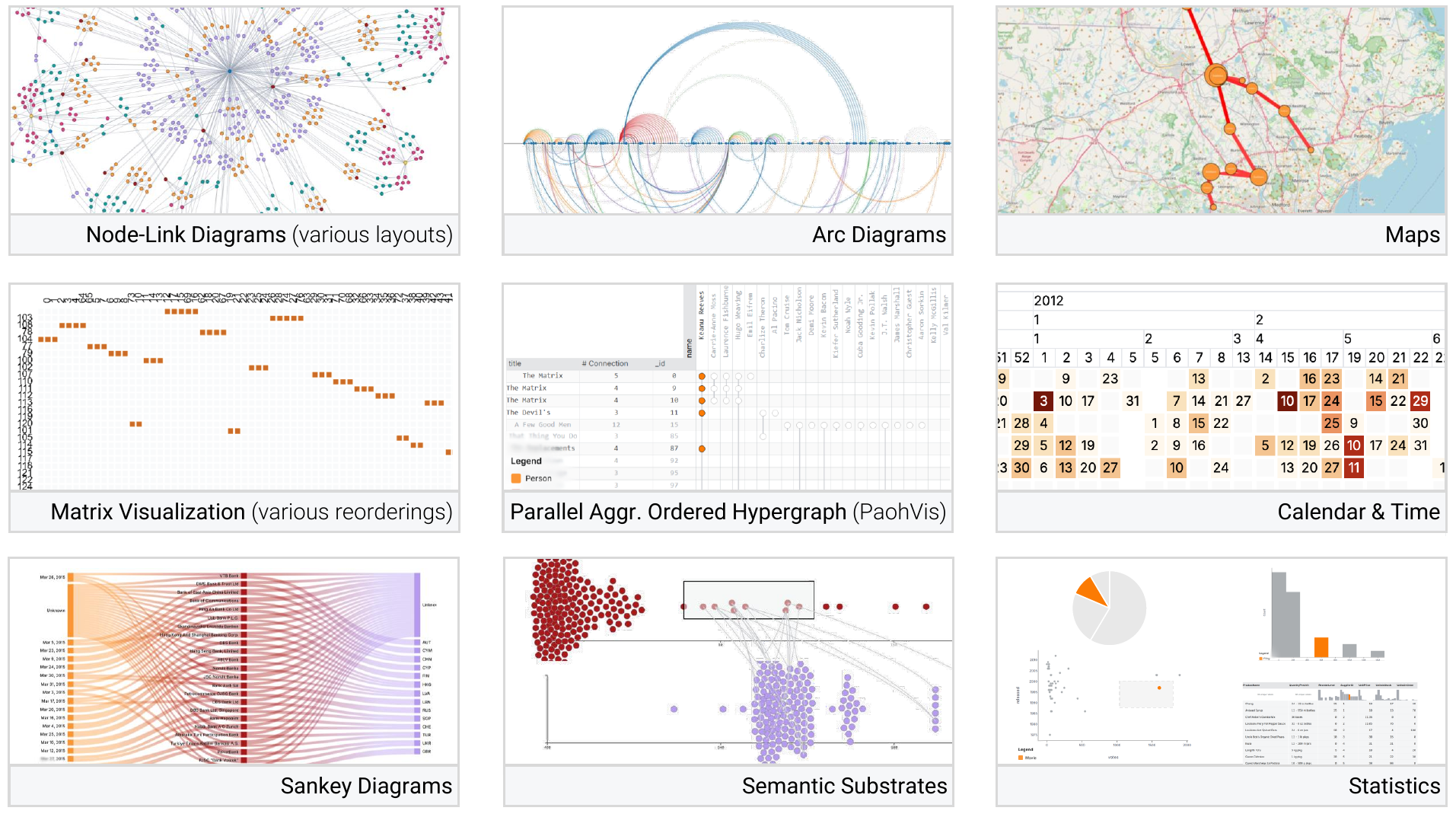}
    \vspace{-2.0em}
    \caption{\textbf{Showcase of visualizations in \toolname.} These examples highlight the system’s versatility across multiple tasks, including network exploration, subgraph comparison, attribute- and flow-analysis.}
    \label{fig:gallery}
    \vspace{-1.5em}
\end{figure}

\vspace{0.35em}\noindent\textbf{Graph Visualizations.}
\toolname includes five graph visualizations selected to support the three major task categories in Lee et al.’s graph task taxonomy~\cite{lee2006task}.
For overview and browsing tasks, we provide \textit{node-link diagrams} and \textit{adjacency matrices}. For topology-focused analysis, we include \textit{PaohVis} \cite{buono2021hypergraph} and \textit{arc diagrams}.
Finally, for attribute-focused exploration, we integrate \textit{Semantic Substrates} \cite{shneiderman2006network}.
The task categories are overlapping, and many visualizations can address more than one.
For clarity, we list only the primary task we intend each visualization to support.
Additionally, a range of other graph visualizations, such as TreeMaps~\cite{DBLP:conf/visualization/JohnsonS91}, Hive plots~\cite{DBLP:journals/bib/KrzywinskiBJM12}, and NodeTrix~\cite{DBLP:journals/tvcg/HenryFM07}
are under development to provide more visualization capabilities.

\vspace{0.35em}\noindent\textbf{Non-Graph Visualizations.}
Nine visualizations are available: six for result sets and attributes—\textit{table}, \textit{histogram}, \textit{scatterplot}, \textit{line chart}, \textit{pie chart}, and \textit{bar chart}, and three for specific data types and use cases: \textit{map visualizations}, a \textit{calendar view}, and a \textit{JSON viewer}.

\revised{\vspace{0.35em}\noindent\textbf{Visualization Scalability.}
Node-link diagrams use GPU-accelerated rendering to display hundreds of thousands of nodes and millions of edges interactively. Map visualizations mitigate overplotting via R-tree spatial indexing~\cite{DBLP:conf/sigmod/Guttman84} and zoom-dependent clustering. Structured visualizations such as PaohVis and Sankey provide built-in aggregation to reduce visual complexity while preserving analytical utility~\cite{buono2021hypergraph,DBLP:conf/infovis/RiehmannHF05}.}

\vspace{0.35em}\noindent\textbf{Selecting and Sharing Visualizations.}
Practitioners select a visualization by choosing a display type in the Visualization Panel. \revised{\toolname does support automated graph visualization recommendations via a predicate-learner approach, which is beyond the scope of this paper and is treated in our complementary work~\cite{DBLP:conf/grivapp/VinkMLCB26}.}
Once a visualization is selected, analysts can adjust settings such as color schemes and axes, tailored to the chosen visualization type. All settings are stored to ensure reproducibility. Practitioners can share session-encoded URL links to facilitate collaboration.

\revised{\subsection{Integrated Analytical Workflow} \label{chap:insightsharing}
The query–visualization process in \toolname is iterative and interactive: analysts construct an initial query, visualize results, and refine queries based on emerging findings. Over time, this leads to \textit{actionable insights} that can be shared to support decision-making (R4). Although \textit{insight} lacks a universally agreed definition in visualization research~\cite{sacha2014knowledge, chang2009defining, battle2023exactly, north2006toward}, we define it here as the outcome of iterative retrieval, exploration, and interpretation of a graph database that is sufficiently meaningful to be preserved or communicated.}

\revised{\toolname captures these insights as \emph{shareable knowledge artifacts}: each \GPQL query, together with its visualization configuration, forms a reusable unit that preserves both structure and interpretation. Sharing these artifacts via URLs lets analysts reproduce and extend analyses and collaborate asynchronously, similar to state-sharing mechanisms in systems like Many Eyes~\cite{viegas2007manyeyes} and Voyager~\cite{heer2007voyagers}.}
\vspace{-0.5em}
\section{Use-Cases} \label{chap:applications}


To demonstrate \toolname's applicability in real-world settings, we present use cases from two large-scale enterprise partners\footnote{Both organizations have more than 10,000 employees; we withhold their names for anonymity.} in telecommunications and in supply-chain management. Because of NDAs, we abstract sensitive details while focusing on how \toolname meets their analytical needs while satisfying our design priorities (Sec. \ref{sec:user-requirements}). Our examples preserve the partners' investigative workflows and key relationships, providing a faithful account of how \toolname supports subject-matter experts in practice.

\subsection{Analyzing Mobile Network Handovers}

Our partner in the telecommunications sector employs \toolname to enhance network resiliency and improve customer satisfaction.
A critical operational challenge is the monitoring and diagnosis of issues in mobile network handovers, where an active call transitions from one cell tower (antenna) to another as a subscriber moves through the coverage area.
Brief inefficiencies or failures in these transitions can result in degraded call quality, increased latency, or dropped connections.
The relevant dataset is modeled as a property graph, in which \emph{connection segments} (nodes) represent individual antenna sessions, and \emph{handover events} (edges) represent the transfer from one antenna to another.
Each connection segment includes attributes such as \texttt{start\_cell\_name}, \texttt{end\_cell\_name}, latency metrics, and termination reasons.

\revised{As Fig.~\ref{fig:combined} shows,} the analysis begins by retrieving all handovers where \texttt{start\_cell\_names} differ, demonstrating R1 and R2. Because each such handover reflects a transition between distinct antennas, these cases reveal how calls traverse the network as subscribers move across coverage areas (e.g., when commuting or traveling).


To narrow the focus to high-impact events, a filter is added on the termination reason to select only \emph{handover failures}, revealing their (spatial) distribution across the network.
Beyond explicit failures, certain structural patterns may indicate network instability. A \emph{cyclical handover}, where a device moves from Antenna~A to Antenna~B and quickly back to Antenna~A, can signal coverage gaps or interference and is directly expressible as a cyclical path in the visual query builder.

The resulting subgraphs are examined in the node-link diagram, with nodes color-encoded by performance metrics (R3). \revised{For instance, coloring by avg. latency can reveal whether handover patterns correlate with poor service quality, while coloring by dropped-call rate points to antenna pairs that require technical intervention (see \autoref{fig:vis_handovers}).}

\setcounter{figure}{5}
\begin{figure}[H]
    \centering
    \vspace{-0.5em}
    \includegraphics[width=\linewidth]{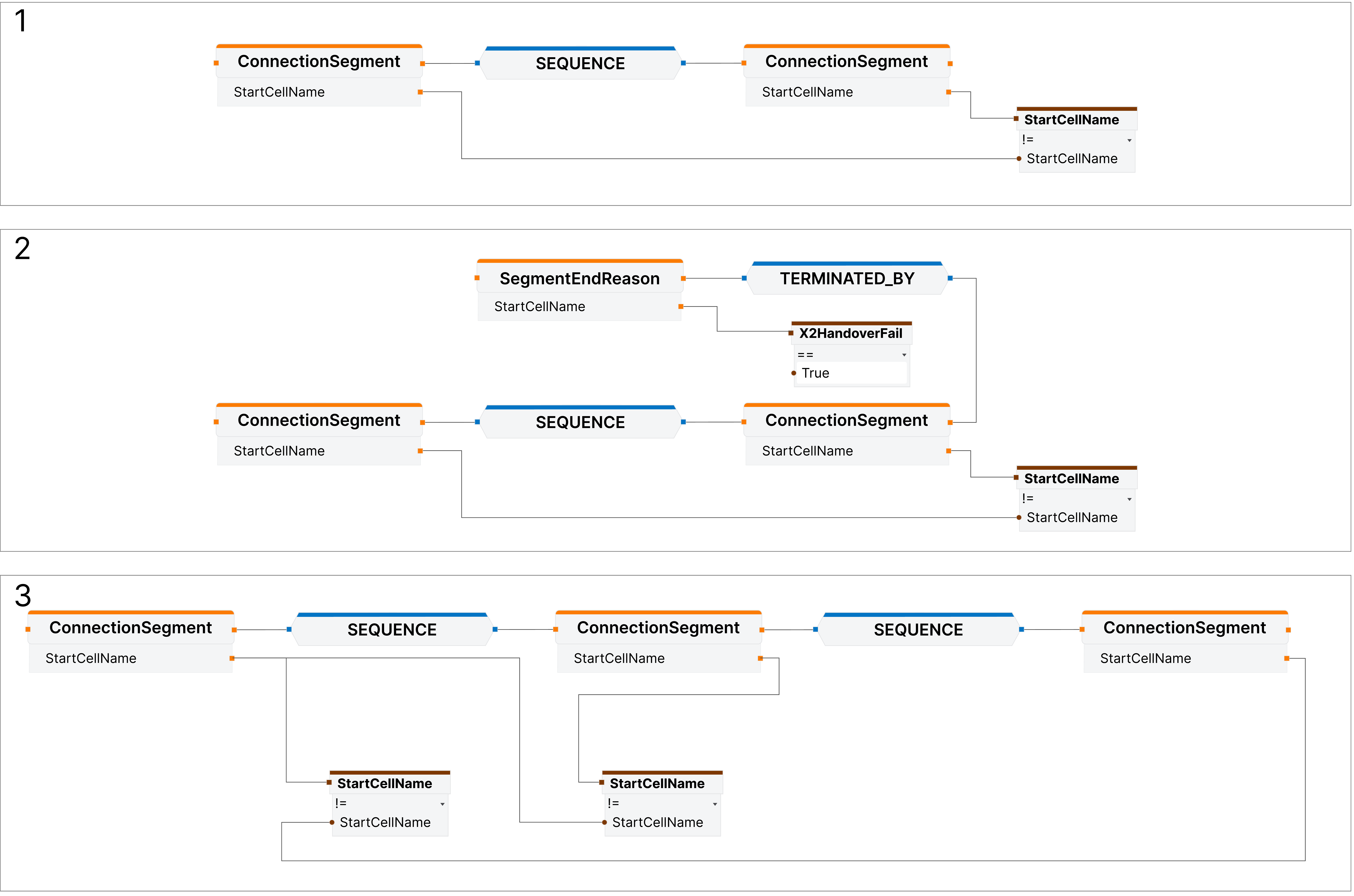}
    \vspace{-2em}
    \caption{\textbf{Iterative analysis of mobile network handovers.} 
    (1) Initial query retrieves all handovers between distinct antennas. 
    (2) Filtering isolates failed handovers to focus on high-impact events. 
    (3) A cyclical path pattern reveals unstable handover behavior. 
    Each query is constructed visually and executed immediately (R1–R2), with results explored through attribute-driven visualization (R3).}
    \vspace{-1.0em}
    \label{fig:combined}
\end{figure}

These findings have been shared directly with antenna engineering teams, who can prioritize inspections or adjustments for high-risk locations. 
These types of analyses are not feasible in existing commercial VA platforms such as Tableau or PowerBI, which are built on tabular abstractions, nor in graph visualization libraries, which lack live querying and analytical integration. Even graph databases like Neo4j can expose such comprehensive workflows only through code-based interfaces. \toolname bridges this gap by supporting iterative no-code exploration of graph data while integrating querying, visualization, and provenance in a single environment.

\begin{figure}[t]
  \centering
  \begin{subfigure}[t]{0.48\linewidth}
    \centering
    \includegraphics[width=\linewidth,height=0.32\textheight,keepaspectratio]{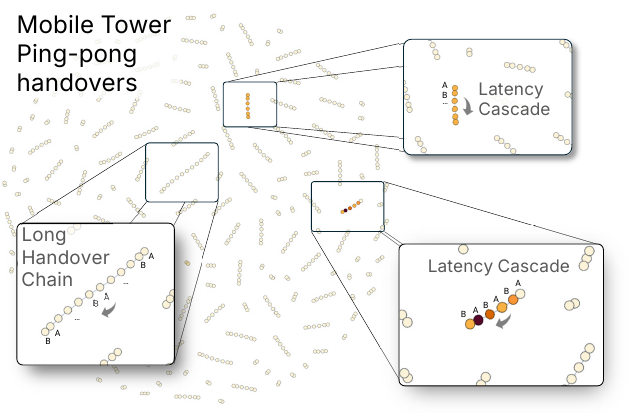}
    \caption{Cyclical handovers visualized with nodes color-encoded by average latency, revealing correlations between structural instability and degraded performance.}
    \label{fig:vis_handovers}
  \end{subfigure}\hfill
  \begin{subfigure}[t]{0.49\linewidth}
    \centering
    \includegraphics[width=\linewidth,height=0.32\textheight,keepaspectratio]{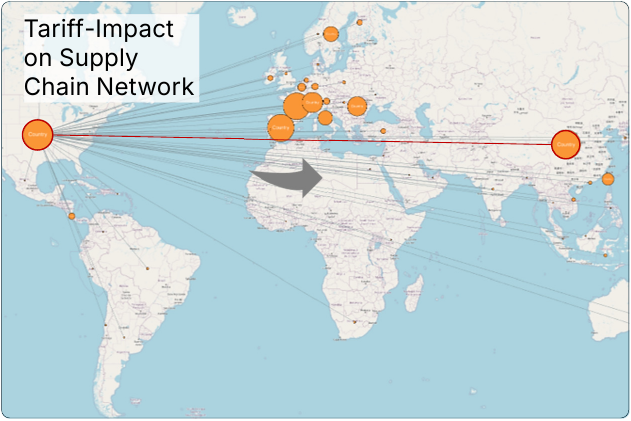}
    \caption{Map-based visualization of tariff-affected trade flows, with node size encoding shipment volume and edge color depicting tariff-change impact.}
    \label{fig:philips_result_map}
  \end{subfigure}
  \vspace{-0.75em}
  \caption{\revised{Selection of results from long-term case study (MILC). 
  (a) Cyclic mobile phone cell-tower handovers cause call instability. (b) Exploring tariff-affected trade flows to assess supply chain vulnerabilities.}}
  \label{fig:showcase_results}
  \vspace{-1.75em}
\end{figure}

\subsection{Analyzing Global Trade Flows and Tariff Impacts}
Our second partner, in supply-chain management, uses \toolname to analyze international trade flows and assess the impact of changing tariff policies.
A  critical operational challenge is identifying not only tariffed direct imports and exports, but also products indirectly affected via tariff-exposed components embedded in manufactured goods.
For example, steel sourced from China is incorporated into EU-produced equipment.
The dataset is represented as a property graph in which \emph{countries} and \emph{products} (nodes) are connected by \emph{trade relationships} (edges) that encode the movement of goods across borders.
Each relationship is annotated with attributes such as product codes (HS codes), shipment volumes, and declared monetary values.

The analysis begins with a visual query that retrieves all traded products, aggregating them to provide both a list and count of unique product categories (R1 and R2). A subsequent filter isolates products whose HS codes indicate tariff-sensitive components (e.g., Chinese steel).
Focusing on downstream implications, the query is refined to retain only shipments arriving in the United States. This reveals the subset of goods most directly exposed to recently imposed U.S. tariffs.
To contextualize these flows geographically, country-of-origin attributes are added and projected onto a map visualization (R3). Shipment volume is encoded by node size and trade intensity by edges, creating an overview of how tariffs propagate through global supply chains.

\vspace{-0.5em}
\begin{figure}[h]
    \centering
    \includegraphics[width=\linewidth]{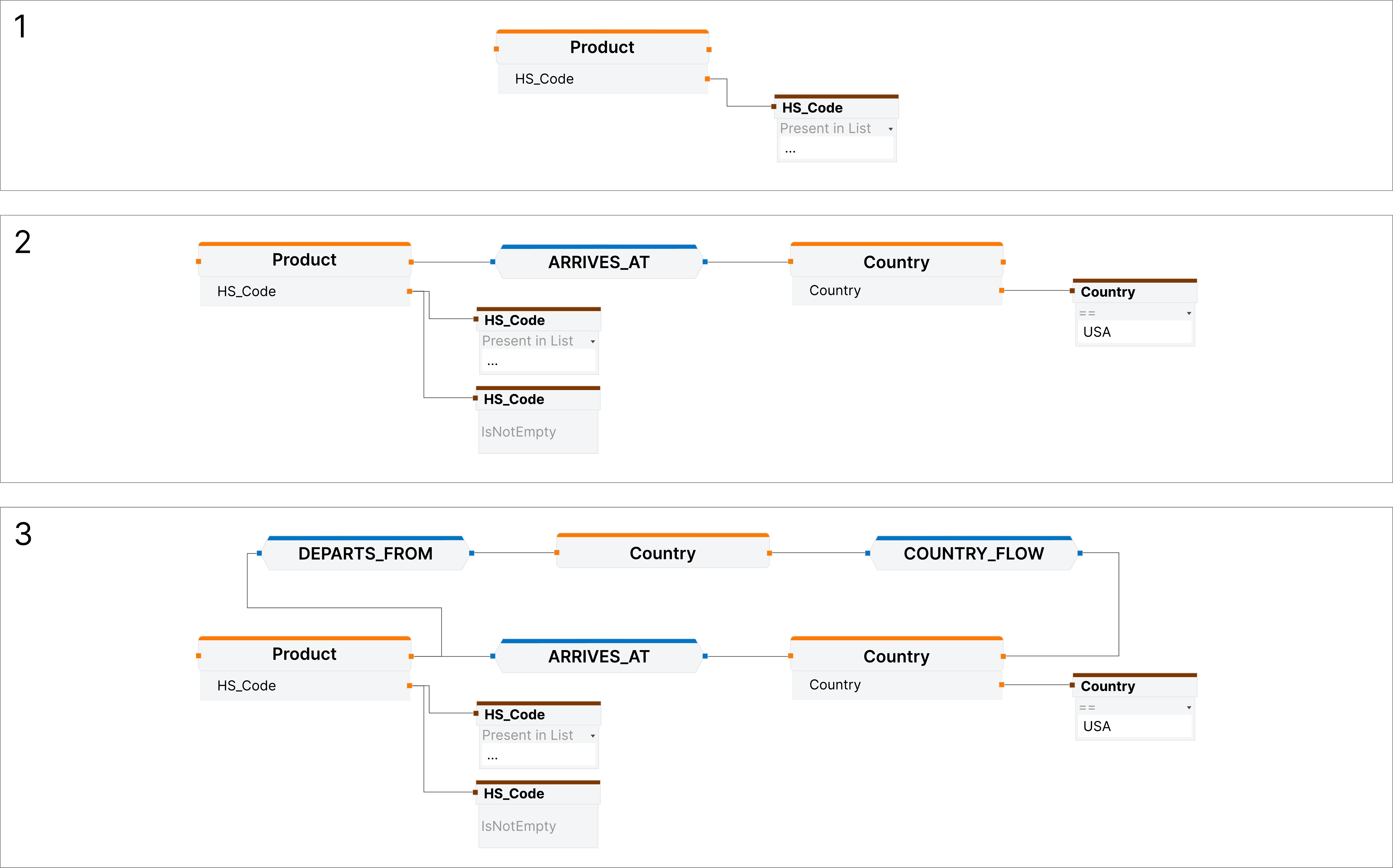}
    \vspace{-1.0em}
    \caption{\textbf{Iterative analysis of global trade flows and tariff exposure.} 
    (1) Initial query retrieves traded products and filters for tariff-sensitive components based on HS codes. 
    (2) Filtering isolates shipments arriving in the United States to assess direct tariff exposure. 
    (3) Extending the query with country-of-origin attributes enables a geographic visualization of trade flows, revealing indirect dependencies and propagation effects. 
    Each step incrementally refines the query through visual interactions (R1–R2), with results explored through multi-view visualization (R3).}
    \label{fig:philips_products}
    \vspace{-1em}
\end{figure}
\smallskip

\revised{The resulting visualizations (see: Fig~\ref{fig:showcase_results}(b)) highlight the disproportionate impact on European trade routes, where high-volume shipments to the U.S. depend on components sourced from tariff-targeted regions.}
By iteratively refining queries and adjusting visual encodings, supply-chain experts can trace both direct and indirect tariff exposures, uncover hidden vulnerabilities, and estimate the likely cost implications for different product lines. These insights are disseminated to supply chain managers and policy teams, who use them to re-evaluate sourcing strategies, negotiate with suppliers, and anticipate pricing shifts (R4).
Through this human-in-the-loop workflow, \toolname helps organizations turn complex trade flows into tractable and actionable questions, strengthening supply-chain resilience in a volatile policy environment. Unlike commercial VA tools that flatten trade into tabular aggregates or graph libraries that offer only static overviews, analysts here must trace multi-hop dependencies and link them to tariff-sensitive attributes in real time. Graph databases expose this only through manually crafted, complex queries, creating a high barrier for domain experts. Interactive query refinement and multi-view visualization together make this end-to-end supply-chain analysis accessible.
\section{Longitudinal Practitioner Experiences and Observations from Real-World Deployments}
\label{sec:milc}

\revised{
Our ongoing iterative design process is grounded in a collaboration of more than two years with four enterprise partners and one academic (sustainability-related) project, during which we co-developed and evaluated \toolname in operational workflows. To structure this evaluation, we followed Shneiderman et al.'s multidimensional in-depth long-term case study (MILC) methodology~\cite{DBLP:conf/avi/ShneidermanP06}, embedding \toolname in authentic analytical practice over a 16-month on-premise deployment and using weekly meetings to capture challenges and guide improvements. Across this period, eight industry (2 medium/6 senior) analysts and one sustainability researcher without coding skills used \toolname for tasks such as schema inspection, pattern retrieval, filtering and aggregation, and sharing results with their teams. We collected usage information, meeting notes, and semi-structured interview data, and analyzed them qualitatively through MILC's lenses, focusing on analytical efficacy, user experience, collaboration and knowledge transfer, and speed-to-insight as a practical evaluation of R1--R4. We structure our observations around three recurring topics:
}

\smallskip
\noindent\textbf{Translating Analytical Intent.}
For participants from non-technical backgrounds, “thinking in graphs” initially posed a considerable hurdle.
Participants with prior experience in relational databases frequently attempted to formulate questions in terms of tables and joins, rather than nodes and edges.
However, after a brief one-hour onboarding session and one to two weeks of intermittent use, most participants successfully transitioned to a graph-native mindset.
This shift enabled them to translate analytical questions into the corresponding graph operations with steadily decreasing cognitive effort. 

\vspace{0.5em}\noindent\textbf{Expressing Queries Visually.}
Consistent with prior findings~\cite{borner2019data, pandey2023mini}, participants displayed a mixed understanding of the graph schema visualization, which underpins the query building process.
Following a short introduction, nearly all participants were able to correctly interpret node types and their relationships, particularly grasping the concept of pills (see \autoref{sec:schemapanel}).
However, mastering the broader visual query language proved more challenging.
Many participants required multiple attempts to construct complex queries successfully.
The concept of meta pills, which represent terminal functions, often caused confusion at first.
However, once participants were introduced to concrete examples of terminal functions (e.g., degree, centrality), they were able to apply them more effectively/successfully.
\revised{\toolname provides textual Cypher editing functionality. Consistent with our empirical expressiveness study (\autoref{sec:grammar-level-limitiations}), we can map about 70–80\% of textual cypher queries back into our visual language, exposing formulation issues and supporting visual language learning.}
Throughout our longitudinal study and in line with \cite{DBLP:journals/ws/Vega-GorgojoSGH16}, we observe a general tendency among practitioners to favor \toolname's no-code functionality over the textual query editing, reducing their time-to-insight from (oftentimes) days or weeks of investigation to hours or minutes of research. Nonetheless, experienced practitioners occasionally refer back to the Cypher representation to validate and build trust in the resulting query.

\vspace{0.5em}\noindent\textbf{Interpreting Query Results.}
When assessing whether \toolnames responses matched their intended questions, participants could generally verify the correctness of simple queries.
However, more complex queries involving multiple relationships or interdependent filters often prove to be more difficult for our participants.
In these cases, participants struggled to reconcile the visualization with their original intent, suggesting a gap in their mental model of how abstract graph operations map onto visual output.
These difficulties point to a gap between R2 (immediate feedback) and participant expectations: while the system provides instant visualizations, some struggled to interpret whether the rendered subgraph faithfully represented their query intent.
This suggests the need for richer explanation mechanisms, such as query-to-visualization annotations or stepwise previews, that can strengthen participants’ mental model of how queries map onto results. We will investigate techniques for better matching visualizations with analysts' query intents as future work.


\vspace{-0.2em}
\section{Discussion and Limitations} \label{chap:limitations}

\toolname's inherited constraints can be distinguished into grammar-level and system-level limitations.

\vspace{-0.2em}
\subsection{Grammar-Level Limitations}
\label{sec:grammar-level-limitiations}

\noindent\textbf{Grammar Expressiveness.}
\revised{
We quantify \GPQL's expressive coverage using the \textit{Intense Queries} of the LDBC Social Network Benchmark~\cite{angles2020ldbc}, which, although designed for performance evaluation, provide a representative set of analytical queries. \GPQL expresses $10/13$ queries, covering the core pattern-matching, filtering, and aggregation tasks we target; the remaining three require negation, optional edges, or search-style expressions, i.e., queries that rank or filter results based on textual relevance such as full-text similarity or keyword matching across node properties. We intentionally excluded these features to keep the grammar tractable and the interaction model predictable, a loss of expressiveness inherent to any opinionated query interface that trades language completeness for usability. Because the “congruency” of two languages cannot be exhaustively tested in practice, we treat this benchmark as an empirical approximation of coverage: achieving $10/13$ ($\approx 77\%$) suggests that \GPQL directly supports the main analyst use cases captured by our MILC, while excluded functionality remains accessible in \toolname{} via custom UDFs or terminal operations.
}
\noindent\textbf{Ensuring Type-Preserving Predicate Semantics.}
\GPQL enforces that predicates over entities return entities, and predicates over relations return relations. This ensures well-typed intermediate results, but limits structural queries that require type transformations. For instance, filtering entities by topological properties such as degree is only possible through UDFs. Supporting such queries would require either relaxing the type constraints or introducing higher-order operators.

These limitations reflect a trade-off inherent in our composability commitment: every \GPQL query is designed to be incrementally constructed and composed with other query patterns, while functionality beyond this must be delegated to UDFs/terminal functions, potentially sacrificing composability and reverting interaction to more tabular operations. Extending expressiveness without undermining composability remains an open challenge in graph analytics design.


\subsection{System-Level Limitations}


\noindent\textbf{Visual Language-Grammar Gap.}
The Visual Query Builder exposes only a subset of \GPQL. 
As in systems such as VizQL~\cite{hanrahan2006vizql}, only common constructs are directly accessible. 
As a result, some queries that are expressible in \GPQL cannot be constructed through the visual interface. 
This differs from the grammar-level limitations, where certain queries cannot be expressed in \GPQL at all. 
Current limitations include the absence of nested predicates, intermediate pattern reuse, and multi-query composition, which would restrict rapid query construction and immediate feedback. 
Determining how much of the grammar to expose in the interface remains an open design question.


\vspace{0.33em}\noindent\textbf{Visualization and Schema Complexity.}
Graph representations impose higher cognitive load than tabular encodings~\cite{8354901}. 
Practitioners also prefer node-link diagrams even when alternative representations, such as matrices, may be more effective~\cite{DBLP:conf/infovis/GhoniemFC04}. 
Effective systems must therefore guide users in selecting and transitioning between representations.
Large schemas further increase complexity.
\revised{We learned that our Schema panel can overwhelm analysts and hinder query formulation. 
We are currently experimenting with alternative interaction methods, such as query recommendations/augmentations for GPQL/Cypher, and chat-based interfaces,
enabling natural language-based to GPQL/Cypher analysis.}
Alternative representations, such as hierarchical layouts or semantic zoom, may improve navigation. 
In the Visual Query Builder, the visual language is very flexible but also introduces visual complexity. 
These challenges motivate improved schema representations, interaction guidance, and visualization recommendations.

\vspace{0.33em}\noindent\textbf{Visualization Scalability.}
\revised{\toolname supports interactive exploration of graphs with hundreds of thousands of nodes and millions of edges, and has not emerged as a performance bottleneck in our enterprise case studies.}
However, for larger graphs (>1M nodes), approaches such as visual or subgraph aggregation \cite{DBLP:journals/tvcg/HenryFM07, 6875972} and server-side rendering \cite{kerpedjiev2018higlass} offer promising post-processing solutions.
Yet, while some aggregation is necessary to make large graphs interpretable, it can also obscure important topological structures~\cite{liu2018graph}. 

\revised{\vspace{0.33em}\noindent\textbf{Backend and ML Scalability.}
\toolname has been deployed on graphs with up to millions of nodes/edges and tens of attributes per element, where it maintains interactive performance; in practice, the primary bottleneck is the underlying database resolver. While our architecture, detailed in \autoref{chap:implementation}, is designed to integrate high-performance graph analytics backends (e.g., GPU-accelerated and distributed frameworks) that can yield order-of-magnitude speedups over CPU-only libraries, these integrations are not yet part of the current release. 
Memory usage scales with the size of the visible subgraph rather than the full database, enabling analysis of very large graphs via strategic data paging.
}

\subsection{Broader Perspectives}
Relative to our design requirements, \toolname supports rapid query construction and immediate feedback through its grammar guarantees and interactive design. 
However, we observe three broader challenges that extend beyond \toolname and define open directions for graph analytics research.


\vspace{0.33em}\noindent\textbf{Missing Algebra of Graph Queries.}
Relational algebra provides a composable foundation for tabular analysis. 
Graph analytics lacks an equivalent framework. 
Existing languages differ in expressiveness and semantics, and no minimal operator set balances composability and structural reasoning. 
In \GPQL, restricting the operator set preserves composability but limits expressiveness. 
Without a unifying algebra, system design requires trade-offs, and users lack a consistent mental model of query behavior.

\vspace{0.33em}\noindent\textbf{Coupling Querying with Visualization for Exploration.}
\GPQL defines data retrieval but not visualization. 
In contrast, systems such as VizQL and APT couple queries with how the data should be visualized. 
To the best of our knowledge, no comparable framework exists for graphs. 
Bridging retrieval and visualization would enable a unified exploration process and remains an open problem.

\vspace{0.33em}\noindent\textbf{When Is Graph Analytics the Right Tool?}
Graph analysis is most effective when tasks require joint reasoning over attributes and topology. 
Purely structural tasks are handled by existing graph techniques, and purely attribute-based tasks are better served by tabular systems. 
Many real-world problems lie between these extremes, but simpler tasks are often inefficient in graph systems. 
Tools must therefore support both paradigms and guide users toward appropriate representations. 
While \toolname takes a step in this direction by integrating graph and attribute-based visualizations within a unified workflow, effectively guiding users between representations remains an open challenge.

\section{Conclusion}

As structurally complex, interconnected datasets become central to domains such as fraud detection, infrastructure resilience, and knowledge discovery, the limitations of aggregation-based analytics become increasingly evident.
Addressing these challenges requires tools that support the retrieval, exploration, and interpretation of graph databases.
We presented \toolname, a no-code Visual Analytics system that integrates graph querying and visualization into a unified workflow.
At its core, \GPQL formalizes analyst interactions as executable, vendor-agnostic queries, enabling interactive exploration without manual query construction.
Through two real-world use cases, we demonstrated how \toolname supports iterative analysis, reveals structural patterns, and enables the communication of results in practical settings.
These examples show how integrated querying and visualization facilitate analysis of complex relational data.
Limitations remain in scalability, query expressiveness, and interaction design.
These challenges define directions for future work.
\toolname establishes a foundation for accessible graph analytics and contributes toward broader adoption of graph-based analysis in Visual Analytics.

\bibliographystyle{abbrv-doi-hyperref}

\bibliography{template}


\appendix
\clearpage 
\section{Formative Study Questionnaire} \label{appendix:survey}

This appendix provides the full set of survey questions used in our formative study (\autoref{sec:user-requirements}). The survey consisted of 17 questions, combining multiple-choice, Likert-scale, ranking, and open-ended formats. Question wording has been lightly edited for brevity and readability. Full raw survey schema available on request.

\subsection{Background and Demographics}
\begin{itemize}
    \item What best describes your primary role?
    \item How many employees does your company, university, or research institute have?
    \item What is your company's primary industry or domain?
\end{itemize}

\subsection{Technical Experience}
\begin{itemize}
    \item Please rate your current experience in the following areas:
    \begin{itemize}
        \item Data analysis tools (e.g., Excel, Jupyter Notebooks, Tableau)
        \item Data visualization tools (e.g., Tableau, Power BI)
        \item Programming languages (e.g., Python, JavaScript)
        \item Graph database queries (e.g., Cypher, Gremlin)
    \end{itemize}
\end{itemize}

\subsection{Organizational Capabilities}
\begin{itemize}
    \item Please rate your organization's current capabilities in the following areas:
    \begin{itemize}
        \item Extracting data from graph databases
        \item Analyzing data relationships
        \item Visualizing complex data connections
        \item Real-time data analytics
        \item Cross-system data integration
        \item Collaborative data exploration
    \end{itemize}
\end{itemize}

\subsection{Time Allocation}
\begin{itemize}
    \item Approximately how much of your work time is spent on the following graph-related tasks?
    \begin{itemize}
        \item Extracting data from graph databases
        \item Visualizing data connections
        \item Integrating cross-system data
        \item etc.
    \end{itemize}
\end{itemize}

\subsection{Pain Points and Challenges}
\begin{itemize}
    \item Rank the following challenges from most to least problematic:
    \begin{itemize}
        \item Learning complex query languages
        \item Tool complexity and steep learning curves
        \item Dependence on technical teams
        \item Lack of real-time data exploration
    \end{itemize}
\end{itemize}

\subsection{Experience with Graph-Based Data}
\begin{itemize}
    \item Which of the following best describe your experience with graph-based data?
    \item What is the typical size of the graph networks your organization works with?
    \item How frequently do you use graph-based data in your work?
\end{itemize}

\subsection{Motivational Factors}
\begin{itemize}
    \item For each statement below, categorize it based on how much it would motivate you to use no-code analytics tools:
    \begin{itemize}
        \item Ability to create visualizations without programming knowledge
        \item Faster time-to-insight for business decisions
        \item Reduced dependence on IT/technical teams
        \item Interactive exploration of complex data relationships
    \end{itemize}
    \item Then rank each item within its category.
\end{itemize}

\subsection{Feature Importance}
\begin{itemize}
    \item Rate the importance of the following features:
    \begin{itemize}
        \item Visual query building (drag-and-drop interface)
        \item Real-time results and instant feedback
        \item Easy collaboration and sharing capabilities
        \item Integration with existing databases
        \item Export capabilities for presentations
    \end{itemize}
    \item Then rank your top 5 most desired features.
\end{itemize}

\subsection{Open-Ended Questions}
\begin{itemize}
    \item Please briefly describe your biggest frustration when working with data visualization tools.
    \item Is there anything else you'd like to share about your experience with graph-based analytics?
\end{itemize}

\section{System Implementation} \label{chap:implementation}

This appendix provides additional technical details on the implementation of \toolname, including component responsibilities, deployment configuration, and optimization-related design choices that complement the high-level description in~\autoref{chap:system}. We intentionally place these details in the appendix, as not all readers will be equally interested in low-level implementation aspects and the core contribution of this paper lies elsewhere.

\subsection{Extensible System Architecture}
\toolname is a modular, extensible Visual Analytics system designed to support iterative development and integration with evolving graph technologies (supporting R4). It follows an open-core approach\footnote{\href{https://en.wikipedia.org/wiki/Open-core_model}{https://en.wikipedia.org/wiki/Open-core\_model}} to foster community engagement and align with emerging academic and industry needs. This section highlights key design decisions that enable the system’s functionality, extensibility, and database interoperability.

Each backend module encapsulates a single analytical responsibility, such as query compilation, result transformation, or visualization configuration, allowing new functionality to be integrated without affecting core system behavior. The system is language-agnostic, enabling modules to be implemented in the most suitable environment (e.g., Python for machine learning, TypeScript for schema management), which supports rapid prototyping and research contributions.

The frontend of \toolname is built in React.js and uses a plugin-based design for visualizations. Each visualization module defines its configuration schema in JSON and can be independently contributed or modified, ensuring compatibility with diverse rendering libraries (including D3, Vega-Lite, and Cytoscape) without changes to the system core. This design facilitates rapid experimentation and adoption of new graph visualization techniques.

\subsection{Backend Architecture and Service Responsibilities}
The \toolname backend is structured as a collection of interconnected microservices, categorized into core, operational, and auxiliary microservices. This modular design ensures scalability, open-source and research contributions, including those from student theses and community contributions, and maintainability across analytical, collaborative, and infrastructural workflows.

\begin{figure}[h]
    \centering
    \includegraphics[width=1.0\linewidth]{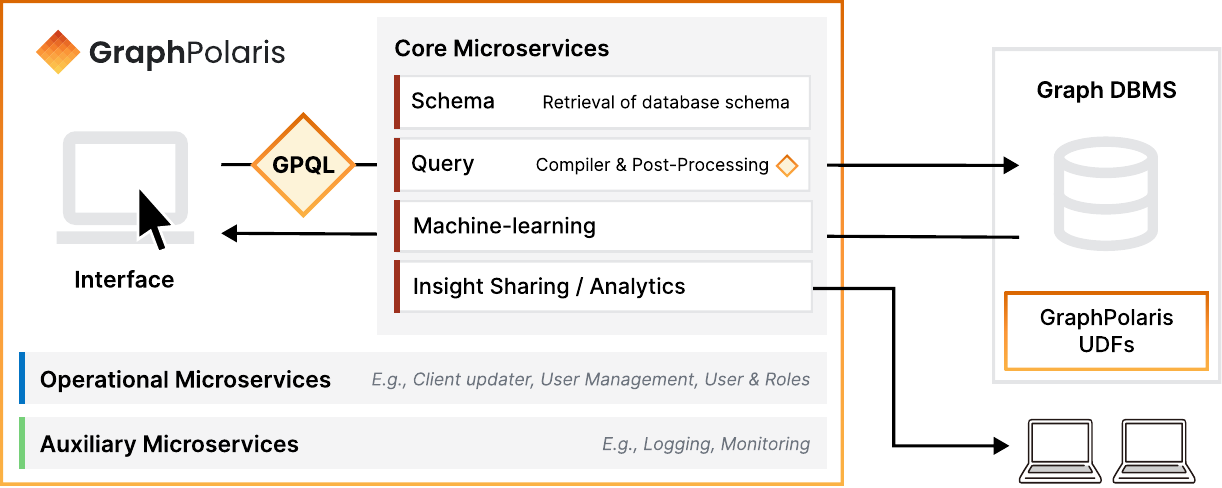}
    \caption{\textbf{User interaction with the \toolname interface} generates a \GPQL graph query. The analytical engine's query handler transpiles this query into a database-specific query language and post-processes the result before returning it to the interface for user interpretation. Additionally, we are enhancing the analysis workflow through a schema retrieval engine, insights-sharing/analytics, and machine-learning functionality.}
    \label{fig:system_diagram}
    \vspace{-1.2em}
\end{figure}

\smallskip

\sectiontitle{\textcolor{systemred}{\rule{2pt}{8pt}}~Core Microservices} form the analytical engine of the system. They transpile user-generated \GPQL queries into database-native graph query languages (e.g., Cypher, GQL), manage execution including \toolname-specific UDFs, and perform post-processing to prepare results for visualization.

\sectiontitle{\textcolor{systemblue}{\rule{2pt}{8pt}}~Operational Microservices} support essential system functionalities beyond query execution. These include user authentication and authorization, persistent session and state management, and the generation of shareable insight artifacts that preserve both query structure and visualization configuration for reproducibility.

\sectiontitle{\textcolor{systemgreen}{\rule{2pt}{8pt}}~Auxiliary Microservices} handle background infrastructure tasks such as logging, diagnostics, and communication tracing. While crucial to system robustness, they do not directly contribute to the analytical workflow and are therefore not discussed in this paper.

\smallskip

The microservice-based design of \toolname enables each service to be implemented in the programming language best suited to its function. For example, services involving machine learning or statistics are often implemented in Python to leverage its rich ecosystem of scientific libraries. Meanwhile, services focused on interface communication or schema management may be written in TypeScript for a type-safe integration of the frontend and backend. Although the current implementation of \toolname primarily uses Python and TypeScript, the system imposes no restrictions on language choice. This polyglot architecture ensures that new services, including those developed in the context of student theses or external research contributions, can be integrated into the existing ecosystem.

\subsection{Backend Advanced Analytics Functionalities}

\toolname supports advanced analytical functionality such as topological metrics, community and substructure detection, and representation learning (R3) through two strategies: (i) direct invocation of database-native functions where supported, and (ii) delegation to external analytical modules for operations not natively available.
Whenever possible, computations are executed within the graph database to minimize data transfer, reduce latency, and take advantage of native optimizations such as indexing and caching.

A key design decision concerns execution strategy: deciding when to rely on database-native functions versus external analytical modules.
Database-native functions are generally preferable for lightweight or moderately complex tasks such as degree computation, clustering coefficients, or path enumeration. For example, Neo4J supports the calculation of degree centralities, clustering coefficients, and shortest path functions. These operations benefit from the query engine’s optimizers, leverage pre-built indexes, and scale efficiently with graph size. However, advanced workloads often exceed what current graph engines can support natively.
For example, learning high-dimensional node embeddings, running iterative community detection at scale, or integrating GPU-accelerated machine learning models typically requires offloading to specialized services.
In these cases, external modules provide the flexibility to incorporate algorithms and hardware acceleration, albeit at the cost of higher latency and additional data movement.

Earlier versions of the system implemented UDFs as external microservices, but this approach introduced performance bottlenecks and infrastructure complexity. Our architecture favors tight integration with graph databases via APIs or query extensions. The shift has been enabled by recent advancements in query language standardization, particularly ISO/IEC 39075 GQL~\cite{ISO39075:2024}, which defines core functions such as \texttt{degree()}, \texttt{labels()}, and \texttt{path()} across database engines.


\subsection{Scalability}
\toolname addresses scalability concerns across backend processing, machine learning integration, and visualization rendering. Although we do not conduct a formal performance evaluation in this work, our informal assessment proves the system's ability to handle large-scale datasets while maintaining interactive performance. We attribute this to our central design decision to diverge from the conventional ``Overview First, Filter, then Detail-on-Demand'' model~\cite{DBLP:conf/vl/Shneiderman96} by adopting the search-focused paradigm of ``Search, Show Context, Expand on Demand'' as proposed by van Ham and Perer~\cite{DBLP:journals/tvcg/HamP09}. This approach proves effective for graph analytics, where analysts typically possess domain hypotheses and seek to explore local neighborhoods rather than comprehend global structure. Context expansion operates through degree-of-interest functions, dynamically adjusting visible subgraphs as exploration scope expands.


\subsubsection*{Backend and ML Scalability}
We assessed \toolname using our enterprise clients' medium to large-scale graphs with hundreds of thousands to millions of nodes, millions of edges, and tens of attributes per element. The system showed robust and interactive performance rates. The bottleneck was mostly found in the database resolver, which we addressed by our own strategic query optimization (publication planned). 
While the architectural foundation enables seamless integration with high-performance machine learning frameworks—such as RAPIDS cuGraph for GPU-accelerated computation~\cite{DBLP:conf/hpec/HricikBG20,nvidia2020rapids} and distributed systems, like Gradoop or Google Pregel, for large-scale graph analytics~\cite{DBLP:journals/vldb/RostGTFSCAJR22,DBLP:conf/sigmod/MalewiczABDHLC10}-these capabilities have not yet been implemented in the current version of \toolname but promise 50--500$\times$ performance improvements over NetworkX~\cite{osti_960616}. Memory consumption scales linearly with visible subgraph size rather than total graph size, enabling analysis of massive graphs through strategic data paging.

\subsubsection*{Visualization Scalability Strategies}
As also mentioned in the paper, several visualizations, e.g., node-link diagrams or matrix plots, support interactive exploration of hundreds of thousands of nodes and millions of edges through GPU-accelerated rendering. Map visualizations address overplotting via R-tree spatial indexing~\cite{DBLP:conf/sigmod/Guttman84} with dynamic clustering based on zoom level. Structured visualizations, like PaohVis and Sankey, implement built-in aggregation to reduce visual complexity while preserving analytical utility~\cite{buono2021hypergraph,DBLP:conf/infovis/RiehmannHF05}.

\subsection{Performance Characteristics}

Current implementation establishes practical limits around $10^5$ visible nodes for interactive visualization, with backend support extending to $10^7$ nodes. Certain operations requiring global properties (e.g., complete centrality measures) necessitate backend delegation. While node-link diagrams maintain interactivity with hundreds of thousands of elements, matrix visualizations face quadratic complexity limiting practical application without aggregation. This framework positions GraphPolaris to handle enterprise-scale analytics while preserving the interactive workflow that distinguishes it from batch-processing alternatives.

\begin{table*}[ht]
  \small
  \centering
  \begin{tabularx}{\textwidth}{@{}l X l l l l@{}}
    \toprule
    \textbf{Category} & \textbf{Function(s)} & \textbf{Operand} & \textbf{Composability} & \textbf{Execution} & \textbf{Status} \\
    \midrule
    Attribute Predicates
      & $=,\neq,>,\geq,<,\leq$, \texttt{contains}, \texttt{substring}, \texttt{in\_list}, \texttt{is\_null}
      & Attribute & Predicate ($F_e,F_r$) & Database-native & Implemented \\
    \midrule
    Aggregation \& Statistics
      & \texttt{avg}, \texttt{count}, \texttt{max}, \texttt{median}, \texttt{min}, \texttt{sum}
      & Attribute or Topology & Terminal ($\omega$) & Post-processing & Implemented \\
    \midrule
    \multirow{2}{*}{Topological Metrics}
      & degree ($>,\geq,<,\leq,=,\neq,$ between)
      & Topology & Predicate ($F_e$) & Post-processing & Implemented \\
      & centrality (betweenness, closeness, PageRank)
      & Topology & Terminal ($\omega$) & External (ml-service) & Part. Impl. \\
    \midrule
    \multirow{2}{*}{Community \& Substructure Detection}
      & label propagation (default)
      & Topology & Terminal ($\omega$) & External (ml-service) & Implemented \\
      & Louvain modularity
      & Topology & Terminal ($\omega$) & External (ml-service) & Implemented \\
    \midrule
    \multirow{2}{*}{Path \& Link Functions}
      & shortest path (single source/target)
      & Topology & Terminal ($\omega$) & External (ml-service) & Implemented \\
      & link prediction (Jaccard coefficient)
      & Topology & Terminal ($\omega$) & External (ml-service) & Implemented \\
    \midrule
    \multirow{4}{*}{Representation Learning \& ML}
      & node embeddings (e.g.\ Node2Vec, GraphSAGE)
      & Hybrid & Terminal ($\omega$) & External (planned) & Planned \\
      & node classification
      & Hybrid & Terminal ($\omega$) & External (planned) & Planned \\
      & clustering ($k$-means, DBSCAN)
      & Attribute & Terminal ($\omega$) & External (planned) & Planned \\
      & anomaly detection
      & Hybrid & Terminal ($\omega$) & External (planned) & Planned \\
    \bottomrule
  \end{tabularx}
  
  \vspace{-0.5em}
  \caption{User-defined functions (UDFs) in the current \toolname implementation.
  \emph{Operand} indicates whether the function consumes a stored node/edge attribute or a value derived from graph
  topology. \emph{Execution} indicates whether the operation is pushed into the native database query, computed by
  GraphPolaris over the retrieved subgraph, or delegated to an external service. \emph{Planned} entries are wired
  into the type system and message routing but have no algorithm implementation yet.}
  \label{tab:udfs}
  \vspace{-1.5em}
\end{table*}

\section{Implemented UDF/Terminal functions}\label{appendix:udf_terminalfunctions}

The \autoref{tab:udfs} summarizes the user-defined functions currently supported or planned in \toolname, grouped by analytical category. Attribute predicates and basic aggregations are implemented and either pushed down to the database or computed in the retrieval backend, while more advanced graph-analytic functions, such as centrality, communities, shortest paths, link prediction, and representation-learning–based ML, are executed via external services, with some already implemented and others prepared in the type library but still pending algorithmic integration.

\section{Exploration Workflows}\label{appendix:use_cases}

This additional use-case section focuses on the exploration sequence and demonstrates how \toolname supports an iterative analytical workflow where each step builds on insights from the previous one. It shows how users can progressively refine their understanding of the data through exploration rather than simply constructing a single query.

\subsection{Use-Case 1: Analyzing Publications by Keywords}

\begin{figure}[H]
    \centering
    \includegraphics[width=1.0\linewidth]{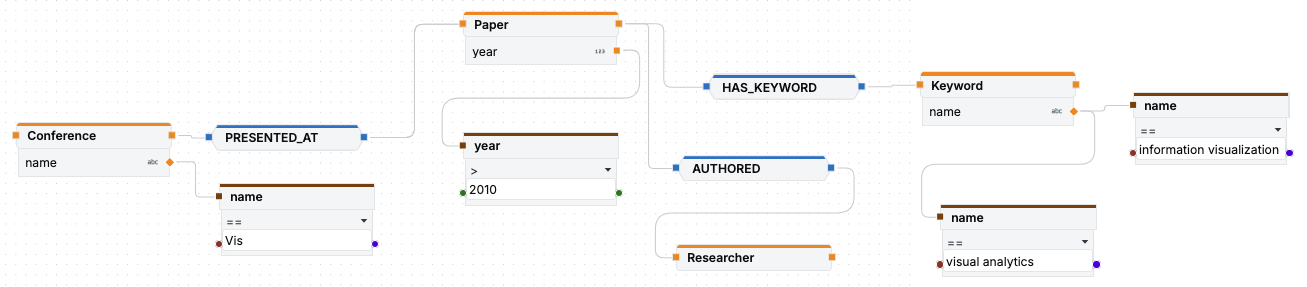}
    \caption{\textbf{Graph query retrieving IEEE VIS papers} published after 2015, their authors, and associated keywords. This analysis maps out topic coverage across contributors.}  
    \label{fig:complex_query} 
\end{figure}

The first showcase investigates multi-hop relationships between authors, papers, and research keywords to understand the topical and geographic distribution of recent IEEE VIS publications. As shown in \autoref{fig:complex_query}, the query identifies papers published at IEEE VIS conferences after 2015, retrieves their authors, and extracts associated keywords.

\subsubsection{Step 1: Initial Exploration Question "VIS Publications over time and conference"}

\begin{figure}[H]
    \centering
    \begin{tcolorbox}[enhanced, drop fuzzy shadow southeast, 
                      boxrule=0.4pt, sharp corners, 
                      colframe=darkgray, colback=white, 
                      width=0.80\linewidth]
        \includegraphics[width=0.9\linewidth]{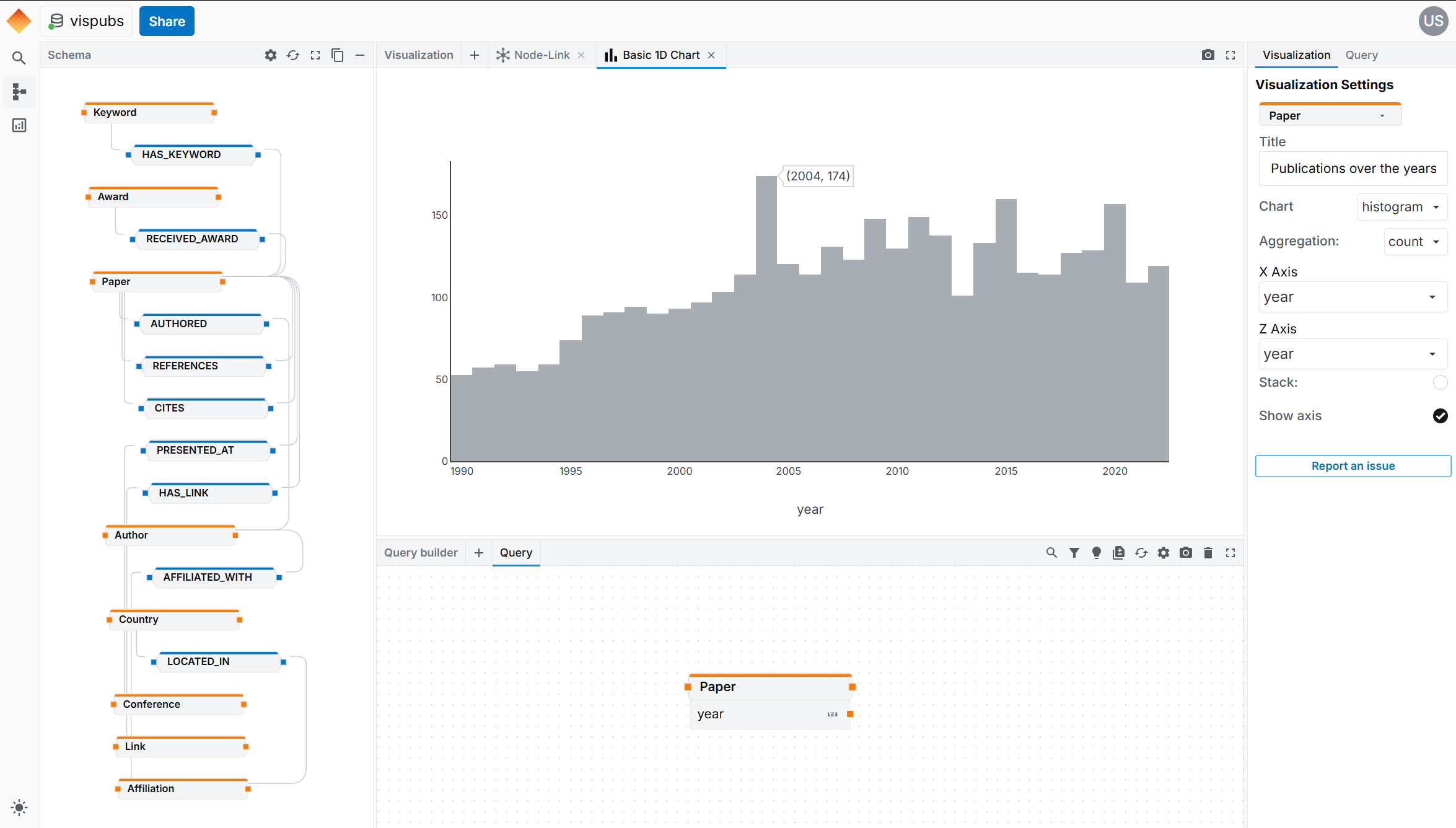}
    \end{tcolorbox}
    \caption{\textit{Let us initially look at Vis publications over time.} In the histogram view, VIS shows a steadily increasing trend, peaking at 174 papers in 2024 compared to slightly fewer than 100 papers in 2013.}
    \label{fig:u1_step1a}
\end{figure}

\begin{figure}[H]
    \centering
    \begin{tcolorbox}[enhanced, drop fuzzy shadow southeast, 
                      boxrule=0.4pt, sharp corners, 
                      colframe=darkgray, colback=white, 
                      width=0.8\linewidth]
        \includegraphics[width=0.9\linewidth]{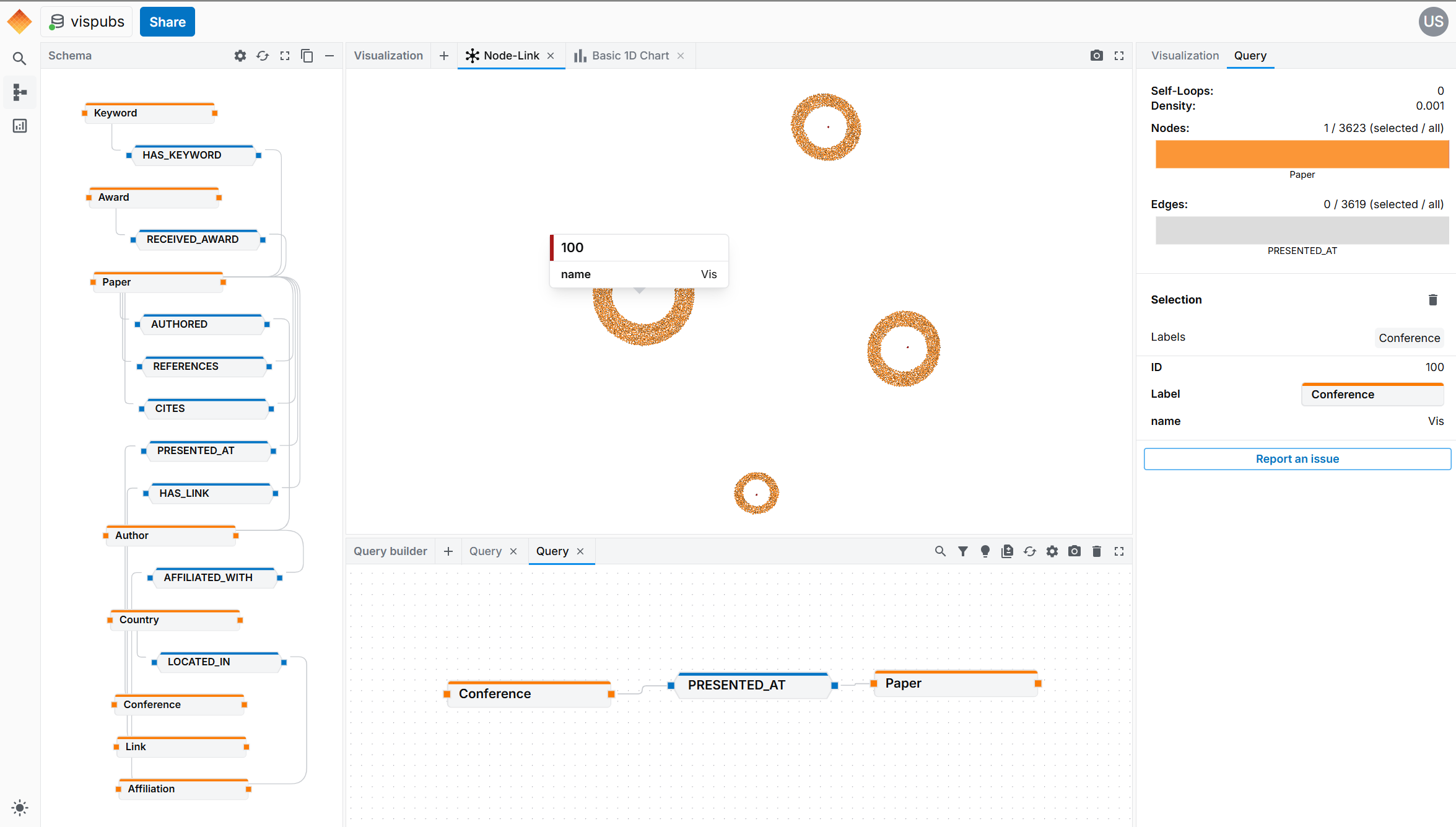}
    \end{tcolorbox}
    \caption{\textit{How are the Vis publications distributed per conference?}; We see a skewed distribution of the total 3623 papers per conference.}
\end{figure}

\begin{figure}[H]
    \centering
    \begin{tcolorbox}[enhanced, drop fuzzy shadow southeast, 
                      boxrule=0.4pt, sharp corners, 
                      colframe=darkgray, colback=white, 
                      width=0.8\linewidth]
        \includegraphics[width=0.9\linewidth]{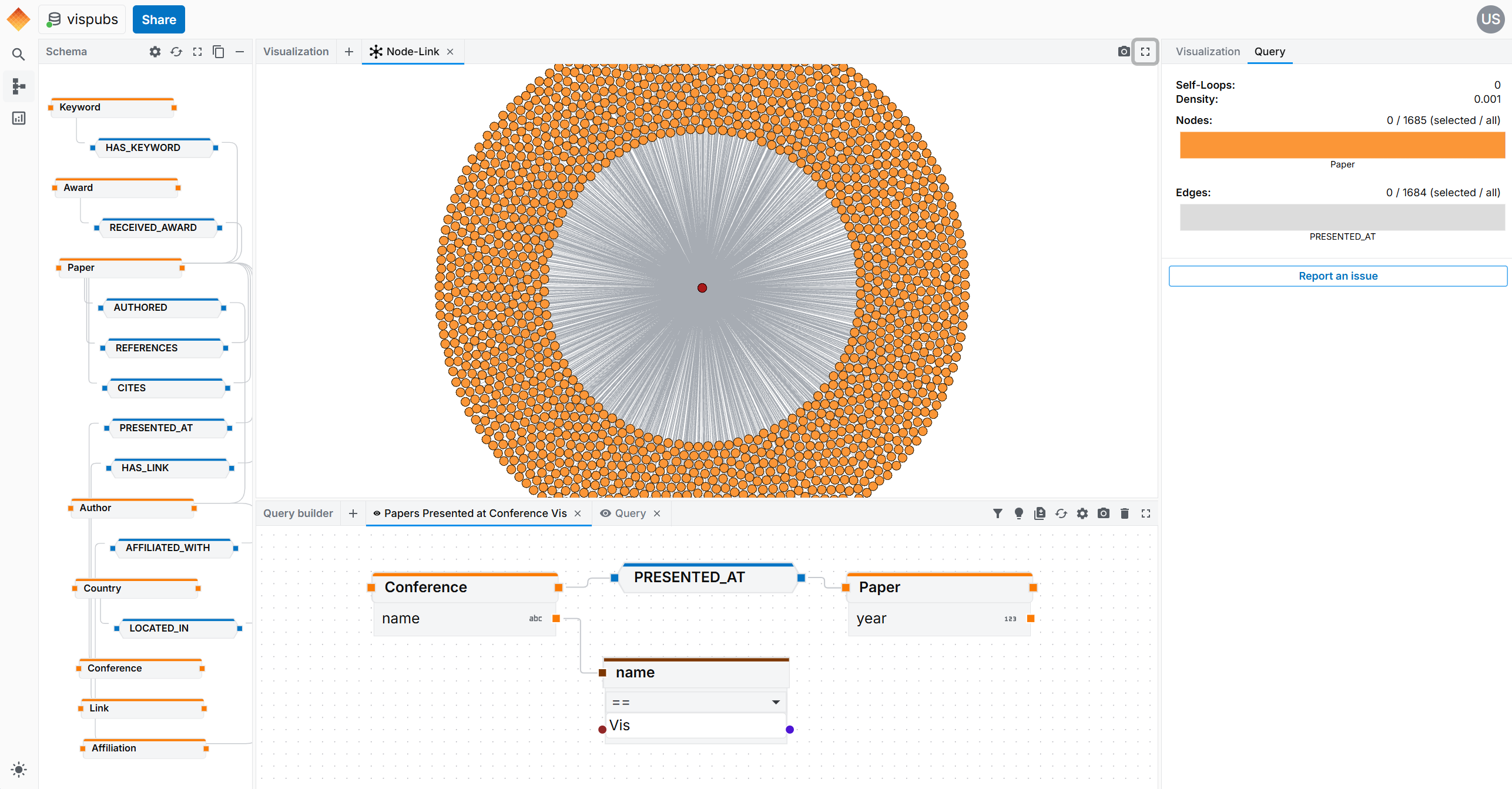}
    \end{tcolorbox}
    \caption{\textit{Let us drill down and focus on Vis publications for the 'Vis' conference only} The detail panel shows that 1685 papers in the database relate to the 'Vis' conference; This is 46,5\% of the papers.}
\end{figure}

\subsubsection{Step 2: Refining the Query "Who are the researchers contributing to these papers?"}

\begin{figure}[H]
    \centering
    \begin{tcolorbox}[enhanced, drop fuzzy shadow southeast, 
                      boxrule=0.4pt, sharp corners, 
                      colframe=darkgray, colback=white, 
                      width=0.9\linewidth]
        \includegraphics[width=1\linewidth]{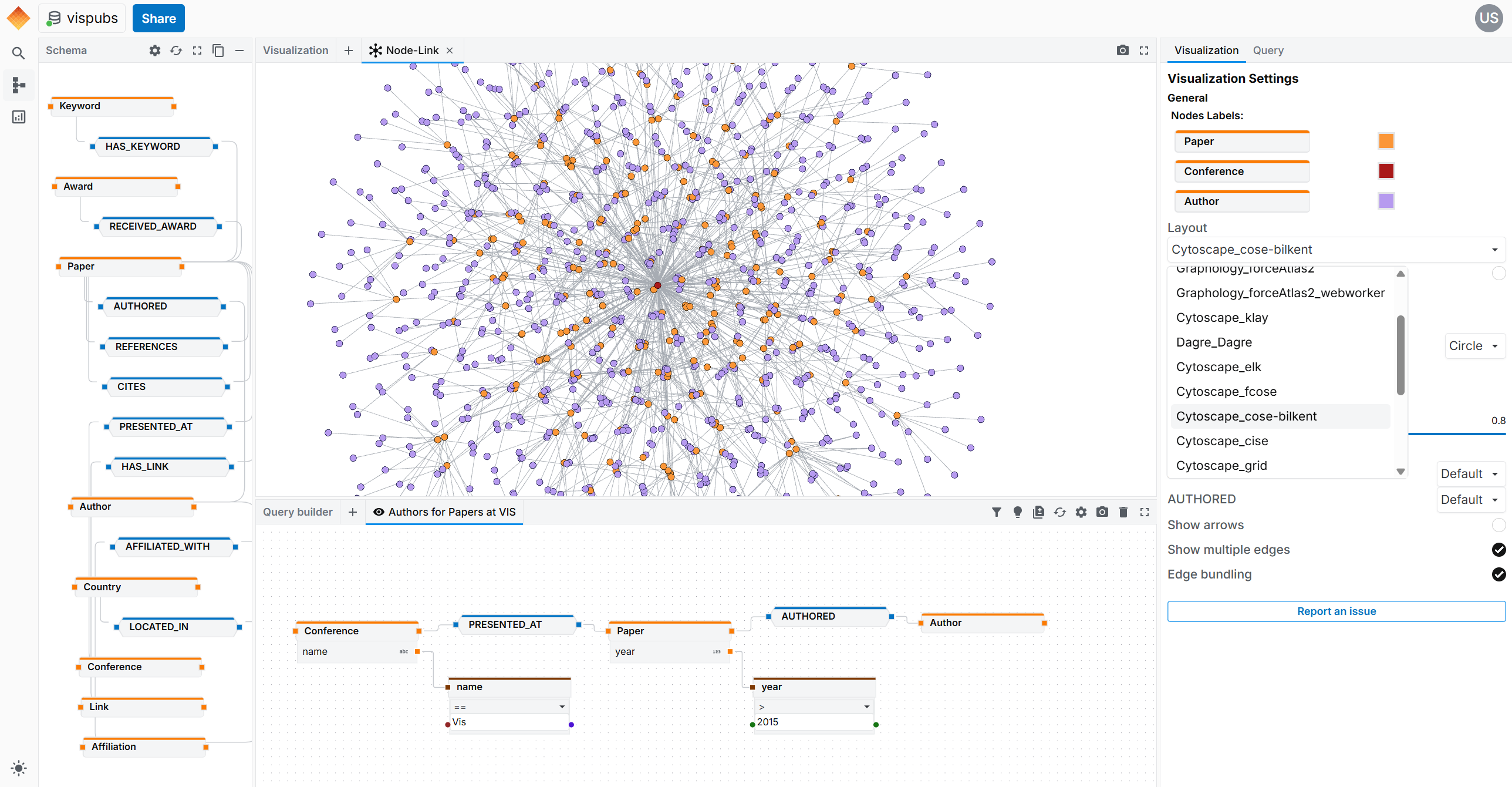}
    \end{tcolorbox}
    \caption{\textit{What if we add authors to the question?} Adding authors to the query does not reveal a clear pattern. The layout \texttt{Cytoscape\_cose-bilkent} appears cluttered. }
\end{figure}

\begin{figure}[H]
    \centering
    \begin{tcolorbox}[enhanced, drop fuzzy shadow southeast, 
                      boxrule=0.4pt, sharp corners, 
                      colframe=darkgray, colback=white, 
                      width=0.9\linewidth]
        \includegraphics[width=1\linewidth]{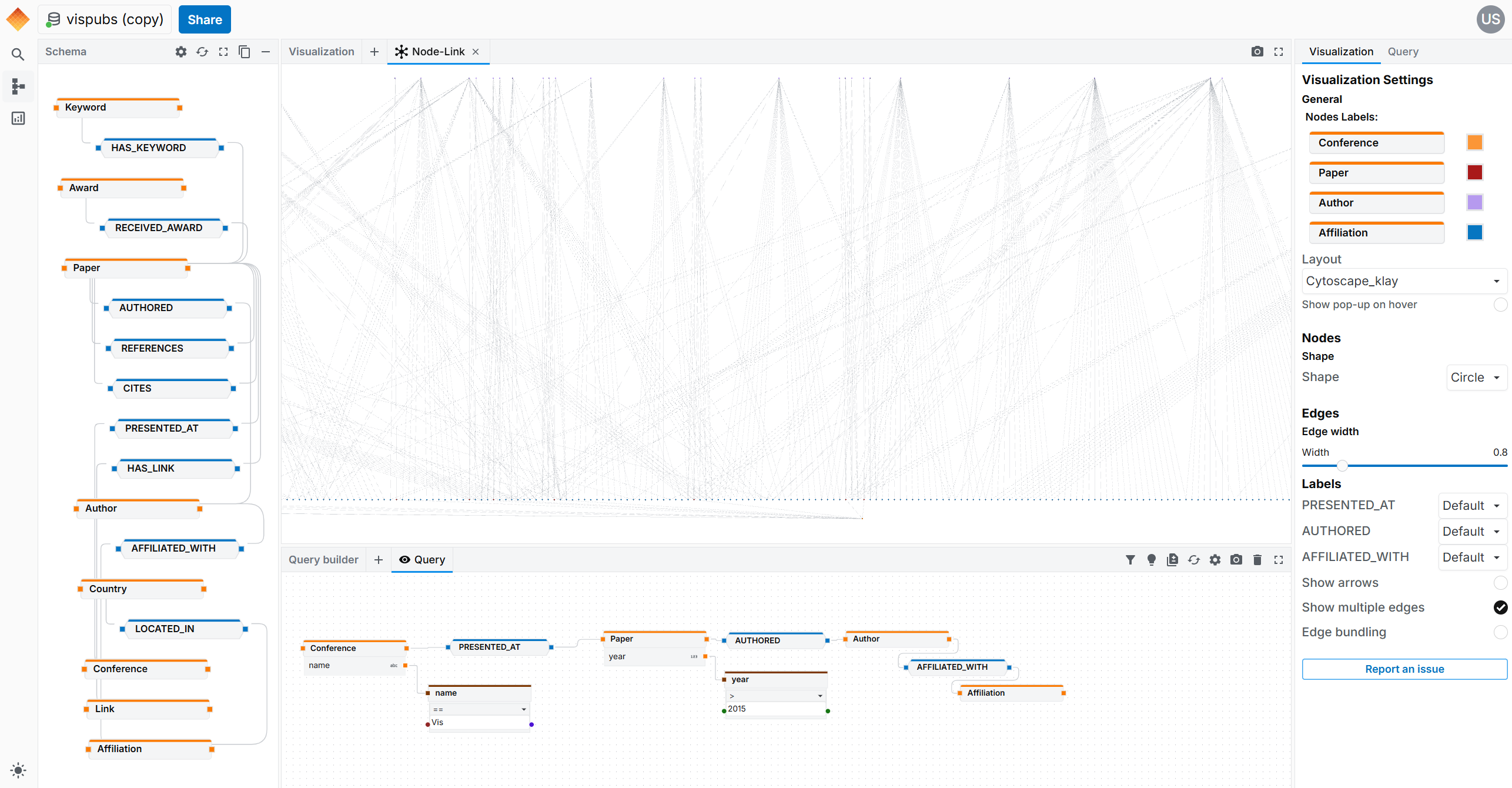}
    \end{tcolorbox}
    \caption{\textit{Maybe authors are too fine-granular? Let us raise the abstraction level and add the author affiliations to distinguish patterns.} The layout \texttt{Cytoscape\_klay} (K-Layers) appears cluttered and has problems distinguishing the node types. This is a general problem we see. The manual intervention required for parameterizing node-link diagram layouts will require a high technical skill set, which starkly contrasts our data democratization goal.}
\end{figure}

\begin{figure}[H]
    \centering
    \begin{tcolorbox}[enhanced, drop fuzzy shadow southeast, 
                      boxrule=0.4pt, sharp corners, 
                      colframe=darkgray, colback=white, 
                      width=1.0\linewidth]
        \includegraphics[width=1\linewidth]{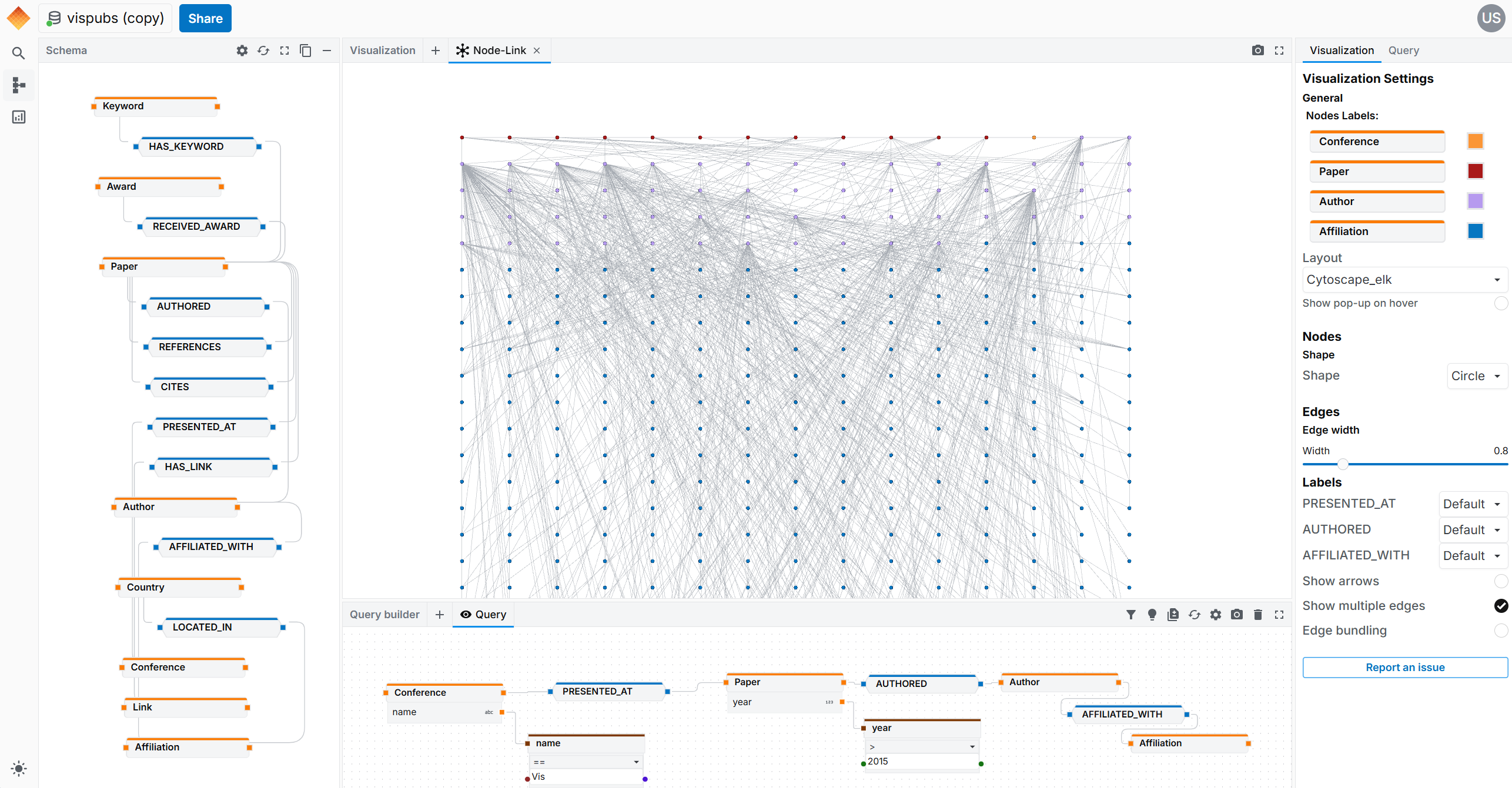}
    \end{tcolorbox}
    \caption{\textit{Maybe this is an algorithmic or layout problem?} To further investigate the meta-finding here, we opt for the \texttt{Cytoscape\_elk} algorithm, initially designed for multi-layer networks. We can generally postulate that most of the node-link layout algorithms are not suited for general-purpose use without strong parameterization effort. Only the force-directed algorithms appear to be giving interpretable results with a low parameterization effort.}
\end{figure}

\begin{figure}[H]
    \centering
    \begin{tcolorbox}[enhanced, drop fuzzy shadow southeast, 
                      boxrule=0.4pt, sharp corners, 
                      colframe=darkgray, colback=white, 
                      width=1.0\linewidth]
        \includegraphics[width=1\linewidth]{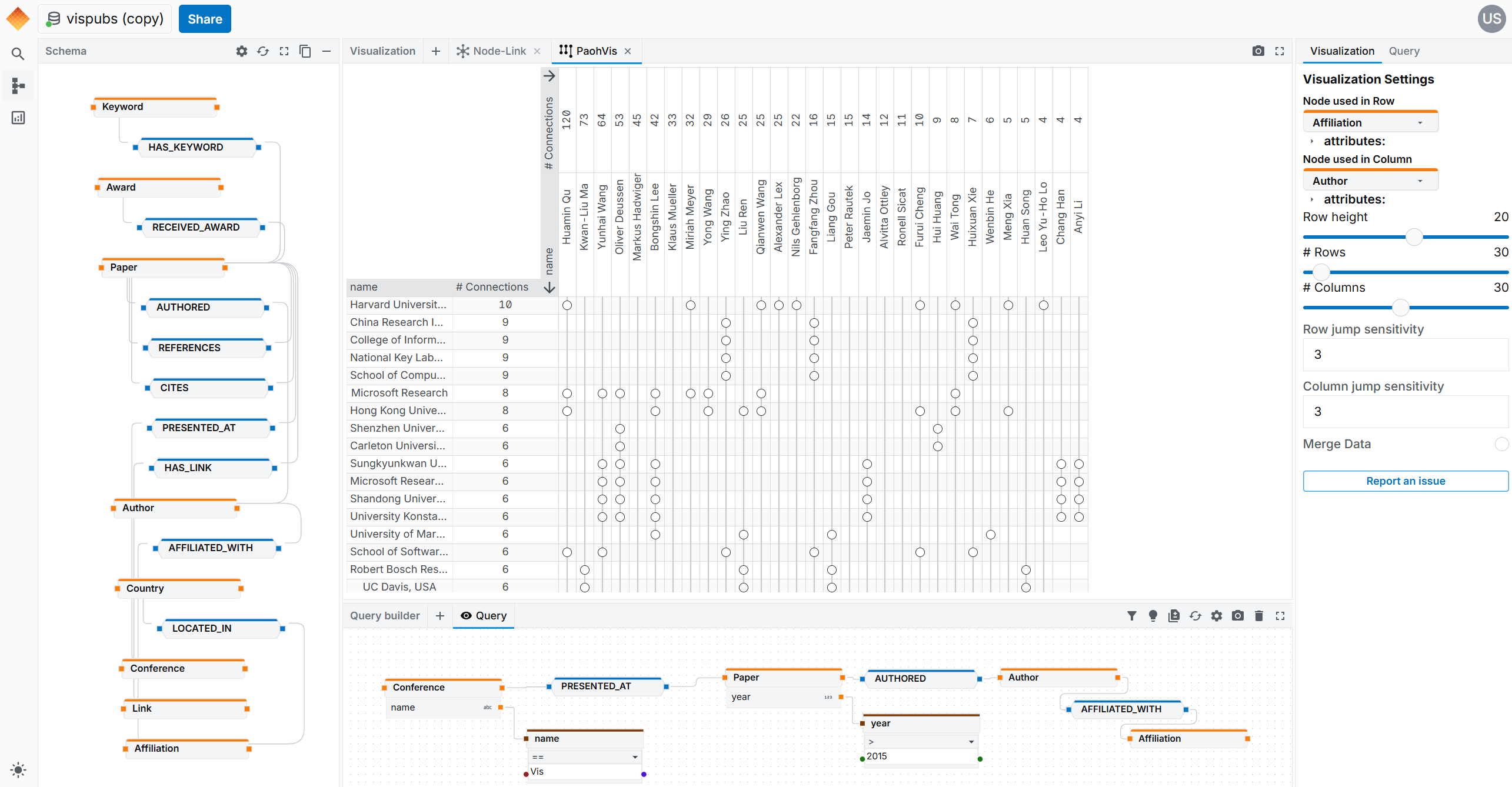}
    \end{tcolorbox}
    \caption{\textit{Maybe the visualization is the problem?} Drilling down into the data with a table-based visualization (like \texttt{PaohVis}), here sorted both times on the x- and y-axis by \textit{\# Connections}, Let us us retrieve the top connected authors in our field along with the top connected affiliations. }
\end{figure}

\subsubsection{Step 3: Pattern Discovery "How do keywords co-occur across publications?"}

Please be aware in this section that those are user-defined keywords. The process is up to the authors to describe their digression. In other words, a paper can be still a classical 'Visual Analytics' paper, even if the author did not put the keyword in the keyword list.

\begin{figure}[H]
    \centering
    \begin{tcolorbox}[enhanced, drop fuzzy shadow southeast, 
                      boxrule=0.4pt, sharp corners, 
                      colframe=darkgray, colback=white, 
                      width=1.0\linewidth]
        \includegraphics[width=1\linewidth]{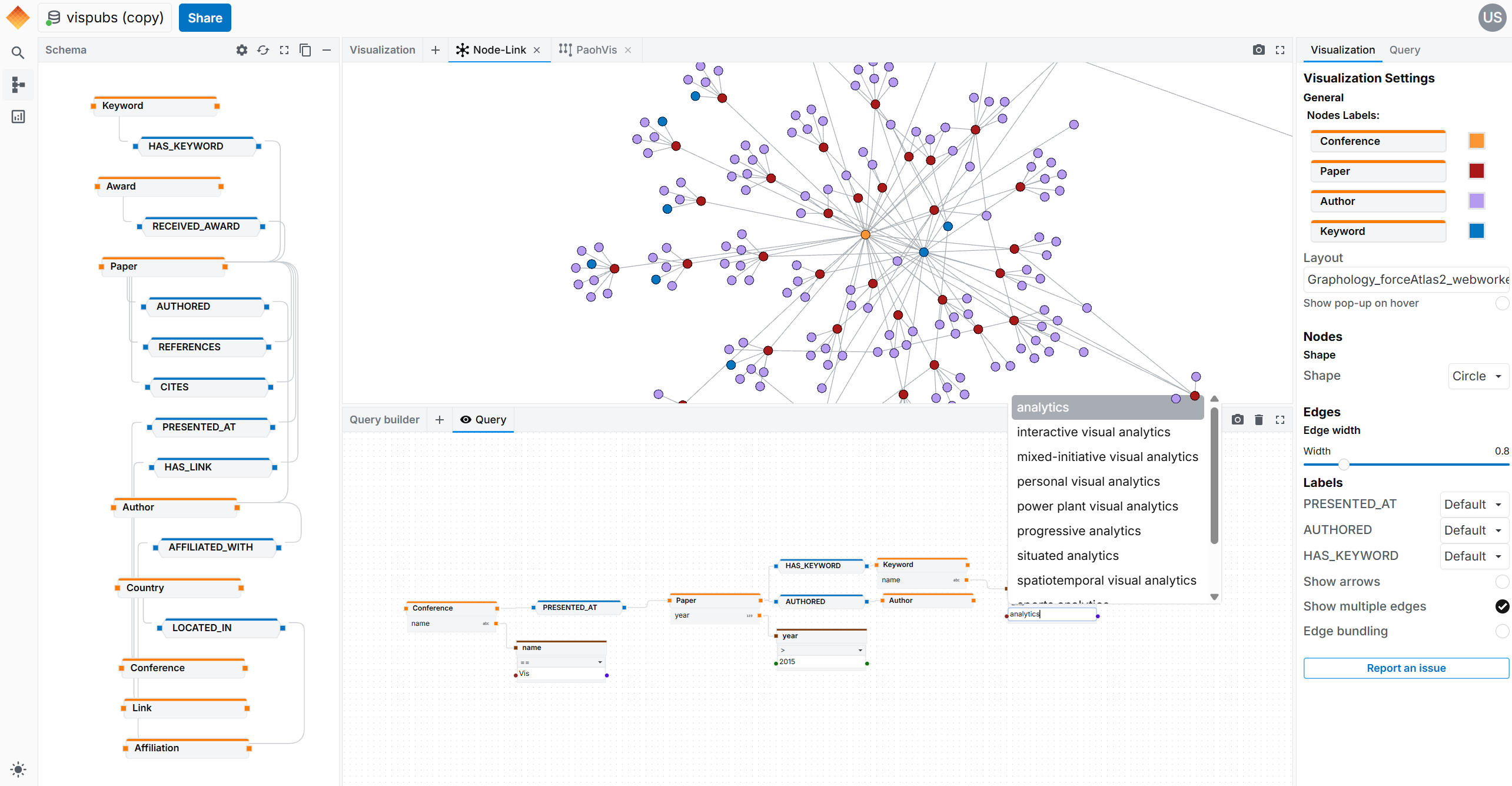}
    \end{tcolorbox}
    \caption{\textit{Which keywords start or end with "analytics"?} By adding a filter on the keyword attribute 'name' we can see which papers contain keywords with "analytics" being a subpart of a keyword. }
\end{figure}

\begin{figure}[H]
    \centering
    \begin{tcolorbox}[enhanced, drop fuzzy shadow southeast, 
                      boxrule=0.4pt, sharp corners, 
                      colframe=darkgray, colback=white, 
                      width=0.7\linewidth]
        \includegraphics[width=1.0\linewidth]{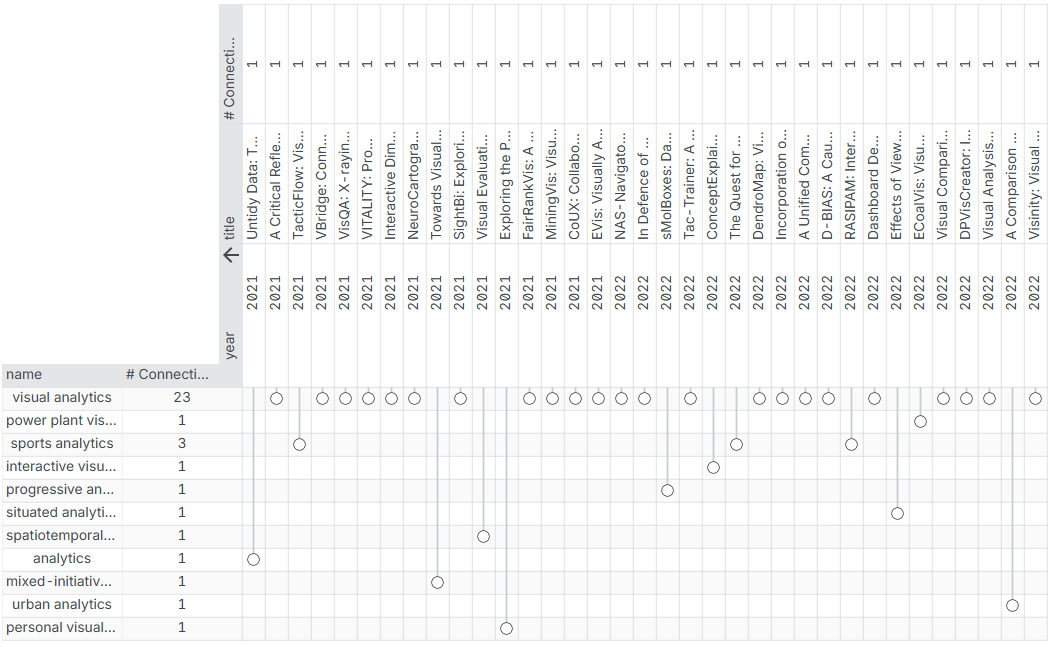}
    \end{tcolorbox}
    \caption{\textit{Which keywords start or end with "analytics"?} Looking at the PaohVis publication shows that 'Visual Analytics' dominates the 'analytics' keywords and never co-occurs with another 'analytics' keyword. }
\end{figure}

\begin{figure}[H]
    \centering
    \begin{tcolorbox}[enhanced, drop fuzzy shadow southeast, 
                      boxrule=0.4pt, sharp corners, 
                      colframe=darkgray, colback=white, 
                      width=1.0\linewidth]
        \includegraphics[width=1\linewidth]{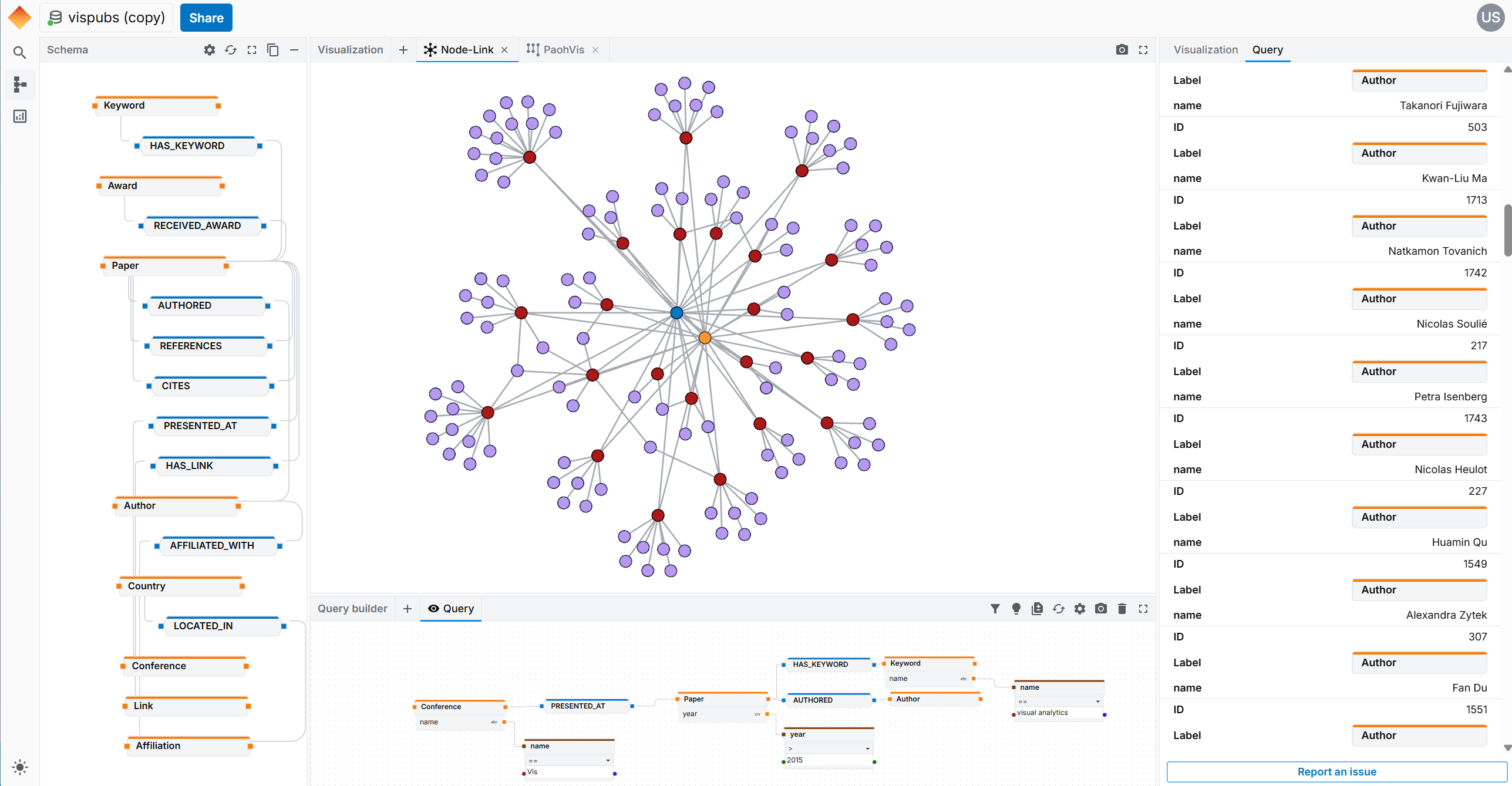}
    \end{tcolorbox}
    \caption{\textit{Who are the authors using "Visual Analytics" to describe their work?} Adding authors to the query makes this a tertiary relationship analysis and requires us to switch to the node-link diagram. Using the detail panel enables us to scroll through the list of 141 authors using this keyword.}
\end{figure}

\begin{figure}[H]
    \centering
    \begin{tcolorbox}[enhanced, drop fuzzy shadow southeast, 
                      boxrule=0.4pt, sharp corners, 
                      colframe=darkgray, colback=white, 
                      width=1.0\linewidth]
        \includegraphics[width=1\linewidth]{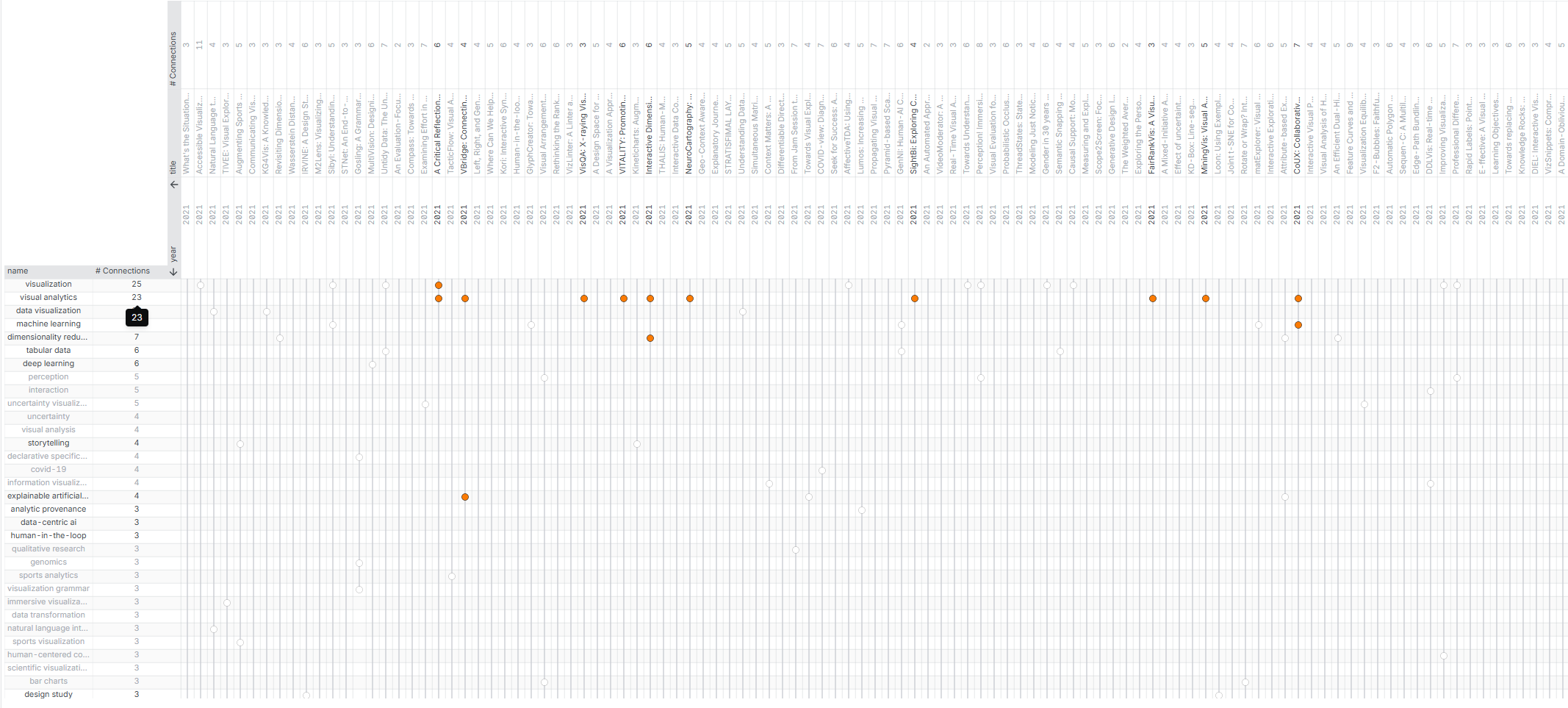}
    \end{tcolorbox}
    \caption{\textit{Which keywords co-occur with 'Visual Analytics'?}  \texttt{PaohVis} lends itself to set analytics questions such as with which keywords 'Visual Analytics' keyword papers co-occur? Here we recognize a wide spread of expected keywords, like 'visualization', 'machine learning', and 'storytelling'. Interestingly, we can see a grouping of keywords around the topic of artificial intelligence with topics like 'deep learning', 'explainable AI', or 'data-centric AI'.}
\end{figure}

\begin{figure}[H]
    \centering
    \begin{tcolorbox}[enhanced, drop fuzzy shadow southeast, 
                      boxrule=0.4pt, sharp corners, 
                      colframe=darkgray, colback=white, 
                      width=1.0\linewidth]
        \includegraphics[width=1\linewidth]{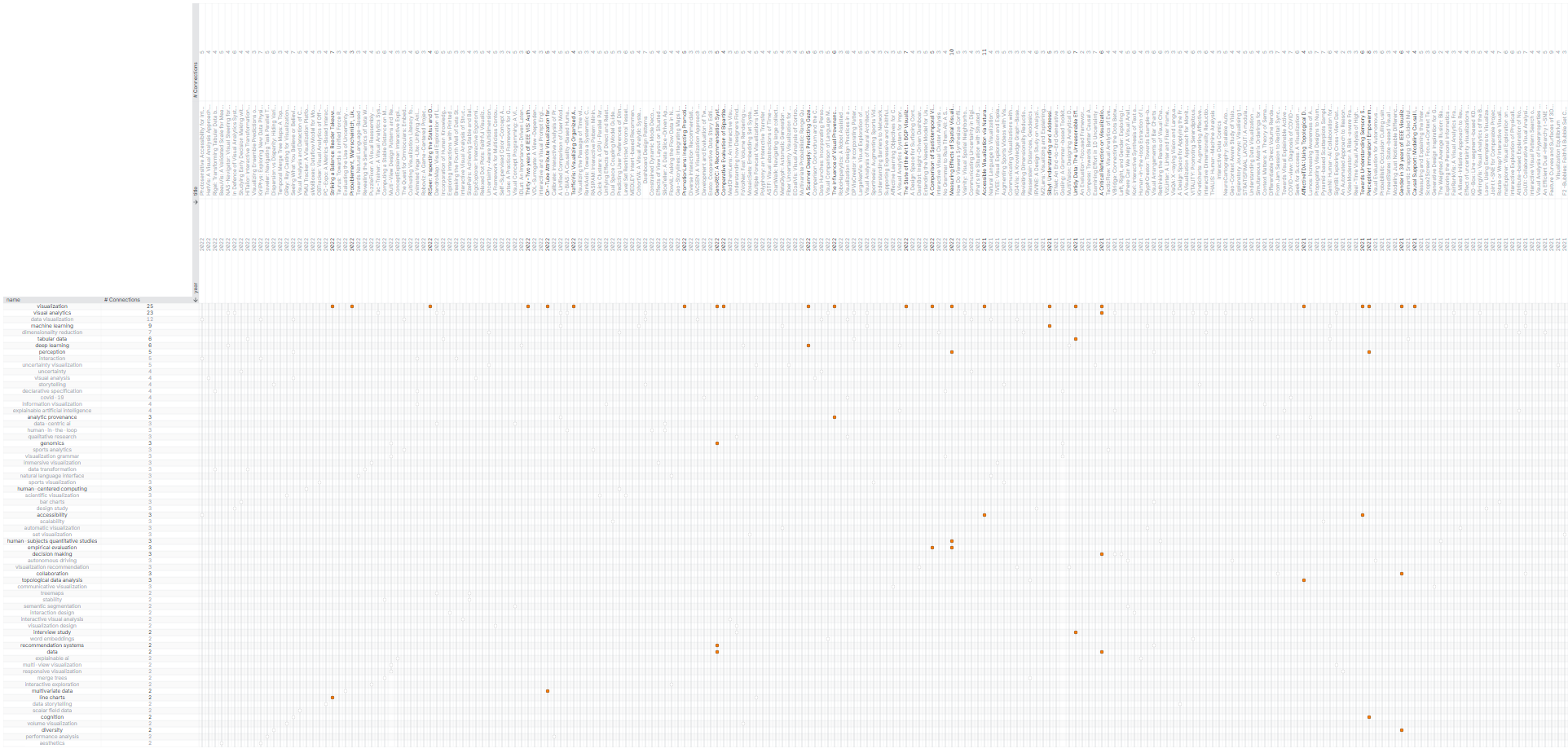}
    \end{tcolorbox}
    \caption{\textit{Follow-up: Which keywords co-occur with 'Visual Analytics'?} We can adapt the \textit{row height} of \texttt{PaohVis}. This adaption reduces the readability of the text but lets us explore this sparse dataset and the extended list of keywords co-occurring with 'Visual Analytics'.}
\end{figure}

\begin{figure}[H]
    \centering
    \begin{tcolorbox}[enhanced, drop fuzzy shadow southeast, 
                      boxrule=0.4pt, sharp corners, 
                      colframe=darkgray, colback=white, 
                      width=0.6\linewidth]
        \includegraphics[width=1\linewidth]{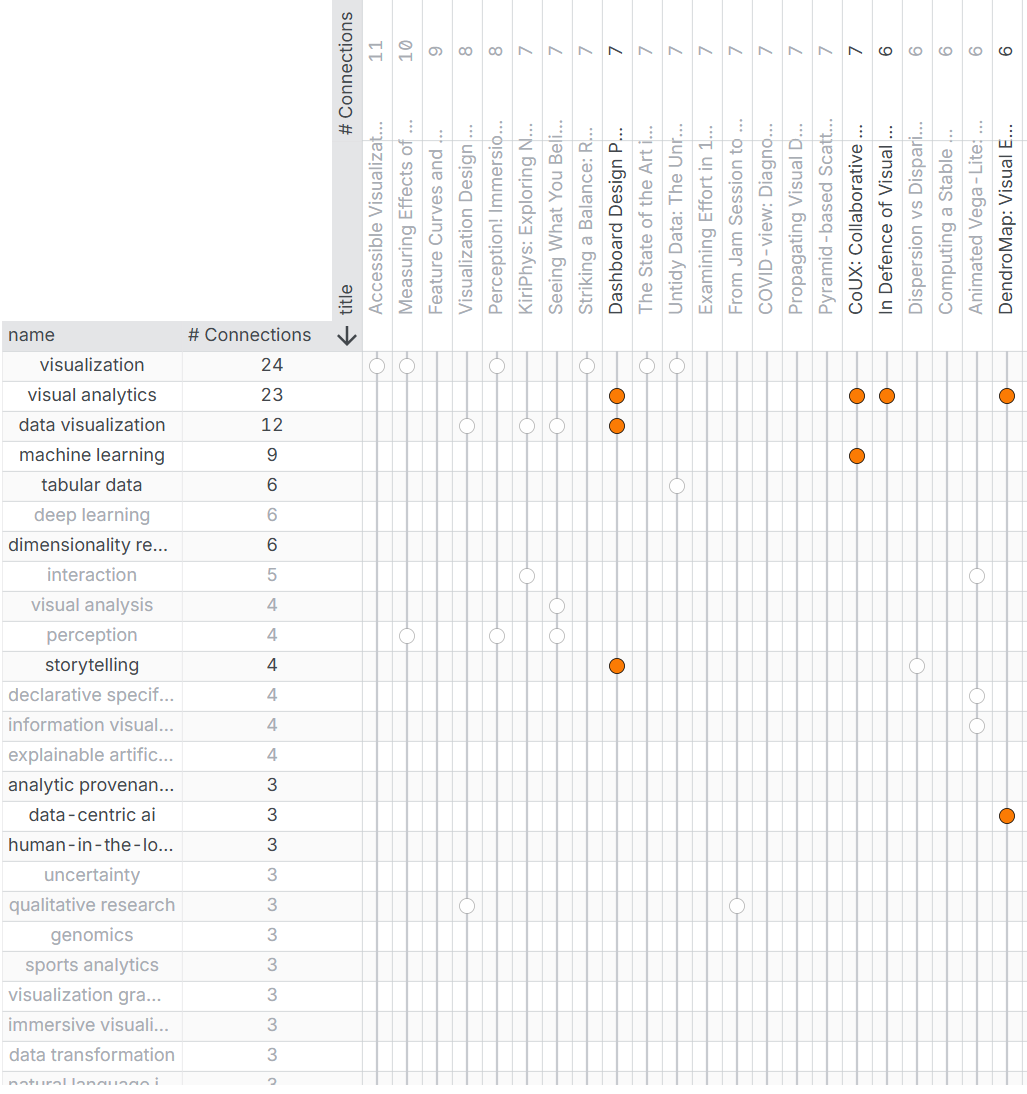}
    \end{tcolorbox}
    \caption{The final result of the exploration reveals key trends in topical coverage. Commonly occurring keywords include 'visual analytics' and 'data visualization', which frequently co-occur with terms such as 'human-in-the-loop', 'benchmark study', and 'machine learning'. Publications with more keywords tend to be broader in scope, while papers with fewer keywords often feature more specific terms such as 'binary sequence' or 'visual abstraction'. This illustrates how even simple path-based queries can yield rich, interpretable insights.}
\end{figure}

\subsection{Detecting Citation-Based Research Communities}
\begin{figure}[H]
    \centering
    \vspace{-1.5em}
    \includegraphics[width=0.8\linewidth]{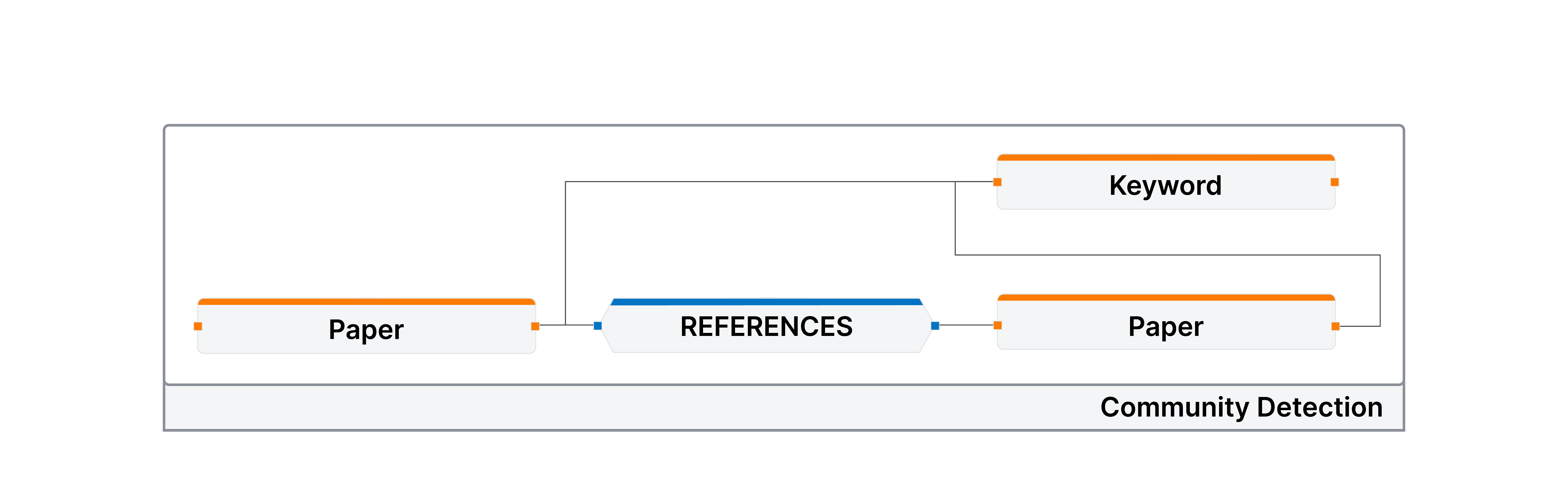}
    \caption{\textbf{Graph query identifying clusters of papers} that reference each other and share at least one keyword. A community detection algorithm groups them into semantically meaningful research clusters.} 
    \label{fig:community_keywords} 
\end{figure}

The second showcase, illustrated by the query in \autoref{fig:community_keywords}, highlights a pattern where groups of papers not only cite each other but also share common keywords. Academic research is inherently interconnected, often resulting in the formation of thematic communities—clusters of papers that reference one another and exhibit shared terminology. Identifying these communities can reveal subfields, emerging trends, and influential works within the visualization research domain. To capture this pattern, we applied a meta pill over a path, utilizing community detection as a terminal function in our graph query. This methodology allows for the identification of clusters of closely related publications based on both citation relationships and shared keywords. 

Generally, we have to acknowledge that the VisPubs dataset might be incomplete with respect to paper cross-references and citations.

\subsubsection{Step 1: Initial Exploration Question: "How does the community cite each other?"}

\begin{figure}[H]
    \centering
    \begin{tcolorbox}[enhanced, drop fuzzy shadow southeast, 
                      boxrule=0.4pt, sharp corners, 
                      colframe=darkgray, colback=white, 
                      width=0.75\linewidth]
        \includegraphics[width=1.0\linewidth]{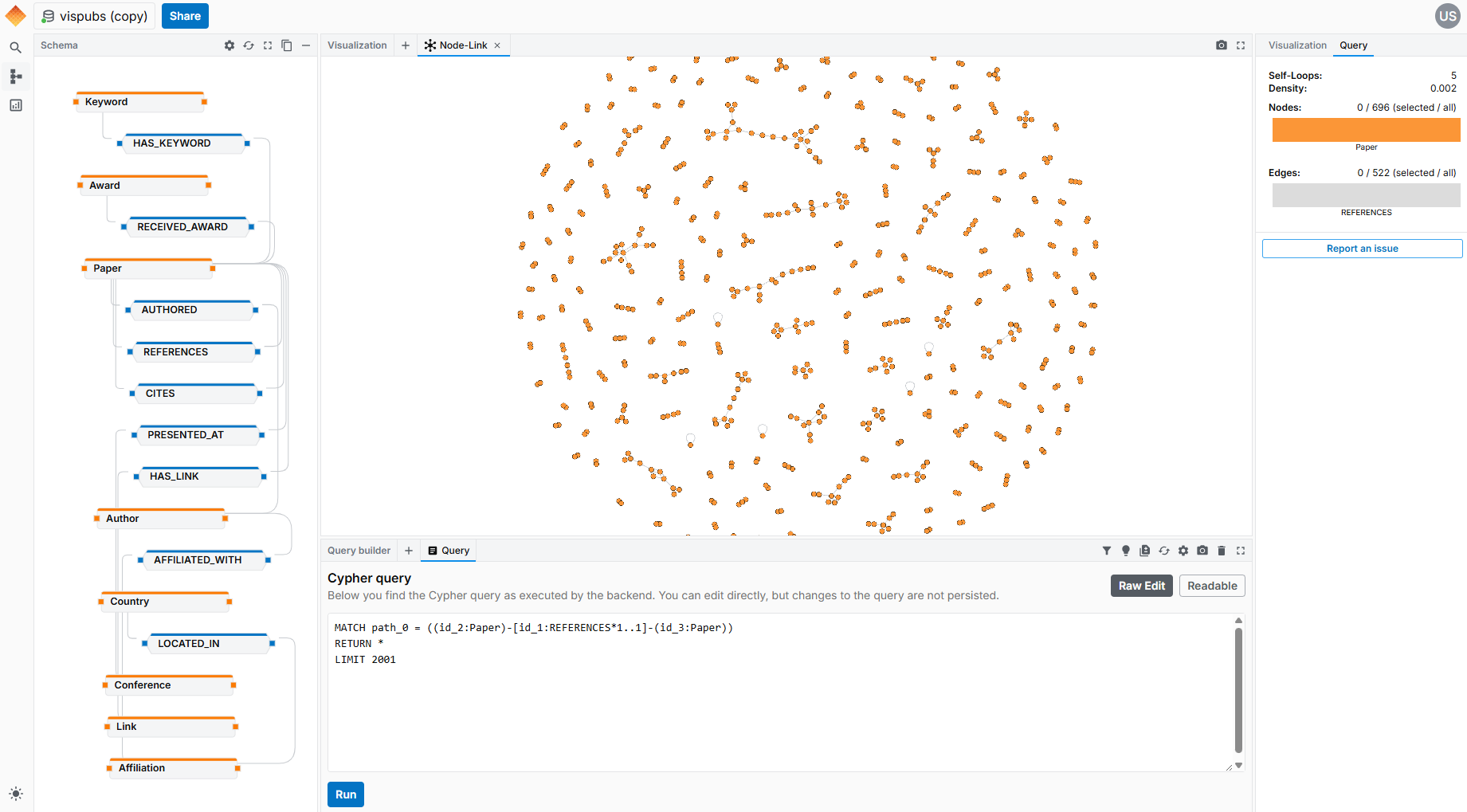}
    \end{tcolorbox}
    \caption{\textit{To what extent do Vis papers cite each other?} Generally, we see that Vis papers mostly reference one other Vis paper.}
\end{figure}

\begin{figure}[H]
    \centering
    \begin{tcolorbox}[enhanced, drop fuzzy shadow southeast, 
                      boxrule=0.4pt, sharp corners, 
                      colframe=darkgray, colback=white, 
                      width=1.0\linewidth]
        \includegraphics[width=1\linewidth]{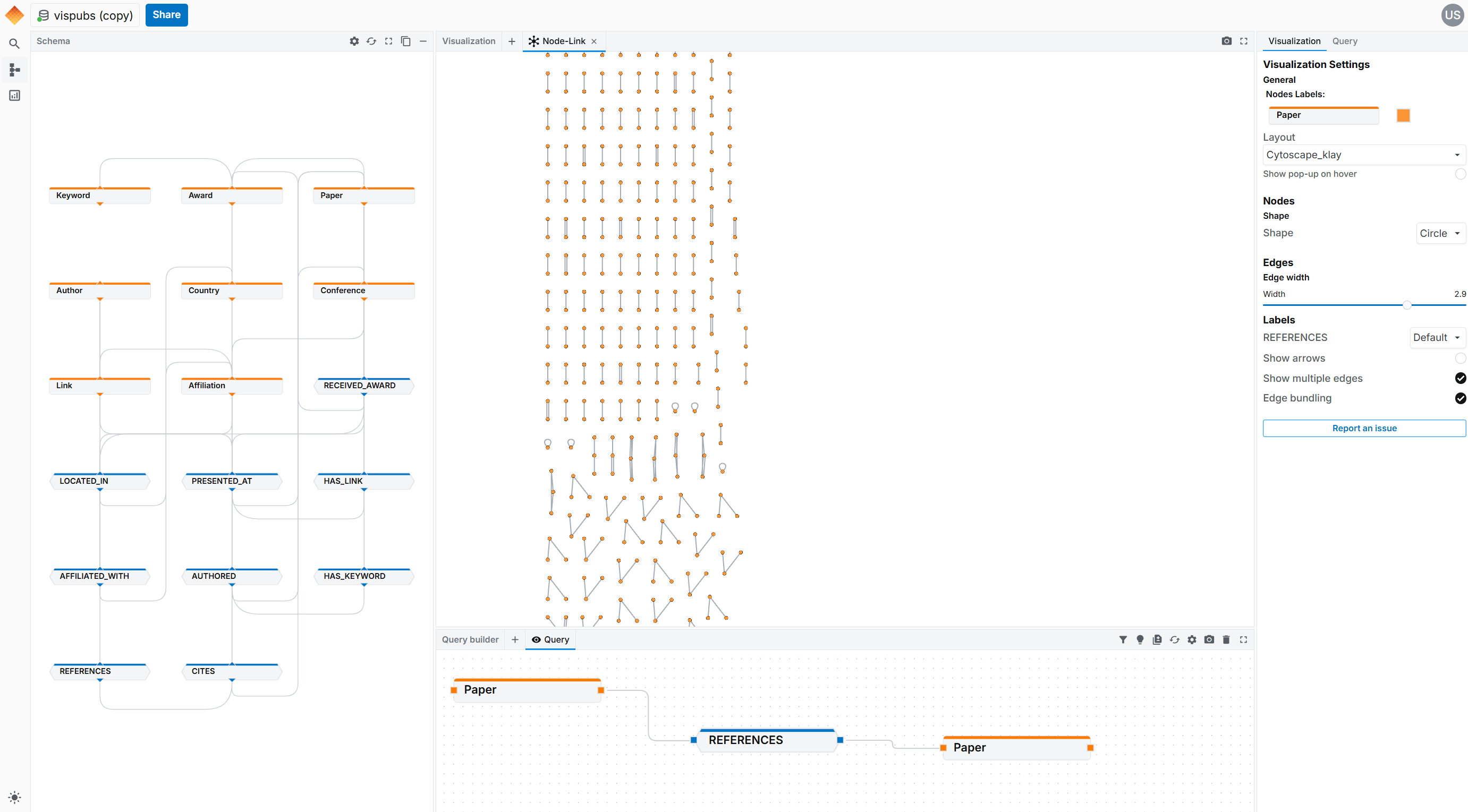}
    \end{tcolorbox}
    \caption{\textit{Follow-Up: To what extent do Vis papers cite each other?} Changing the layout algorithm to \texttt{Cytoscape\_klay} lets us see the graph motifs more clearly.}
\end{figure}

\begin{figure}[H]
    \centering
    \begin{tcolorbox}[enhanced, drop fuzzy shadow southeast, 
                      boxrule=0.4pt, sharp corners, 
                      colframe=darkgray, colback=white, 
                      width=0.9\linewidth]
        \includegraphics[width=1\linewidth]{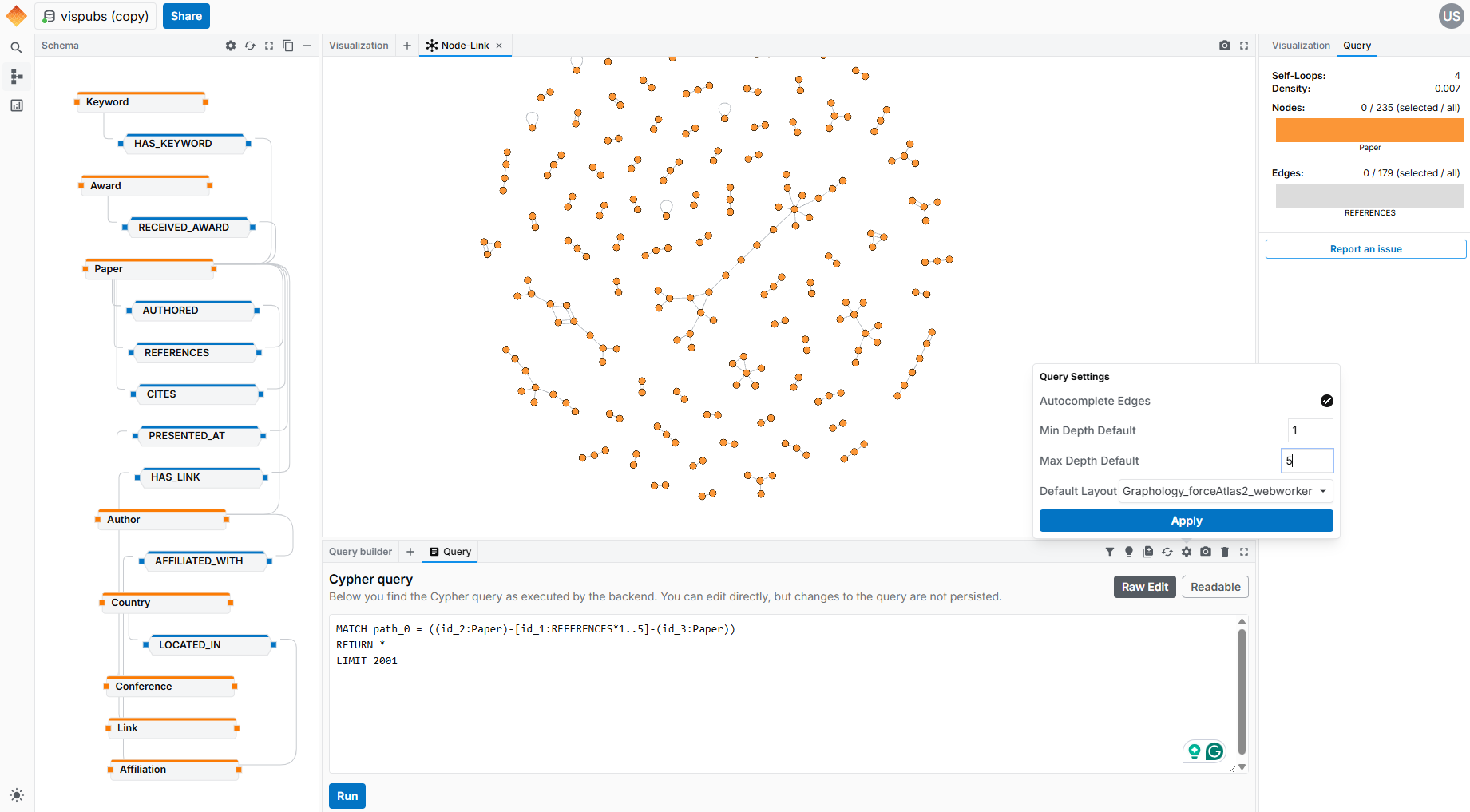}
    \end{tcolorbox}
    \caption{\textit{Follow-up: To what extent do Vis papers cite each other? } Let us make sure the Cypher query is correct, by switching into manual query mode and adapting the query to retrieve multi-hop recursive references. However, the results do not change. Interestingly, we see self-loops here. After a closer investigation outside of GraphPolaris, we found that these papers oftentimes have \texttt{osf.io} links to their supplementary materials in the core paper.}
\end{figure}

\subsubsection{Step 2: Pattern Exploration Question: "What have the cross-referencing papers in common?"}

\begin{figure}[H]
    \centering
    \begin{tcolorbox}[enhanced, drop fuzzy shadow southeast, 
                      boxrule=0.4pt, sharp corners, 
                      colframe=darkgray, colback=white, 
                      width=0.85\linewidth]
        \includegraphics[width=1\linewidth]{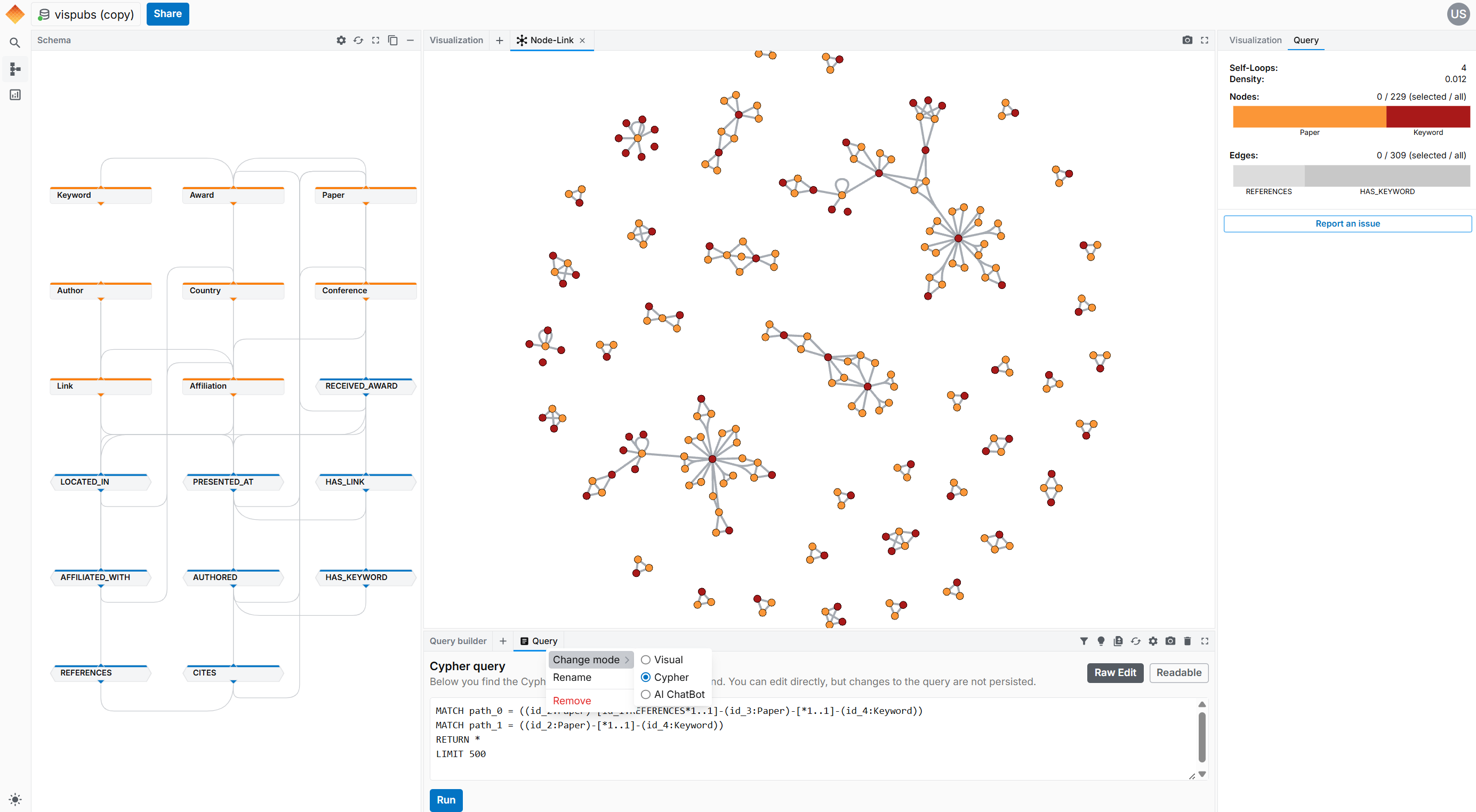}
    \end{tcolorbox}
    \caption{\textit{Why do these papers cite each other?} Let us add the keyword pill as a proxy for describing the paper's topic and link it both times to the papers. By switching over to the textual Cypher query input panel, we can validate that this interaction produces a cypher query, which searches for papers sharing the same Keyword \texttt{id\_4}.}
\end{figure}

\begin{figure}[H]
    \centering
    \begin{tcolorbox}[enhanced, drop fuzzy shadow southeast, 
                      boxrule=0.4pt, sharp corners, 
                      colframe=darkgray, colback=white, 
                      width=0.5\linewidth]
        \includegraphics[width=1\linewidth]{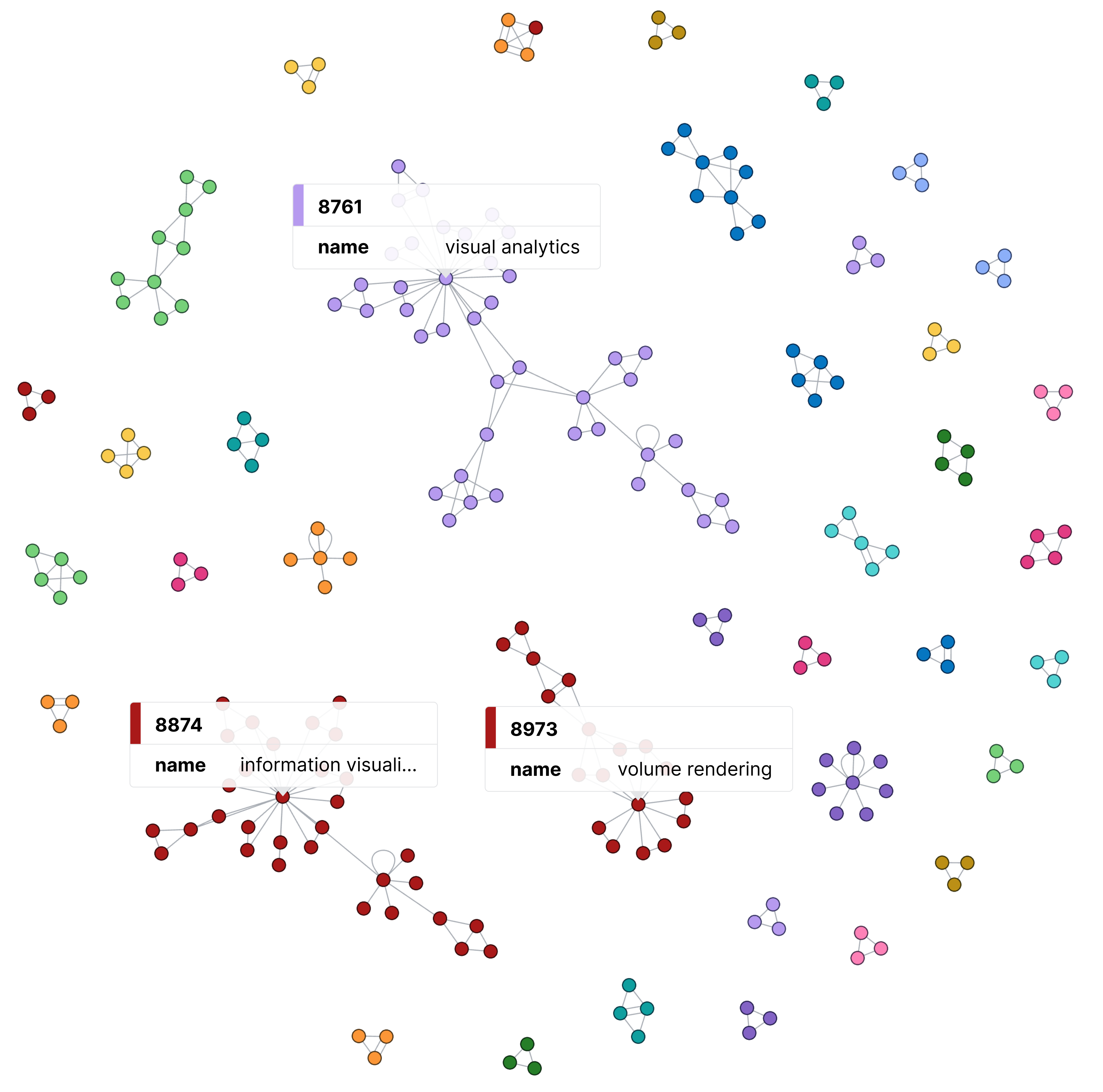}
    \end{tcolorbox}
    \caption{This figure shows the result of executing the query including the machine-learning meta pill depicted in \autoref{fig:community_keywords}. 
Smaller dyads in the visualization highlight tightly coupled work within niche domains, while the three larger components suggest broader thematic clusters.
Semantically, these three large components align with established areas in the visualization literature. 
The lower-left component is characterized by keywords such as visual analytics, dimensionality reduction, and principal component analysis. 
The right-hand cluster is centered around information visualization, graphics, and spreadsheets. 
Finally, the upper-left region contains papers focused on volume rendering, flow visualization, and non-photorealistic rendering. 
}
\end{figure}

\subsection{Frequent Publishers Getting Awards}
The final showcase explores academic recognition by analyzing which authors have consistently published award-winning work. In this query, we use a \includegraphics[height=0.75em]{figs/meta.pdf} \textit{pill} containing a degree predicate to select researchers associated with more than three papers that received awards. This query is illustrated in \autoref{fig:teaser}. The corresponding visualization shows a subgraph centered on nine prolific authors who collectively received 32 Honorable Mentions at IEEE VIS. In the node-link diagram, the purple node is an award, the red nodes represent papers, and the orange nodes represent authors. Edges denote authorship links to the award-winning papers.

\subsubsection{Step 1: Initial Exploration Question: "How many papers and corresponding authors received awards at VIS?"}

\begin{figure}[H]
    \centering
    \begin{tcolorbox}[enhanced, drop fuzzy shadow southeast, 
                      boxrule=0.4pt, sharp corners, 
                      colframe=darkgray, colback=white, 
                      width=0.85\linewidth]
        \includegraphics[width=1\linewidth]{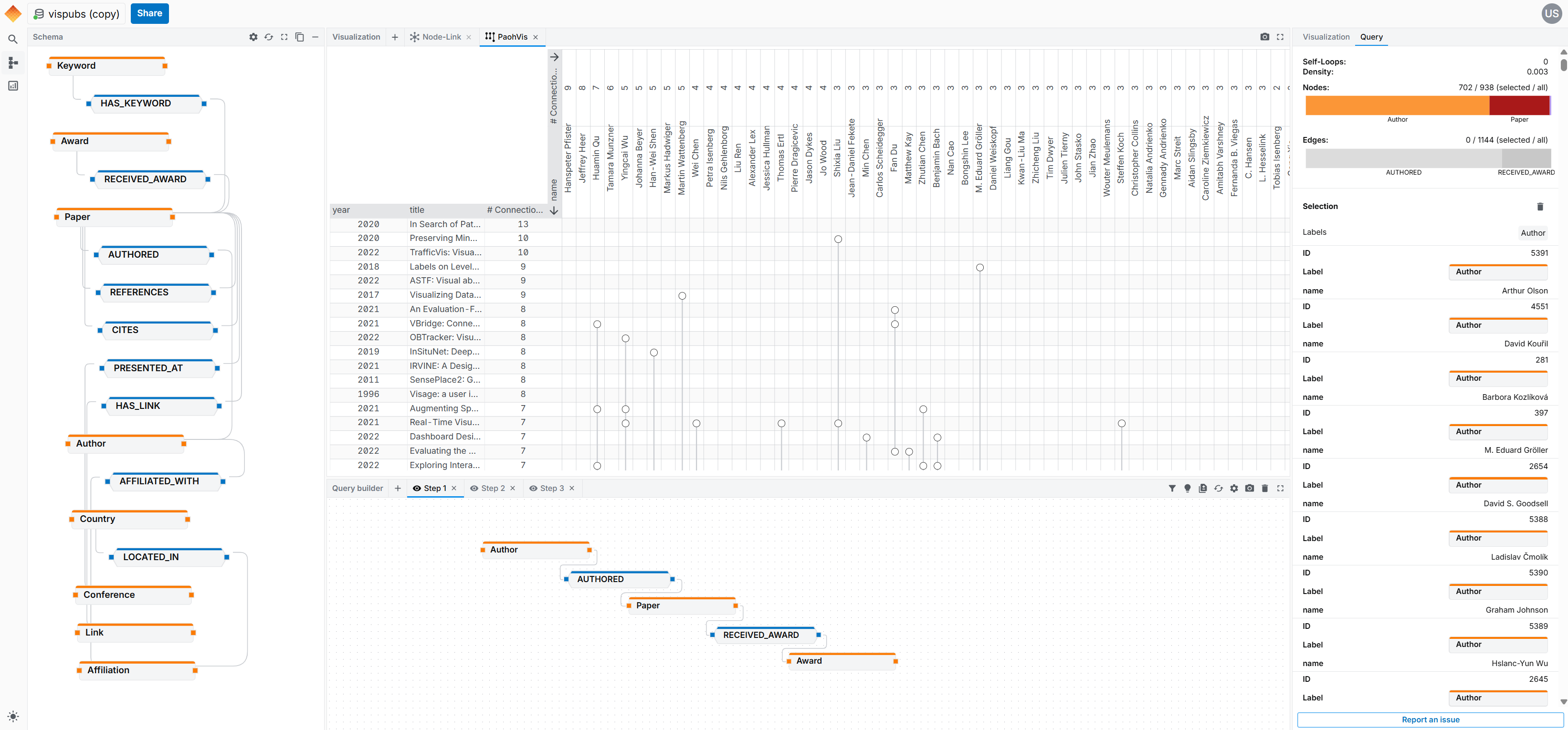}
    \end{tcolorbox}
    \caption{\textit{How many authors got awards? How many papers were recognized?} Our exploration starts with querying all authors that published papers, which received an award. The distribution plot on the right shows that in total 702 authors received a paper recognition at IEEE VIS. In total, 230 papers have been voted BP = best paper, HM = best paper honorable mention (= like a runner-up), TT = test of time award (=an award for a paper that has proved to be influential to the community over time), BA = best application paper award (in the early years of the conference), BCS = best case study award (in early years of the conference).}
\end{figure}

\begin{figure}[H]
    \centering
    \begin{tcolorbox}[enhanced, drop fuzzy shadow southeast, 
                      boxrule=0.4pt, sharp corners, 
                      colframe=darkgray, colback=white, 
                      width=0.85\linewidth]
        \includegraphics[width=1\linewidth]{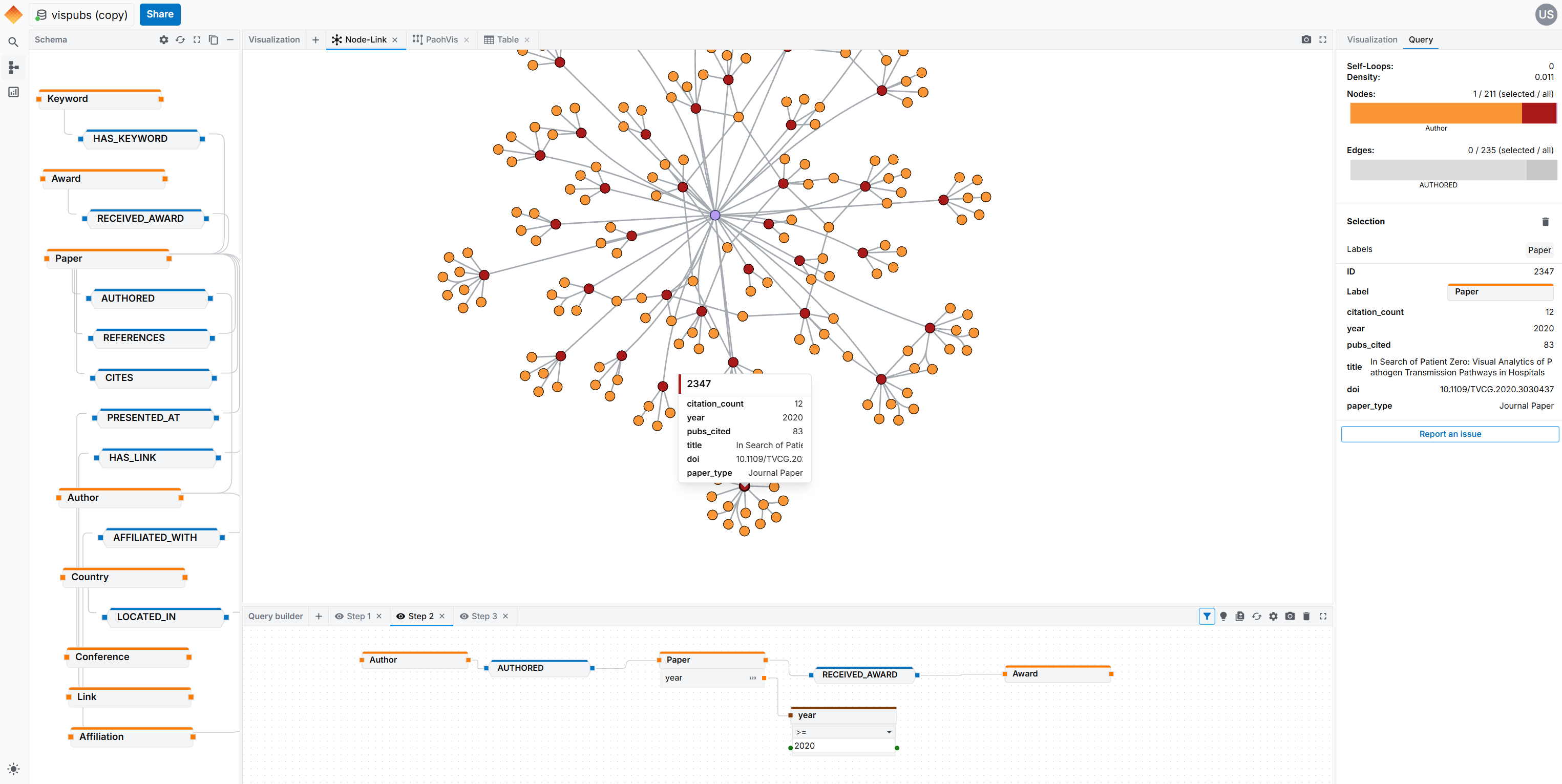}
    \end{tcolorbox}
    \caption{\textit{Follow-Up: How many authors got awards? How many papers were recognized?} Interestingly, the paper "\textit{In Search of Patient Zero: Visual Analytics of Pathogen Transmission Pathways in Hospitals}" (DOI: 10.1109/TVCG.2020.3030437; awarded at VAST 2020) was by far the paper with the most (13) co-authors. All of the authors have received exactly one award at VIS. On the other hand, ten awarded papers have a single author.}
\end{figure}

\subsubsection{Step 2: Refining the Query: "Did the Corona time influence successful authors? Restricting papers between 2020 and 2022."}

\begin{figure}[H]
    \centering
    \begin{tcolorbox}[enhanced, drop fuzzy shadow southeast, 
                      boxrule=0.4pt, sharp corners, 
                      colframe=darkgray, colback=white, 
                      width=1.0\linewidth]
        \includegraphics[width=1\linewidth]{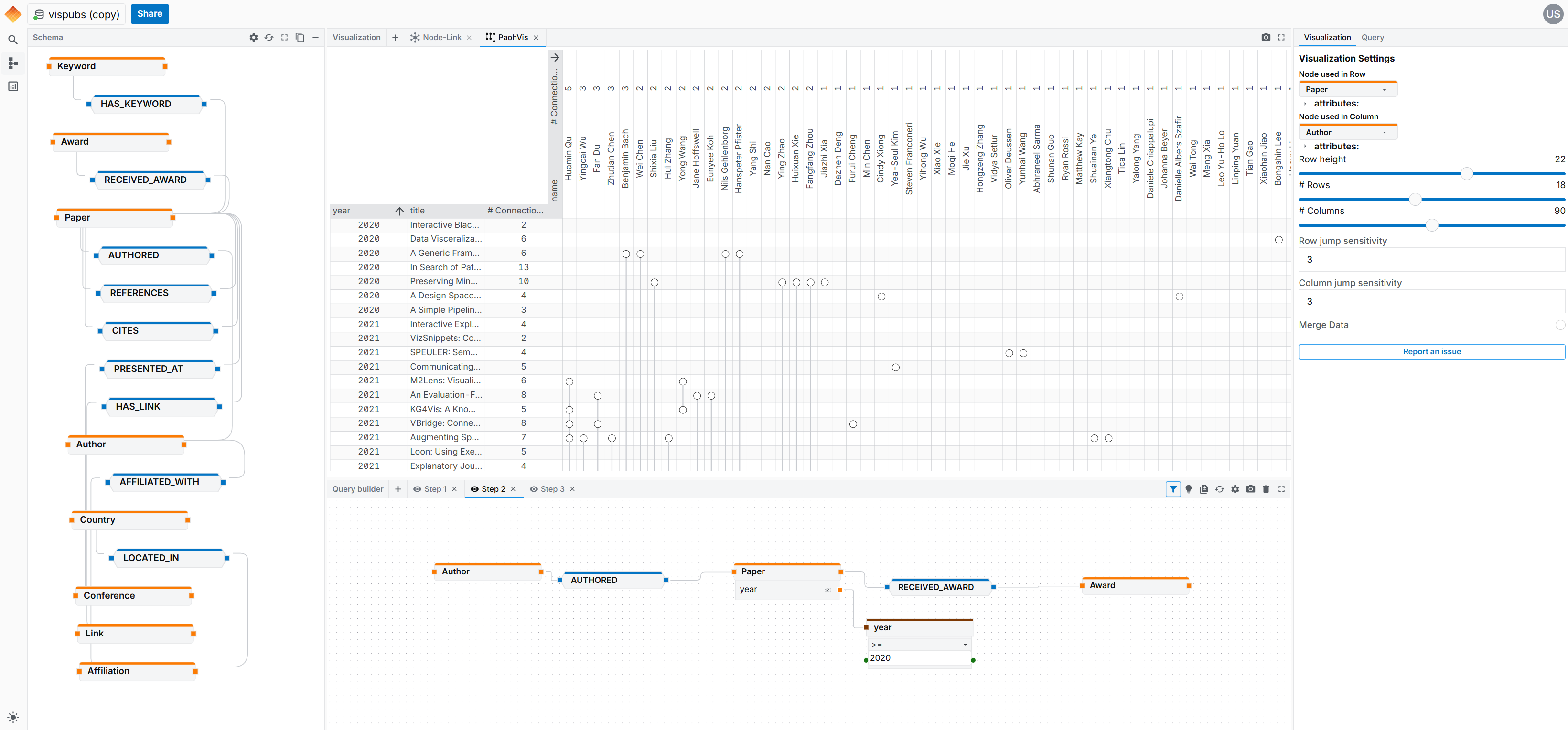}
    \end{tcolorbox}
    \caption{\textit{Can we recognize a pattern between 2020 and 2022?} To filter for award papers from the same author after 2020 and before 2022 we have to make two disjoint filters on the same paper. One for award papers after 2020 ...}
\end{figure}

\begin{figure}[H]
    \centering
    \begin{tcolorbox}[enhanced, drop fuzzy shadow southeast, 
                      boxrule=0.4pt, sharp corners, 
                      colframe=darkgray, colback=white, 
                      width=1.0\linewidth]
        \includegraphics[width=1\linewidth]{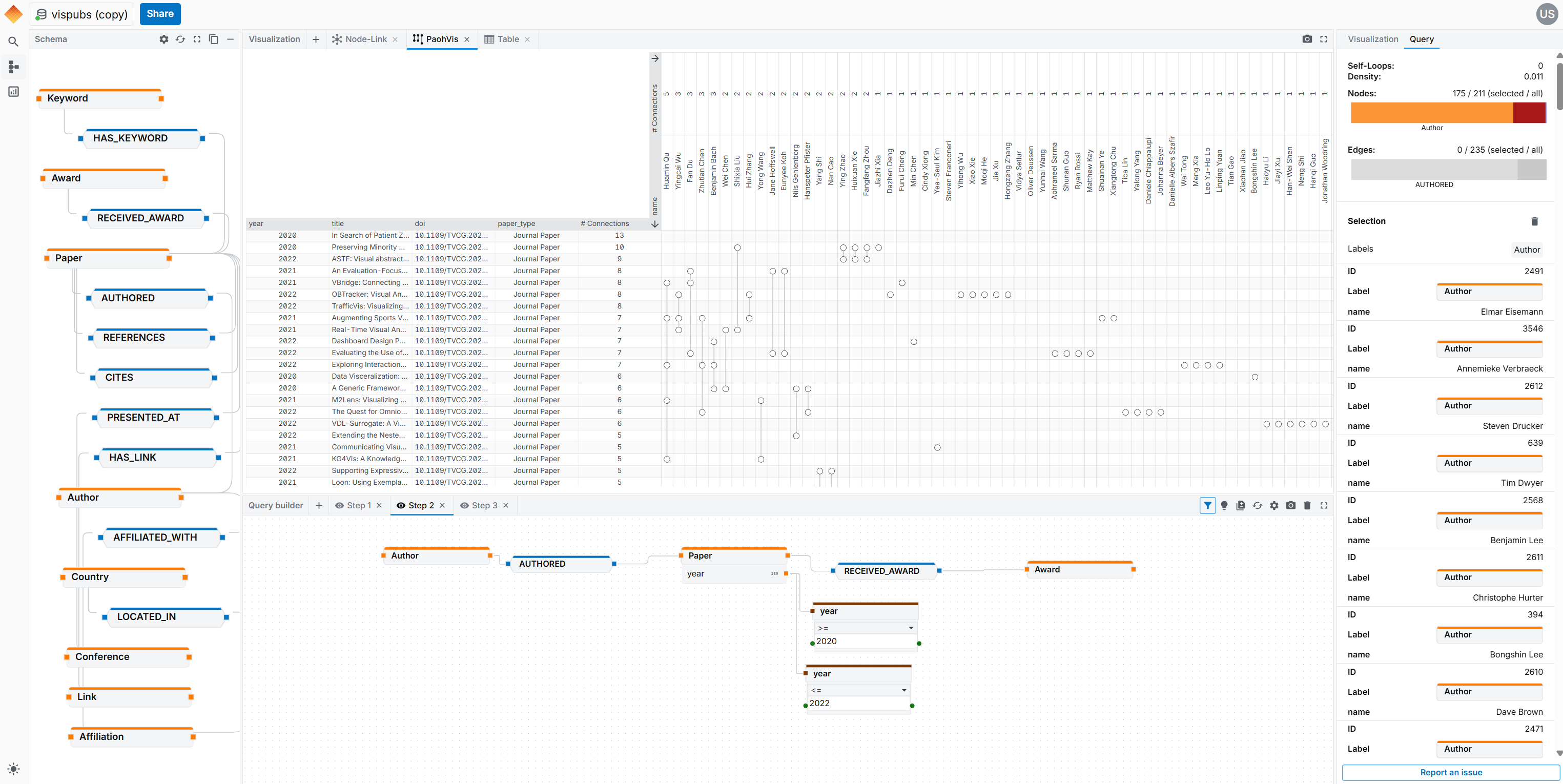}
    \end{tcolorbox}
    \caption{\textit{Follow-Up: Can we recognize a pattern between 2020 and 2022?} ... and another filter restricting this path's papers until the year 2022. Overall, we can see that with this filter setting the most recognized author is \textit{Huamin Qu} with five awards. }
\end{figure}

\subsubsection{Step 3: Drilling into the results: "Which awards did the most recognized authors win?"}

\begin{figure}[H]
    \centering
    \begin{tcolorbox}[enhanced, drop fuzzy shadow southeast, 
                      boxrule=0.4pt, sharp corners, 
                      colframe=darkgray, colback=white, 
                      width=1.0\linewidth]
        \includegraphics[width=1\linewidth]{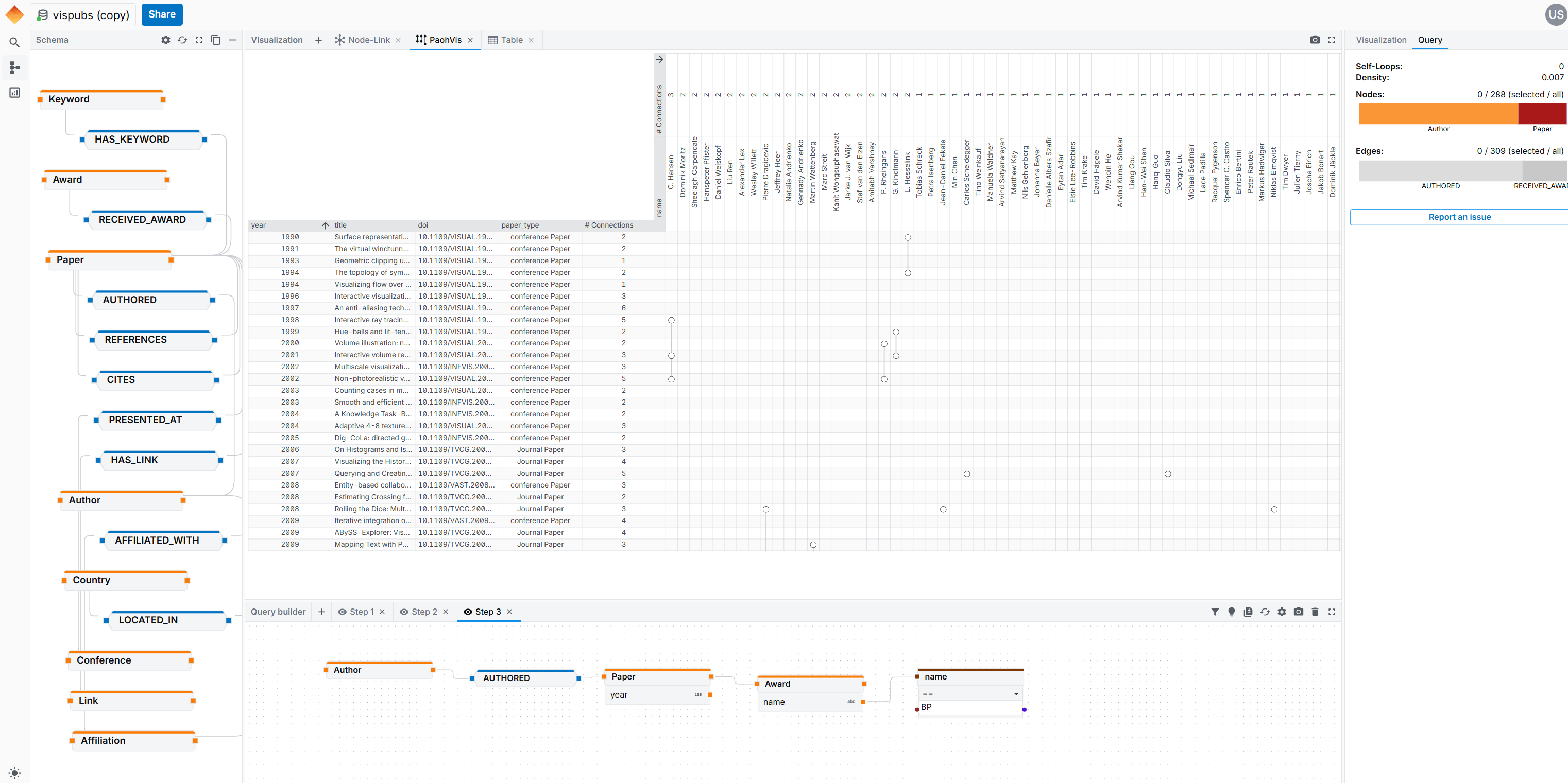}
    \end{tcolorbox}
    \caption{\textit{Which awards did the most recognized authors win?} To drill into the paper award type, we can filter by award name. Here 'BP' filters only by best papers awards. We can see a different set of top performers than before, with \textit{Chuck Hansen} having received three best paper awards (1998, 2001, 2002).}
\end{figure}

\begin{figure}[H]
    \centering
    \begin{tcolorbox}[enhanced, drop fuzzy shadow southeast, 
                      boxrule=0.4pt, sharp corners, 
                      colframe=darkgray, colback=white, 
                      width=1.0\linewidth]
        \includegraphics[width=1\linewidth]{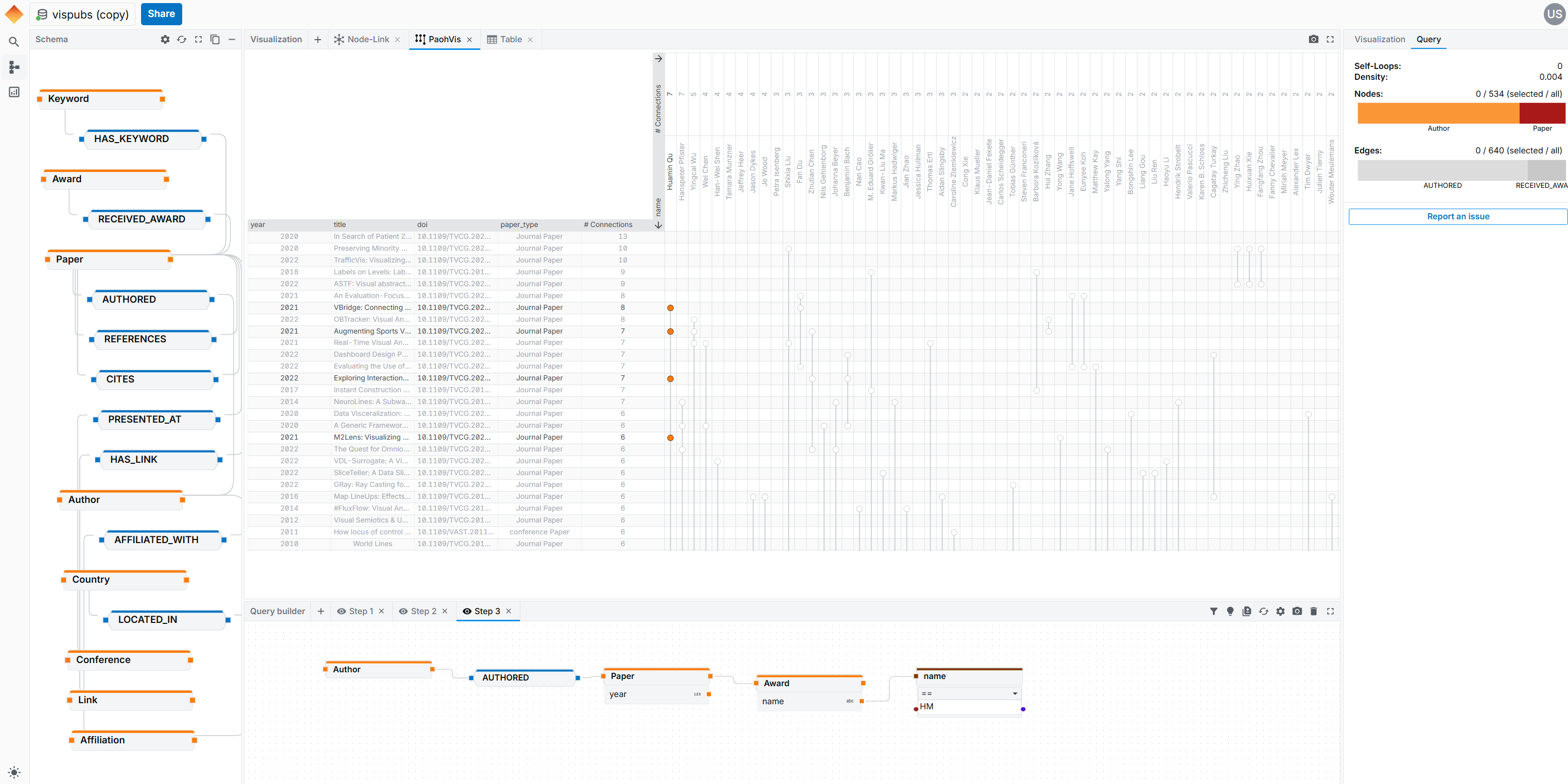}
    \end{tcolorbox}
    \caption{\textit{Follow-Up: Which awards did the most recognized authors win?} 
The analysis finds notable clusters of collaboration: some researchers co-authored multiple award-winning papers, suggesting a strong correlation between collaboration and impact. Huamin Qu and Hanspeter Pfister emerge as the most frequently awarded individuals in the dataset, each associated with seven Honorable Mentions. This query highlights how graph-based analysis can uncover patterns of academic productivity, influence, and recognition.}
\end{figure}

\section{Evaluating GPQL Expressiveness Using Benchmark Queries} \label{appendix:query_benchmark}

To evaluate the expressive power of \GPQL, we applied it to the LDBC Social Network Benchmark~\cite{angles2020ldbc}, which contains a set of representative queries over social graph data. For each query, we attempted to construct a semantically equivalent \GPQL. Below, we list all queries and their corresponding \GPQL translations. Despite covering a broad range of tasks, \GPQL currently lacks the expressive constructs required for a small subset of benchmark queries. Specifically:

\begin{itemize}
    \item \textbf{IC 3 — Friends and friends of friends that have been to given countries}: Not expressible due to the need for \textit{negation}, i.e., filtering based on the absence of an edge (e.g., people who have not been to certain countries).
    \item \textbf{IC 4 — New topics}: Not expressible due to the need for \textit{negation or set difference}, such as identifying posts with tags that have not previously appeared in the user’s neighborhood.
    \item \textbf{IC 7 — Recent likers}: Not expressible due to the lack of support for \textit{optional edges}, i.e., relationships that may or may not exist and should not constrain the result.
\end{itemize}

\subsection{Benchmark Queries to GPQL}

\subsubsection*{IC 1 - Transitive friends with a certain name}

{\ttfamily
(Person$_{1}$[id=input] KNOWS$_{1}$ Person$_{2}$[firstname=input]) (Person$_{2}$ LOCATED$_{2}$ City$_{3}$) (Person$_{2}$ WORKS$_{3}$ Company$_{4}$) (Company$_{4}$ LOCATED$_{4}$ Country$_{5}$) (Person$_{2}$ STUDIES$_{5}$ University$_{6}$) (University$_{6}$ LOCATED$_{6}$ City$_{7}$) (Person$_{1}$ KNOWS$_{7}$ Person$_{2}$) (Person$_{2}$ KNOWS$_{8}$ Person$_{3}$[firstname=input]) (Person$_{3}$ LOCATED$_{9}$ City$_{4}$) (Person$_{3}$ WORKS$_{10}$ Company$_{5}$) (Company$_{5}$ LOCATED$_{11}$ Country$_{6}$) (Person$_{3}$ STUDIES$_{12}$ University$_{7}$) (University$_{7}$ LOCATED$_{13}$ City$_{8}$) (Person$_{2}$ KNOWS$_{14}$ Person$_{3}$) (Person$_{3}$ KNOWS$_{15}$ Person$_{4}$[firstname=input]) (Person$_{4}$ LOCATED$_{16}$ City$_{5}$) (Person$_{4}$ WORKS$_{17}$ Company$_{6}$) (Company$_{6}$ LOCATED$_{18}$ Country$_{7}$) (Person$_{4}$ STUDIES$_{19}$ University$_{8}$) (University$_{8}$ LOCATED$_{20}$ City$_{9}$)
}

\subsubsection*{IC 2 - Recent messages by your friends}

{\ttfamily
(Person$_{1}$[id=input] KNOWS$_{1}$ Person$_{2}$) (Person$_{2}$ CREATED$_{2}$ Message$_{3}$[creation\_date < input])
}

\subsubsection*{IC 3 – Friends and friends of friends that have been to given countries}

{\ttfamily
(Person$_{1}$[id=input] KNOWS$_{1}$ Person$_{2}$) (Person$_{2}$ CREATES$_{2}$ Message$_{3}$[start\_date <= input])[Message$_{3}$.degree = input] (Person$_{2}$ CREATES$_{3}$ Message$_{4}$[start\_date <= input])[Message$_{4}$.degree = input] (Person$_{2}$ LOCATED$_{4}$ City$_{5}$) (Message$_{3}$ LOCATED$_{5}$ Country$_{7}$) (Message$_{4}$ LOCATED$_{6}$ Country$_{8}$) (Person$_{1}$[id=input\_id] KNOWS$_{7}$ Person$_{2}$) (Person$_{2}$ KNOWS$_{8}$ Person$_{3}$) (Person$_{3}$ CREATES$_{9}$ Message$_{4}$[start\_date <= input]) (Person$_{3}$ CREATES$_{10}$ Message$_{5}$[start\_date <= input]) (Person$_{3}$ LOCATED$_{11}$ City$_{6}$) (Message$_{4}$ LOCATED$_{12}$ Country$_{7}$) (Message$_{5}$ LOCATED$_{13}$ Country$_{8}$)
}

\subsubsection*{IC 4 – New topics}

{\ttfamily
(Person$_{1}$ KNOWS$_{1}$ Person$_{2}$[id=input]) (Person$_{2}$ KNOWS$_{2}$ Person$_{3}$) (Person$_{1}$ CREATES$_{3}$ Post$_{4}$[creation\_date = input]) (Person$_{3}$ CREATES$_{4}$ Post$_{5}$[creation\_date = input])[Post$_{5}$.degree = input] (Post$_{5}$ HAS$_{5}$ Tag$_{6}$)
}

\subsubsection*{IC 5 – New groups}

{\ttfamily
(Person$_{1}$[id=input] KNOWS$_{1}$ Person$_{2}$) (Person$_{2}$ HAS$_{2}$[min\_date = input] Forum$_{3}$) (Forum$_{3}$ CONTAINER$_{3}$ Post$_{4}$) (Person$_{2}$ CREATES$_{4}$ Post$_{4}$) (Person$_{1}$[id=input] KNOWS$_{5}$ Person$_{2}$) (Person$_{2}$ KNOWS$_{6}$ Person$_{3}$) (Person$_{3}$ HAS$_{7}$[min\_date = input] Forum$_{3}$) (Forum$_{3}$ CONTAINER$_{8}$ Post$_{4}$) (Person$_{3}$ CREATES$_{9}$ Post$_{4}$)
}

\subsubsection*{IC 6 – Tag co-occurence}

{\ttfamily
(Person$_{1}$ KNOWS$_{1}$ Person$_{2}$) (Post$_{3}$ HAS\_CREATOR$_{2}$ Person$_{2}$) (Post$_{3}$ HAS\_TAG$_{3}$ Tag$_{4}$[name = input]) (Post$_{3}$ HAS\_TAG$_{4}$ Tag$_{5}$[name $\neq$ input]) (Person$_{1}$ KNOWS$_{5}$ Person$_{2}$) (Person$_{2}$ KNOWS$_{6}$ Person$_{3}$) (Post$_{4}$ HAS\_CREATOR$_{7}$ Person$_{2}$) (Post$_{4}$ HAS\_TAG$_{8}$ Tag$_{5}$[name = input]) (Post$_{4}$ HAS\_TAG$_{9}$ Tag$_{6}$[name $\neq$ input])
}

\subsubsection*{IC 7 – Recent likers}

{\ttfamily
(Person$_{1}$[id = input] KNOWS$_{1}$ Person$_{2}$) (Person$_{2}$ LIKES$_{2}$ Message$_{3}$) (Message$_{3}$ HAS\_CREATOR$_{3}$ Person$_{1}$)
}

\subsubsection*{IC 8 – Recent replies}

{\ttfamily
(Message$_{1}$ HAS\_CREATOR$_{1}$ Person$_{2}$[id = input]) (Message$_{1}$ REPLY\_OF$_{2}$ Comment$_{3}$) (Comment$_{3}$ HAS\_CREATOR$_{3}$ Person$_{4}$)
}

\subsubsection*{IC 9 – Recent messages by friends or friends of friends}

{\ttfamily
(Person$_{1}$[id = input] KNOWS$_{1}$ Person$_{2}$) (Message$_{3}$[date $\leq$ input] HAS\_CREATOR$_{2}$ Person$_{2}$) (Person$_{1}$[id = input]  KNOWS$_{3}$ Person$_{2}$) (Person$_{2}$ KNOWS$_{4}$ Person$_{3}$) (Message$_{4}$[date $\leq$ input]  HAS\_CREATOR$_{5}$ Person$_{3}$)
}

\subsubsection*{IC 10 – Friend recommendation}

{\ttfamily
(Person$_{1}$[id = input] KNOWS$_{1}$ Person$_{2}$) (Person$_{2}$ KNOWS$_{2}$ Person$_{3}$[month = input and day $\geq$ input, or month = input + 1 and day $\textless$ input]) (Person$_{3}$ LOCATED\_IN$_{3}$ City$_{4}$) (Person$_{5}$ HAS\_INTEREST$_{4}$ Tag$_{6}$) (Post$_{7}$ HAS\_TAG$_{}5$ Tag$_{6}$) (Post$_{7}$ HAS\_CREATOR$_{6}$ Person$_{3}$))
}

\subsubsection*{IC 11 – Job referral}

{\ttfamily
(Person$_{1}$[id = input] KNOWS$_{1}$ Person$_{2}$) (Person$_{2}$  WORKS\_AT$_{2}$[year $\textless$ input] Company$_{3}$) (Company$_{3}$ IS\_LOCATED$_{3}$ Country$_{4}$[name = input]) (Person$_{1}$[id = input] KNOWS Person$_{2}$) (Person$_{2}$ KNOWS Person$_{3}$) (Person$_{3}$  WORKS\_AT$_{3}$[year $\textless$ input] Company$_{4}$) (Company$_{4}$ IS\_LOCATED$_{4}$ Country$_{5}$[name = input])
}

\subsubsection*{IC 12 – Expert search}

{\ttfamily
(Person$_{1}$[id = input] KNOWS$_{1}$ Person$_{2}$) (Comment$_{3}$ HAS\_CREATOR$_{2}$ Person$_{2}$) (Comment$_{3}$ REPLY\_OF$_{3}$ Post$_{4}$) (Post$_{4}$ HAS\_TAG$_{4}$ Tag$_{5}$) (Tag$_{5}$ HAS\_TYPE$_{5}$ TagClass$_{6}$) (TagClass$_{6}$ SUBCLASS$_{6}$ TagClass$_{7}$[name = input])
}

\subsubsection*{IC 13 – Single shortest path}

{\ttfamily
(Person$_{1}$ KNOWS$_{1}$ Person$_{2}$)[shortest\_path()]
}


\end{document}